\documentclass[review]{elsarticle}

\usepackage[colorlinks,citecolor=blue,linktoc=all,linkcolor=cyan]{hyperref}
\usepackage{graphicx}

\usepackage[T1]{fontenc}
\usepackage{dsfont}
\usepackage{mathrsfs}
\usepackage{slashed}
\usepackage{amsmath,amssymb,amsbsy,amsfonts}
\usepackage{bm}

\numberwithin{equation}{section}
\numberwithin{table}{section}
\numberwithin{figure}{section}

\newcommand{\ff}[2]{F^{*}_{#1}F_{#2}}

\usepackage{titlesec}
\usepackage{sectsty}
\titleformat{\section}{\normalfont\Large\bfseries}{\thesection}{1em}{}
\titleformat{\subsection}{\normalfont\large\bfseries}{\thesubsection}{1em}{}
\titleformat{\subsubsection}{\normalfont\normalsize\bfseries}{\thesubsubsection}{1em}{}
\usepackage{graphicx}

\journal{Progress in Particle and Nuclear Physics}

\begin{document}

\begin{frontmatter}

\title{Hadron photoproduction experiments in LEPS2}

\author[IMP]{N.~Muramatsu\corref{mycorrespondingauthor}}
\cortext[mycorrespondingauthor]{Corresponding author}
\ead{mura@impcas.ac.cn}
\author[RARiS]{A.~O.~Tokiyasu}
\author[RCNP]{T.~A.~Hashimoto}

\address[IMP]{Institute of Modern Physics, Chinese Academy of Sciences, Lanzhou, 730000, China}
\address[RARiS]{Research Center for Accelerator and Radioisotope Science, Tohoku University, Sendai, Miyagi 982-0826, Japan}
\address[RCNP]{Research Center for Nuclear Physics, The University of Osaka, Ibaraki, Osaka 567-0047, Japan}

\begin{abstract}
In the LEPS2 facility constructed at SPring-8, hadron photoproduction experiments have been carried out by using a photon beam with high linear polarization in the tagged energy range of $1.3$--$2.4$~$\mathrm{GeV}$. The BGOegg and Solenoid experiments are alternately running with large-acceptance detector setups, which have been prepared by focusing on the measurement of $\gamma$-rays and charged particles, respectively. These two projects have covered a wide range of hadron physics research, including the search for in-medium meson mass modification, the baryon resonance spectroscopy, the studies on exotic hadrons, and so on. This article reviews the physics programs, the obtained results, and the future prospects in the LEPS2 facility.
\end{abstract}


\end{frontmatter}

\newpage
\thispagestyle{empty}
\tableofcontents


\newpage
\section{Introduction}\label{sec:first}

   Understanding the low-energy quantum chromodynamics (QCD) is one of the modern physics subjects widely pursued for several decades. This subject is important to elucidate the origin of the universe with hadrons, where quarks are confined, while the non-perturbative nature of the low-energy QCD makes it difficult to quantitatively describe a variety of phenomena related to hadrons. For instance, quark models with two or three constituent quarks explain low-lying hadron properties, but it is also known that there are missing resonances and mass-reversed states in the current experimental data \cite{pdg, PPNP.125.103949, RMP.82.1095}. Moreover, exotic hadron candidates like tetra- and penta-quark states, hadronic molecules, and gluonic states have been found accumulatively. These facts suggest the rich structure of hadrons beyond the constituent quark models and recall the diversity of important degrees of freedom in strong interactions. In addition, the hadron mass cannot be explained by summing the bare masses of constituent quarks, which are described by the Higgs mechanism (e.g., $1$--$2$\% for a nucleon) \cite{pdg}. This problem raises a large question about the origin of hadron masses, which would have close connections with the hadron nature and structure. Due to the difficulties mentioned above, many efforts for the hadron physics research using various quantum beams in a wide range of energies have been required on the experimental side simultaneously with the supports of theoretical predictions and interpretations.

   In the progress of the research for the low-energy QCD, hadron photoproduction experiments are attractive because of the unique features of a $\gamma$-ray beam, as described below.
\begin{enumerate}
   \item A $\gamma$-ray beam can be polarized either linearly or circularly with relatively easy control. The $\gamma$-ray beam polarization is quite effective to provide new information in addition to unpolarized cross sections: spin-dependent asymmetries in the photoproduction of a hadron relevant to the exclusive reaction that is under investigation. For example, the angular asymmetries of hadron production using a linearly or circularly polarized beam help to decompose the reaction amplitudes for $s$-channel resonance contributions that interfere with each other, as well as $t$- and $u$-channel components. If enough number of different-type asymmetry data are collected, a so-called complete experiment \cite{PPNP.125.103949} can be carried out to solve those amplitudes \footnote{In the case of pseudoscalar meson photoproduction, the amplitudes can generally be solved by measuring 16 observables. However, the ways to reduce the number of necessary observables have been proposed as so-called complete experiments. For instance, Ref.~\cite{TPWA} discusses the selection of five observables to solve the amplitudes in a truncated partial wave analysis.}. In the case of vector meson photoproduction, the angular asymmetry of decay products with a linearly polarized beam differentiates a reaction mechanism in the $t$-channel diagram.

   \item A $\gamma$-ray beam is suitable for the excitation of a target nucleon to produce nucleon or Delta resonances ($N^*$ and $\Delta^*$, respectively). The spin-parity of decay products as well as their angular asymmetries mentioned above are sensitive to the properties of the $s$-channel baryon resonance.
   
   \item A photon couples with a quark and anti-quark pair. The reactions that occur with the pair production of $s\bar{s}$ can be produced with reasonably good signal-to-noise (S/N) ratios if the photon beam energy exceeds their production thresholds. This feature is obvious, for example, in $\phi$ photoproduction, although it is also affected by the absence of OZI suppression. In the case of open $s\bar{s}$ production, the detection of $K^+$ or $K^-$ enables to tag the strangeness of a counterpart baryon (namely, a hyperon or a possible pentaquark baryon). Neutral hyperons ($Y^0$s) can be analyzed with good S/N ratios and resolutions in the reaction $\gamma p \to K^+ Y^0$, enabling the use of a rest proton in a fixed target and the clean and precise measurement of a charged kaon in the final state.

   \item The quantum number of a photon is unique with the vector nature. As a typical feature of the photon quantum number, photoproduction of vector mesons in the $t$-channel becomes dominant at higher energies. In addition, photoproduced mesons can possess exotic quantum numbers that are impossible for a conventional combination of quark and anti-quark (e.g., $J^{PC} = 1^{-+}$). Thus, the detection of hybrid mesons, containing a gluonic component, is also expected in photoproduction experiments. 

   \item A $\gamma$-ray beam is attractive for investigating the in-medium nature of hadrons because the $\gamma$-rays reach and interact deeply inside a target nucleus. Thus, the distribution of reaction points can be considered to simply follow the nuclear density. In contrast, hadron beams tend to interact at the surface region of a nucleus. This feature originates from the low reaction rate of a $\gamma$-ray beam, so increasing the beam intensity is desired to accumulate high-statistics data at the same time.

   \item There is no hadronic contamination in a $\gamma$-ray beam, so that backgrounds due to unexpected hadron interactions can be suppressed. Only $e^+e^-$ conversions at materials on the beam path are contaminated, but they can be easily removed and identified by putting a dipole magnet and a thin plastic scintillator, respectively.

\end{enumerate}

\begin{table}[t]
  \label{photoproexp}
  \caption{Summary of the major facilities conducting hadron photoproduction experiments with a real photon beam in the recent three decades. Facility and collaboration names, operated period, $\gamma$-ray beam production method, tagged photon energies, utilized beam polarization, and typical reference are listed for each project.}
  \centering
  \begin{tabular}{llllll}
  \hline
  Facility / Collaboration & Operation & Method & $E_\gamma$ (GeV) & Polarization & Ref. \\
  \hline
  MAMI / A2        & 1991--     & brems. & 0.04--1.6  & linear / circular & \cite{PhysRevC.97.055212} \\
  ELSA / SAPHIR    & 1991--1999 & brems. & 0.5--2.6   & linear            & \cite{Schwille:1994vg} \\
  ELSA / CBELSA(/TAPS) & 2000-- & brems. & 0.3--3.2   & linear / circular & \cite{Elsner2009epja} \\
  ELSA / BGOOD     & 2013--     & brems. & 0.45-2.9   & linear            & \cite{Alef2020} \\
  JLab / CLAS      & 1998--2012 & brems. & 0.7--5.2   & linear / circular & \cite{SOBER2000263} \\
  JLab / GlueX     & 2016--     & brems. & $\sim$9    & linear            & \cite{ADHIKARI2021164807} \\
  RARiS\footnotemark / FOREST, NKS2 & 2002-- & brems. & 0.8--1.3 & unpolarized & \cite{ISHIKAWA20101, KANETA201888} \\
  BNL / LEGS       & 1990--2001 & LCS    & 0.21--0.33 & linear / circular & \cite{PhysRevC.64.025203} \\
  ESRF / GRAAL     & 1996--2008 & LCS    & 0.35--1.5  & linear            & \cite{BOCQUET1997c124} \\
  SPring-8 / LEPS  & 1999--     & LCS    & 1.5--2.4 / 2.9 & linear        & \cite{MURAMATSU2014184} \\
  SPring-8 / LEPS2 & 2014--     & LCS    & 1.3--2.4 / 2.9 & linear        & \cite{MURAMATSU2022166677} \\
  \hline
  \end{tabular}
\end{table}
\footnotetext{Its former name is ELPH.}
   Since the 1990s, hadron photoproduction experiments have begun in earnest around the world using high-intensity electron accelerators. Table~\ref{photoproexp} shows a list of the facilities for major hadron experiments using a real photon beam in the recent three decades. The features of $\gamma$-ray beams in those facilities are also summarized there. Half of the facilities produce a $\gamma$-ray beam via bremsstrahlung of the electron beam at a thin radiator. A linearly polarized beam is produced with a technique of coherent bremsstrahlung \cite{PhysRev.99.604, RMP.40.611} using a diamond radiator, whereas a circularly polarized beam is generated even with an amorphous radiator by making an electron beam longitudinally polarized. In the other half of the facilities, a $\gamma$-ray beam is produced by backward Compton scattering of laser light from a high-energy electron beam (Laser Compton Scattering, LCS) \cite{PhysRevLett.10.75, PhysLett.4.176}. This method can easily control the linear or circular polarization of a $\gamma$-ray beam by handling laser polarization with optical wave plates. In addition, it should be noted that the polarization degree of $\gamma$-rays in an LCS facility becomes higher than that in a coherent bremsstrahlung facility, particularly for the case of linear polarization as discussed later. LCS has another advantage in cost because such an experimental facility can be prepared at one beamline of the high-current electron storage ring that is commonly used for a wide area of research with synchrotron radiation. Instead, the available energy range of a produced $\gamma$-ray beam is highly dependent on the storage ring energy, so that the freedom to design experiments tends to be limited. Thus, only the LEPS and LEPS2 beamlines in SPring-8, which cover the highest energy range among the LCS facilities, remain under operation at present.

\begin{figure}[tbp]
  \centering
  \begin{tabular}{cc}
    \begin{minipage}{0.45\textwidth}
      \centering
      \includegraphics[width=7.5cm]{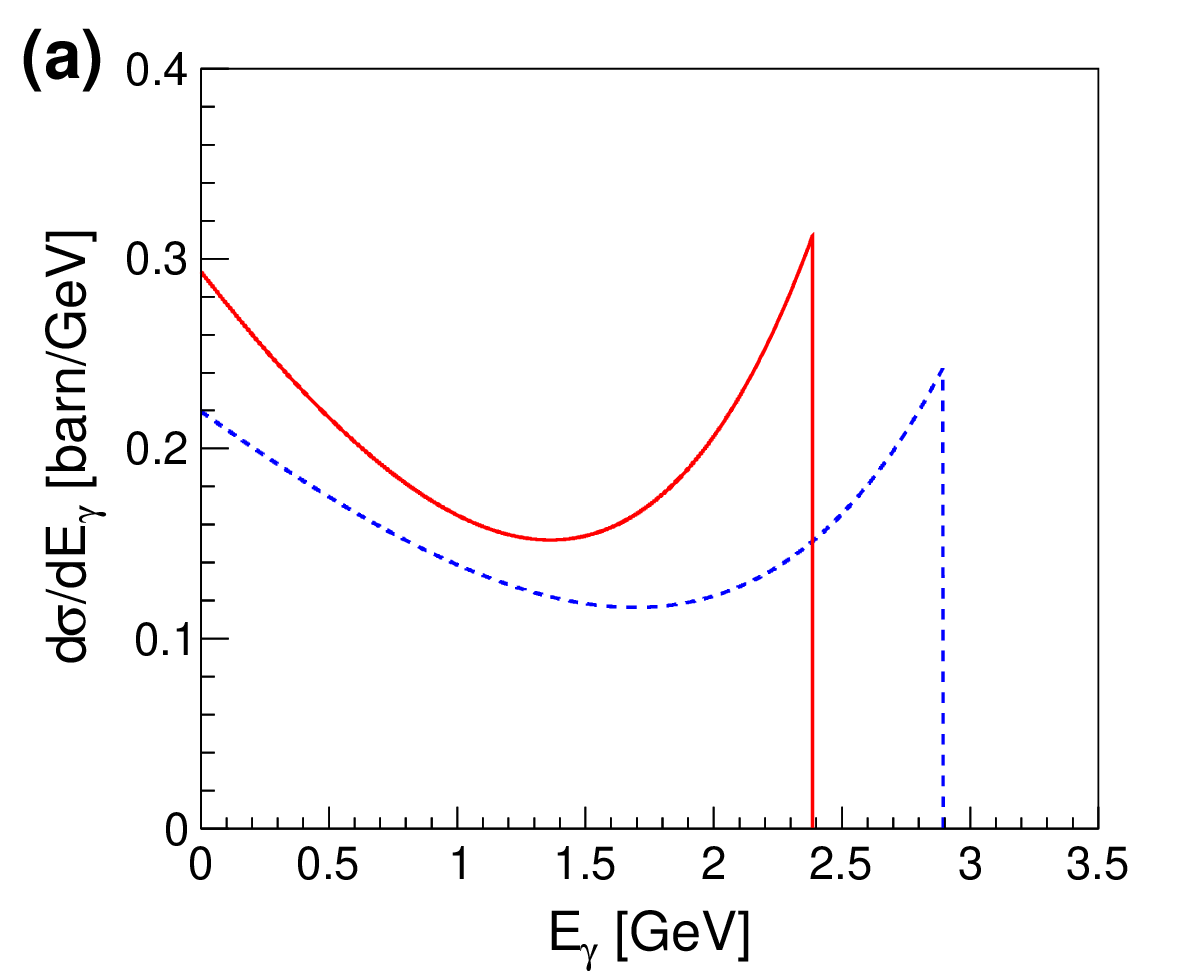}
    \end{minipage}
    \hspace{2mm}
    \begin{minipage}{0.45\textwidth}
      \centering
      \includegraphics[width=7.5cm]{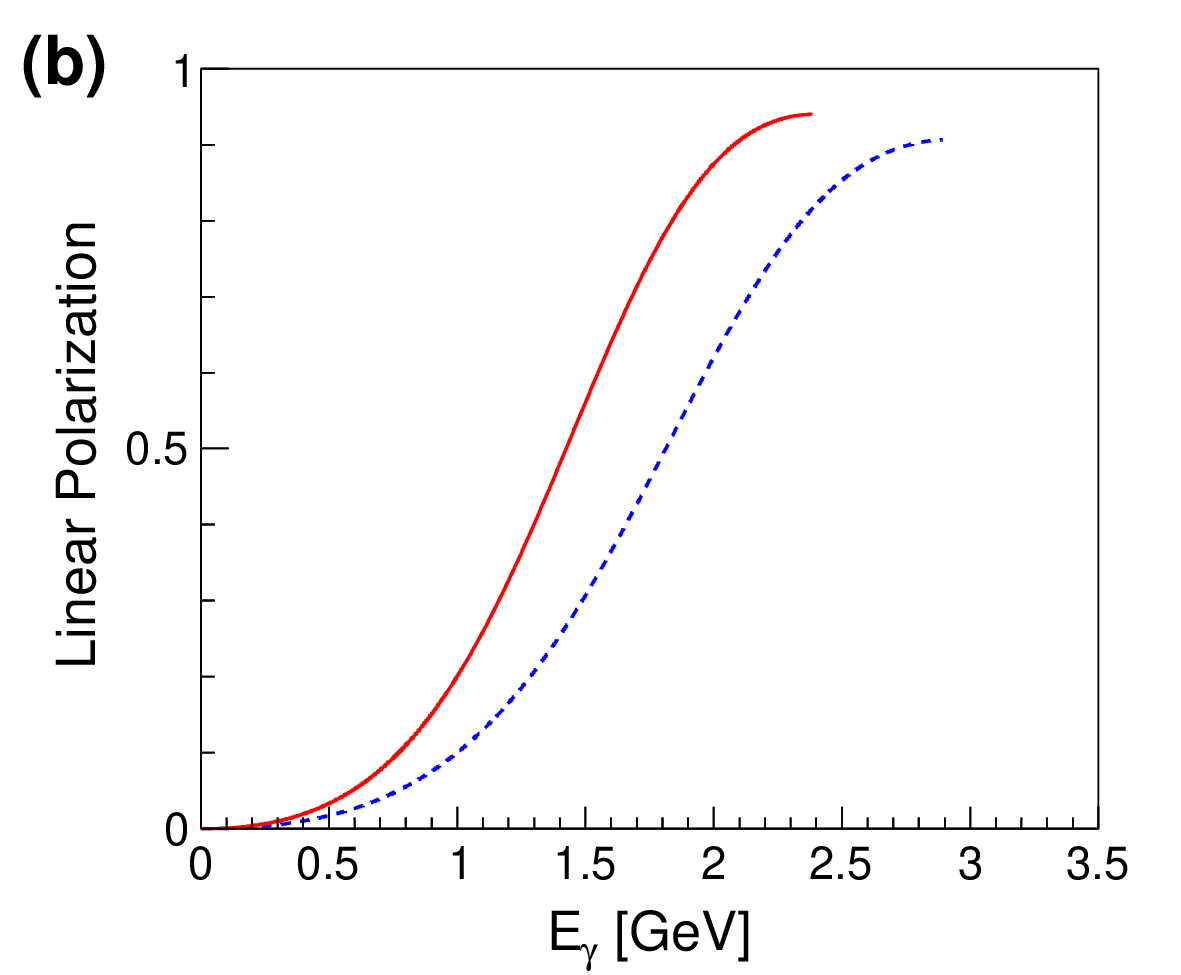}
    \end{minipage}
  \end{tabular}
  \caption{Properties of $\gamma$-ray beams available at LEPS2 beamline. Panel (a) shows the energy spectra calculated in the form of differential cross sections for laser Compton scattering, while Panel (b) indicates linear polarization degrees of $\gamma$-rays when the laser light is 100\% polarized. The red solid and blue dashed lines correspond to the injection of $355$ and $266$~$\mathrm{nm}$ wavelength laser light into SPring-8, respectively. Both quantities are displayed as a function of the $\gamma$-ray beam energy ($E_\gamma$).}
  \label{leps2prop}
\end{figure}

\begin{figure}[tbp]
  \centering
  \includegraphics[width=8.5cm]{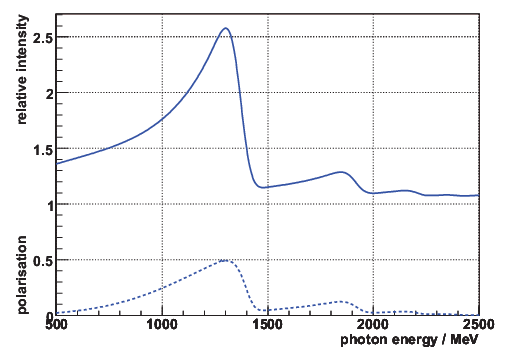}
  \caption{The dashed line shows the linear polarization of a $\gamma$-ray beam produced by coherent bremsstrahlung at the CBELSA/TAPS experiment. The solid line indicates the relative intensity of coherent to incoherent bremsstrahlung radiation. {\it Source}: Figure taken from Fig.~2 of Ref.~\cite{EPJA.33.147}.}
  \label{elsapol}
\end{figure}

   The LEPS2 facility, which is the subject of this review article, is the second $\gamma$-ray beamline using LCS in SPring-8. Since 2014, it has been operated as the main beamline for hadron photoproduction experiments instead of the first facility called the LEPS beamline. The new facility was constructed by aiming the increase of a $\gamma$-ray beam intensity with a new beamline design and the expansion of the experimental space to place large-acceptance detector systems, as described in detail later. In the LEPS2 facility, a $\gamma$-ray beam up to $2.4$ or $2.9$~$\mathrm{GeV}$ is obtained by injecting ultraviolet (UV) or deep-ultraviolet (DUV) laser light, respectively, into the $8$~$\mathrm{GeV}$ electron storage ring. Figure~\ref{leps2prop} shows the properties of $\gamma$-ray beams obtained in the LEPS2 beamline. As shown in the panel (a), the energy spectra calculated by leading-order quantum electrodynamics (QED) \cite{NIMA.455.1, PRSTAB.14.044701} distribute with relatively flat intensities (differential cross sections) up to the maximum energy or the Compton edge. This feature is one of the merits of LCS over the beam production via bremsstrahlung, whose energy spectrum increases considerably at lower energies, particularly in the untagged region that provides backgrounds. The panel (b) shows the linear polarization degrees of $\gamma$-ray beams obtained in the LEPS2 beamline when the injected laser light is assumed to be fully polarized. The high linear polarization, particularly at higher energies, is the biggest advantage of LCS facilities as recognized from the comparison with Fig.~\ref{elsapol}, showing the typical linear polarization of $\gamma$-rays produced by coherent bremsstrahlung\footnote{The energy region of a linearly polarized coherent bremsstrahlung beam is adjustable anywhere within the $20$--$80$\% range of an electron beam energy, but the maximum polarization drops as adjusted to higher energies. Actual physics data with high linear polarization are only available at the photon beam energies below $1.8$~$\mathrm{GeV}$ for coherent bremsstrahlung.}. Therefore, the polarization observables measured with a linearly polarized photon beam in the LEPS2 facility always provide the first-time results, as discussed throughout this article.


   The purpose of this article is a review of physics programs, obtained results, and future prospects in the two experiments running at the LEPS2 beamline. The following part is organized as follows: description of the facility including the beamline and the two experimental setups in Sec.~\ref{sec:second}, overview of the physics achievements, the on-going analyses, and the experimental status in Sec.~\ref{sec:third}, and progress of the development toward a next-generation facility in Sec.~\ref{sec:fourth}. Section~\ref{sec:third} discusses individual physics programs by focusing on the three categories of hadron physics: (1) research for the origin of hadron masses, investigating the possibility of $\eta^\prime$ meson mass reduction inside a nucleus (in-medium mass modification) with several complementary analysis methods, (2) baryon resonance spectroscopy, examining $s$-channel resonances in the two- or three-body photoproduction reactions of various mesons with different quantum numbers and masses, and (3) exotic hadron studies for the $f_0$(980) scalar meson, the $\bar{K}NN$ bound state, the $\Theta^+$ pentaquark, etc. Finally, a summary is described in Sec.~\ref{sec:99th}.

\newpage
\section{LEPS2 facility}\label{sec:second}

This section describes the LEPS2 facility, which contains the LEPS2 beamline (Sec.~\ref{subsec:second-first}), the BGOegg experimental setup (Sec.~\ref{subsec:second-second}), and the Solenoid experimental setup (Sec.~\ref{subsec:second-third}).

\subsection{LEPS2 beamline}\label{subsec:second-first}
\begin{figure}[htbp]
 \centering
 \begin{minipage}{1.0\textwidth}
  \includegraphics[width=18cm]{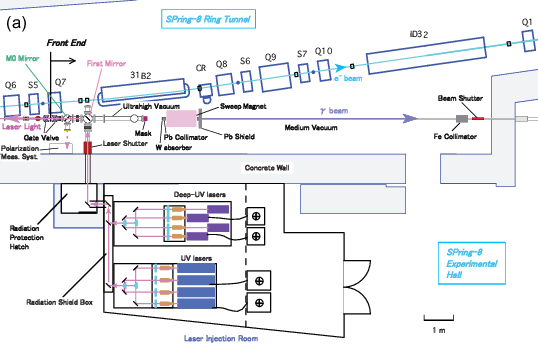}
 \end{minipage} \\
 \vspace{2cm}
 \begin{minipage}{1.0\textwidth}
  \includegraphics[width=18cm]{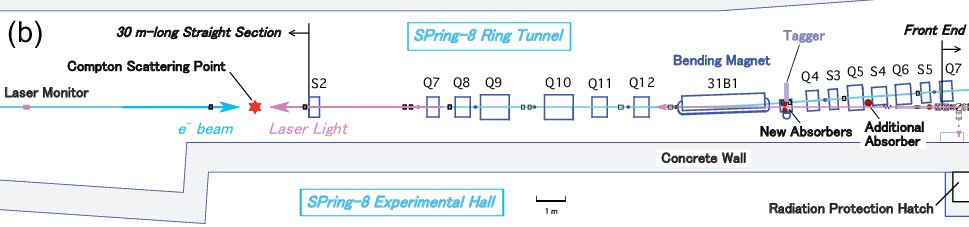}
 \end{minipage} \\
 \vspace{1cm}
 \caption{Plan views around (a) the laser injection system and (b) the straight section of the LEPS2 beamline. {\it Source}: Figures taken from Figs.~3 and 6 of Ref.~\cite{MURAMATSU2022166677}.}
 \label{bl31lep}
\end{figure}

   The LEPS2 beamline was constructed at one of the four long straight sections in SPring-8 \cite{spring8} to produce a GeV $\gamma$-ray beam for hadron photoproduction experiments. (See the detailed descriptions of this beamline and the $\gamma$-ray beam properties in Ref.~\cite{MURAMATSU2022166677}.) The first $\gamma$-ray beam production was confirmed in Jan.~2013. Plan views of the LEPS2 beamline are shown in Fig.~\ref{bl31lep}. In usual operations, ultraviolet (UV) laser light with the wavelength of $355$~$\mathrm{nm}$ ($3.49$~$\mathrm{eV}$) is injected into the storage ring, whose electron energy is $7.975$~$\mathrm{GeV}$, and a part of photons in the laser light obtain energies up to $2.39$~$\mathrm{GeV}$ via backward Compton scattering (laser Compton scattering, LCS), showing the energy spectrum indicated in Fig.~\ref{leps2prop}(a). The energy of each backscattered photon can be measured by analyzing the momentum of a recoil electron at the tagging counter (tagger), which is placed at the exit of the bending magnet located downstream of the straight section. Due to the tagger acceptance for recoil electron detection, $\gamma$-ray energies can be measured in the region greater than $1.3$~$\mathrm{GeV}$ (tagged energy range). Because the long straight section where the LEPS2 facility is constructed has the excellent electron beam divergence of $\sigma_{x'}=8$ and $\sigma_{y'}=0.7$~$\mathrm{\mu rad}$, the angular spreads of both scattered photons and recoil electrons become small. Thus, the $\gamma$-ray beam can be extracted over a distance of about $130$~$\mathrm{m}$ from the Compton scattering point to the experimental site. Additionally, the resolution of the $\gamma$-ray energy measured by the tagger is reduced to $12$~$\mathrm{MeV}$.

   For the purpose to increase a $\gamma$-ray beam intensity, the LEPS2 beamline is designed to allow the injection of at most four laser beams simultaneously. Each two beams are optically arranged side by side at the same height, and two pairs of the neighboring beams are injected with a height difference corresponding to twice the beam radius. The laser beam sizes must be enlarged individually by beam expanders to make focuses at the Compton scattering point which is $31.5$~$\mathrm{m}$ away, so that the inner cavity of beamline chambers is specially made larger than normal chambers \cite{YoritaIPAC2013}. This side-by-side four laser injection method has an advantage to increase the transmittance of the LCS $\gamma$-ray beam during transportation over the length of approximately $\sim 130$~$\mathrm{m}$ up to $77$\% because a laser reflection mirror inside the front-end chamber extended from the straight section can have a hole for the $\gamma$-ray beam path in the small central space between the four laser beam paths whose axes are located at square corners. In the initial operation of the LEPS2 beamline, two to four quasi-CW lasers with the outputs of $8$--$24$~$\mathrm{W}$ at the wavelength of $355$~$\mathrm{nm}$ were used, resulting in the tagged $\gamma$-ray intensity of $1$--$2$~$\mathrm{Mcps}$. Recently, a new type of pulsed laser whose output timing can be adjusted to the external signal has been developed \cite{KatsuraRCNPar2020}, so that two of such lasers which typically output the $355$~$\mathrm{nm}$-wavelength beam power close to $10$~$\mathrm{W}$ at a repetition frequency around several tens $\mathrm{MHz}$ have been introduced to the LEPS2 beamline as well as the control unit to synchronize the laser injection with the electron beam bunch timings. The use of these pulsed lasers in addition to two quasi-CW lasers with $16$~$\mathrm{W}$ outputs has achieved the tagged $\gamma$-ray intensity of $3$--$5$~$\mathrm{Mcps}$. The new type of laser has also been available for the deep-ultraviolet (DUV) wavelength of $266$~$\mathrm{nm}$ ($4.66$~$\mathrm{eV}$). The setup to inject a single laser beam for the DUV wavelength has been prepared in the LEPS2 beamline to provide a $\gamma$-ray beam whose maximum energy reaches $2.89$~$\mathrm{GeV}$. This higher energy beam is expected to have the order of a $\mathrm{Mcps}$ for the tagged $\gamma$-ray intensity, and is useful to exceed the production threshold of higher mass particles.

   The LEPS2 beamline is operated to produce a linearly polarized $\gamma$-ray beam by injecting laser beams which have nearly $100$\% linear polarization thanks to the Brewster's angle of internal laser optics. The laser polarization degree including the influence of reflections at various injection optics is sometimes measured during experiments by extracting laser beams to the polarization measurement system indicated in Fig.~\ref{bl31lep}(a). This system mainly consists of a rotating polarizer prism with a photodiode to measure the polarization extinction ratio from a sine curve fit. Then, the linear polarization of a $\gamma$-ray beam is calculated based on the energy-dependent function shown in Fig.~\ref{leps2prop}(b) with the multiplication of the measured laser polarization. The direction of laser linear polarization is changed by $90^\circ$ using a waveplate \cite{Optics} at appropriate intervals in the physics data taking to enable the estimation of systematic uncertainties for polarization observables, which are the keys of the LEPS2 experiments.

\begin{figure}[t]
 \centering
 \includegraphics[width=18cm]{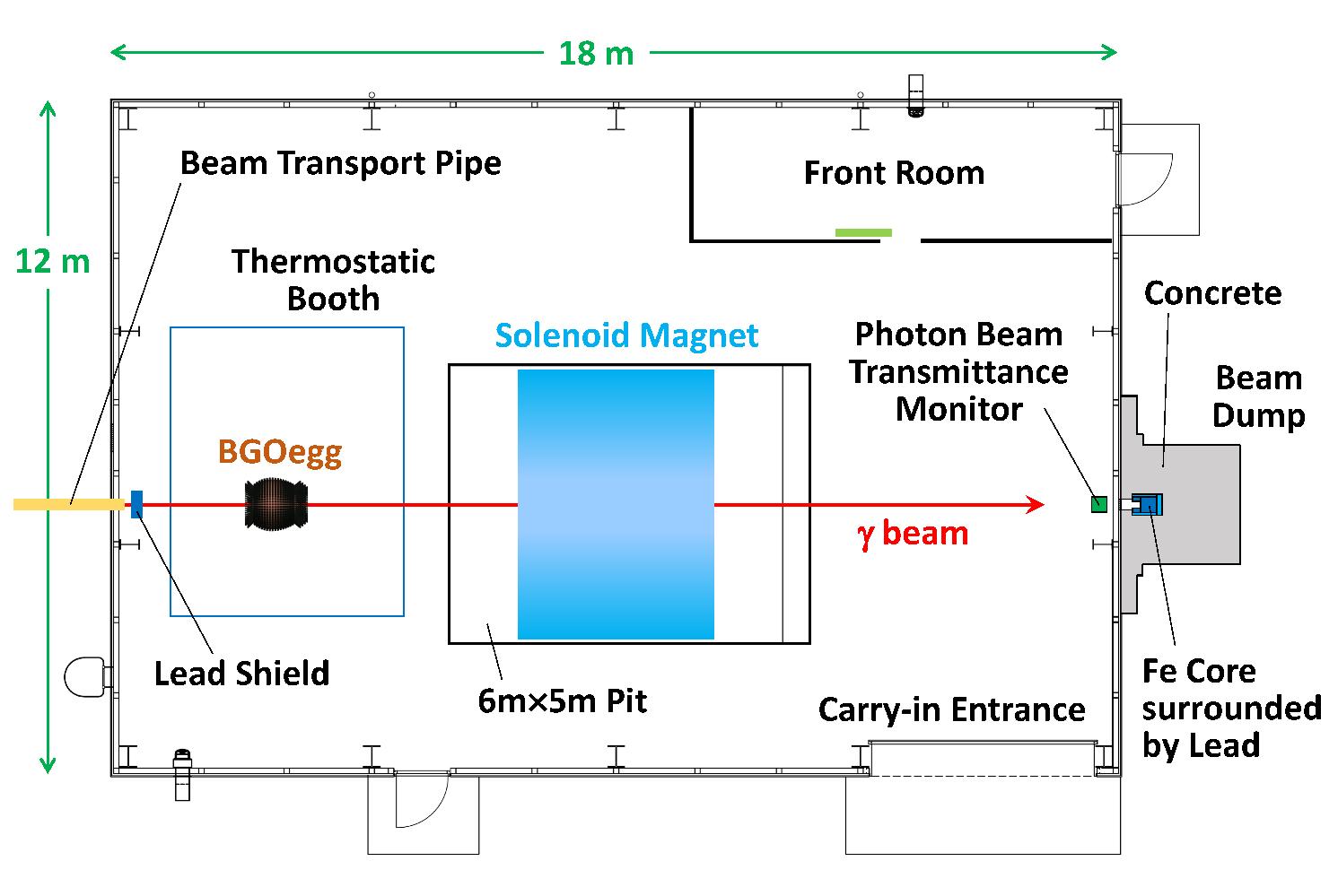}
 \caption{A plan view of the experimental building located at the most downstream of the LEPS2 beamline. {\it Source}: Figure taken from Fig.~8 of Ref.~\cite{MURAMATSU2022166677}.}
 \label{expbuilding}
\end{figure}

\begin{figure}[t]
     \begin{tabular}{cc}
       \begin{minipage}{0.45\textwidth}
         \centering
         \includegraphics[width=7.5cm]{figures/DSC_0281_mod.eps}
       \end{minipage}
       \hspace{5mm}
       \begin{minipage}{0.45\textwidth}
         \centering
         \includegraphics[width=7.5cm]{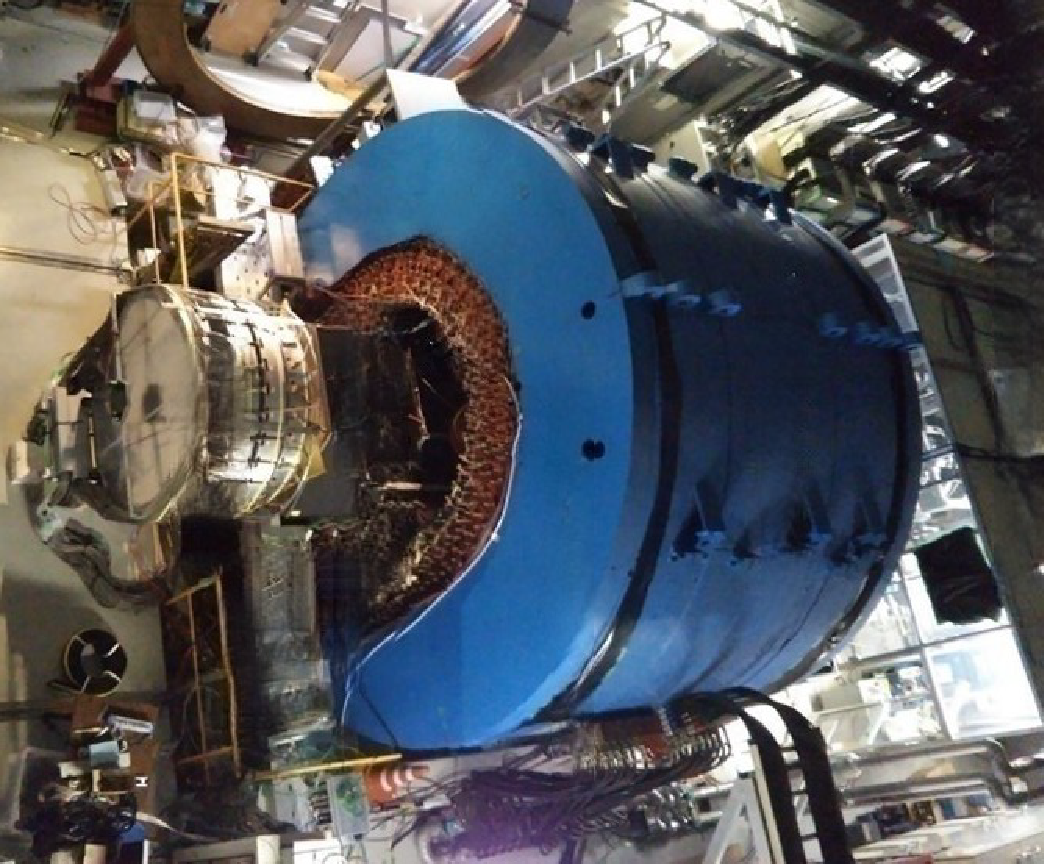}
       \end{minipage}
     \end{tabular}
     \caption{Pictures of detector setups inside the LEPS2 experimental building. The left panel shows a side view of the BGOegg experimental setup from slightly downstream. The right panel is a picture of the setup for the LEPS2 Solenoid experiment, where drift chambers are extracted downstream. A thermostatic booth containing the BGOegg experimental setup is seen in the upstream side of the solenoid spectrometer.}
     \label{leps2pic}
   \end{figure}

   In the most downstream part of the LEPS2 beamline, the experimental building with an area of $12 \times 18$~$\mathrm{m^2}$ and a height of $10$~$\mathrm{m}$ was constructed separately from the SPring-8 ring building to ensure an open space for setting up large detector systems. Although the cone angle of the scattered $\gamma$-ray direction in LCS kinematically increases at lower energies, a high Lorentz boost factor suppresses it to about $70$~$\mathrm{\mu rad}$ or less in the tagged energy range above $1.3$~$\mathrm{GeV}$. In addition, the excellent divergence of the electron beam mentioned above reduces the further spread of a $\gamma$-ray scattering angle. Therefore, the size of a $\gamma$-ray beam in $\sigma$ is about $4$~$\mathrm{mm}$ even in the experimental building. As shown in Fig.~\ref{expbuilding}, two detector systems have been constructed for the different focuses of particle measurement and physics programs. (See also Fig.~\ref{leps2pic}, which shows the pictures of these detector systems inside the LEPS2 experimental building.) In the upstream part of the experimental building, the BGOegg experimental setup mainly using a large-acceptance electromagnetic calorimeter has been built so that the target position should be located at $125$~$\mathrm{m}$ from the Compton scattering point. On the other hand, the LEPS2 Solenoid experiment runs independently with a large charged particle spectrometer inside a magnetic field by using the central part of the experimental building. These two experiments are carried out alternately for the efficient sharing of the LEPS2 beamline, whose beam time is at most about 4500 hours per year. Each of the experiments is conducted by an international collaboration of around 50 people, although there is significant overlap between them. The descriptions of those experimental setups follow in the next subsections.
   
\subsection{BGOegg experiment}\label{subsec:second-second}
The BGOegg experiment runs at the LEPS2 beamline of SPring-8 with the primary aims of exploring the in-medium $\eta^\prime$ mass modification and advancing baryon resonance spectroscopy. The experiment was carried out in two phases, each with a different detector configuration. The BGOegg Phase-I experiment started in 2010 by constructing the setup with a large-acceptance electromagnetic calorimeter (BGOegg calorimeter) and forward charged-particle detectors \cite{PhysRevC.100.055202, PhysRevC.106.035201}. This configuration was designed as a general-purpose detector setup while focusing on the measurement of neutral mesons that decay into multiple $\gamma$'s. In contrast, the BGOegg Phase-II experiment launched in 2021 by replacing the forward charged-particle detectors with a forward electromagnetic calorimeter which was separately prepared. This modification has significantly extended the acceptance for photons from meson decays, enabling the selection of exclusive reactions more efficiently. Together with the upgrades of the photon beam intensity and the maximum allowed target thickness, high-precision and high-statistics studies have become feasible in the Phase-II experiment. In the following sections, the experimental setups of individual phases are described in more detail.

\subsubsection{Phase-I experimental setup}\label{subsubsec:second-seconde-first}
\begin{figure*}[t]
 \centering
 \includegraphics[width=180truemm]{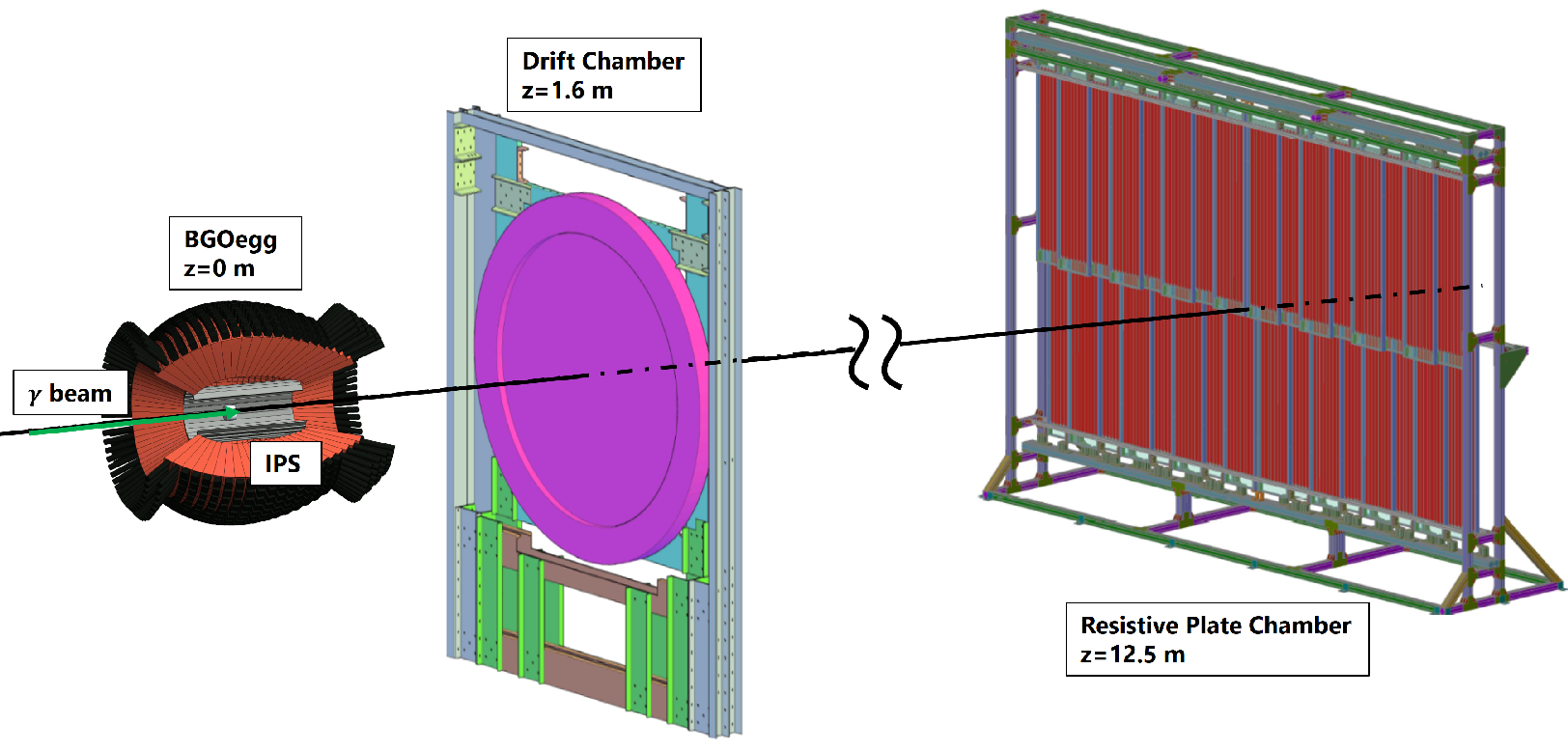}
 \caption{The setup of the BGOegg Phase-I experiment at the LEPS2 beamline.}
 \label{ph1setup}
\end{figure*}
The detector setup used in the Phase-I experiment is shown in Fig.~\ref{ph1setup}. The main detector of this experiment is the BGOegg calorimeter, which consists of 1320 bismuth germanate (BGO) crystals \cite{IR.14.3} assembled in an egg-shaped geometry. It covers polar angles between $24^{\circ}$ and $144^{\circ}$ with full azimuthal acceptance. High granularity has been achieved by integrating 22 layers of 60 BGO crystals that are azimuthally arranged as a ring. Its self-supporting structure minimizes insensitive regions between the neighboring crystals. Each crystal has the shape of a frustum square pyramid with a length of $220$~$\mathrm{mm}$, corresponding to $20$ radiation lengths. This length allows the full energy coverage of an electromagnetic shower. The light output of a crystal was read out by a $30 \times 30$~$\mathrm{mm^2}$ metal package photomultiplier tube (in a majority of layers) or a 3/4 inch linear focus photomultiplier tube (in edge layers). Energy calibration was performed iteratively using the $\gamma \gamma$ invariant mass peak of $\pi^{0}$ decays, and ensured the uniform detector response across all channels. The calorimeter achieves an energy resolution of $1.4$\% for $1$~$\mathrm{GeV}$ photons in Gaussian standard deviations\,($\sigma$) \cite{ISHIKAWA2016109}. Throughout this article, resolutions are quoted as Gaussian $\sigma$. In the case of using a $20$~$\mathrm{mm}$-thick target, the $\gamma \gamma$ invariant mass resolutions of $\pi^{0}$ and $\eta$ are estimated to be $6.7$ and $14.4$~$\textrm{MeV}/c^{2}$, respectively \cite{phdmatsu}. These resolution values represent the world-leading performance in the $\gamma$ energy range around $1$~$\mathrm{GeV}$ or less, as shown in Table~\ref{calcomp}.

The tagging counter, installed at the LEPS2 beamline, plays a key role in measuring the momentum of recoil electrons from the Compton scattering process under the magnetic field of the storage-ring bending magnet. This measurement allows the incident photon energy to be determined on an event-by-event basis. The system consists of two layers of fine scintillating fibers and another two layers of wider plastic scintillator paddles. The fibers provide position information for the recoil electron trajectory, while the plastic scintillators generate fast trigger signals and beam-flux monitoring. This arrangement ensures the accurate tagging of photon energies up to $2.4$~$\mathrm{GeV}$.

Upstream of the BGOegg calorimeter, a veto counter was installed to suppress backgrounds arising from the positron-electron pair production of photons in the materials before reaching the fixed target used for photoproduction reactions. This detector, made of a thin plastic scintillator with the readout by two fine-mesh photomultiplier tubes, achieved a detection efficiency above $99.8$\%, ensuring that only genuine photon interactions on the target were retained.

\begin{table*}[t]
  \normalsize
  \caption{Comparison of the energy and $\gamma \gamma$ invariant mass resolutions for the electromagnetic calorimeters used in the $\gamma$ energy range around $1$~$\mathrm{GeV}$ or less. The BGO-OD mass resolutions indicated by $\ast$ are the result after strong kinematic requirements.
  All the mass resolutions are quoted as the Gaussian standard deviation $\sigma$.}
  \label{calcomp}
  \centering
  \small
  \begin{tabular}{ccccccc}
  \hline
  Experiment & LEPS2/BGOegg & \multicolumn{2}{c}{MAMI-A2} & \multicolumn{2}{c}{CBELSA/TAPS} & BGO-OD \\
  \hline\hline
  Calorimeter & BGOegg & Crystal Ball & \multicolumn{2}{c}{TAPS} & Crystal Barrel & BGO Rugby Ball \\
  Scintillation crystal & {\rm BGO} & {\rm NaI(Tl)} & \multicolumn{2}{c}{BaF$_2$} & {\rm CsI(Tl)} & {\rm BGO} \\
  Number of channels & $1320$ & $672$ & $384$ & $528$ & $1290$ & $480$ \\
  Energy resolution ($\sigma$) at $1$~$\mathrm{GeV}$ & $1.38$\% & $2.0$\% \cite{PhysRevC.79.035204} & \multicolumn{2}{c}{$2.6$\% \cite{PhysRevC.79.035204}} & $2.5$\% \cite{NIMA.321.69, JPCS.160.012006} & $1.3$\% \cite{NIMA.370.396} \\
  $\pi^0$ mass resolution ($\mathrm{MeV/c^2}$) & $6.7$ & \multicolumn{2}{c}{$9$ \cite{MAMIA2proposal}} & \multicolumn{2}{c}{$10$ \cite{EPJA.33.147}} & $12 \ast$ \cite{Alef2020} \\
  $\eta$ mass resolution ($\mathrm{MeV/c^2}$) & $14.4$ & \multicolumn{2}{c}{$21$ \cite{MAMIA2proposal}} & \multicolumn{2}{c}{$22$ \cite{EPJA.33.147}} & $15 \ast$ \cite{Alef2020} \\
  \hline
  \end{tabular}
  \normalsize
\end{table*}

The target is placed at the center of the BGOegg calorimeter, through which the central axes of all the crystals pass. The BGOegg Phase-I experiment employed two types of targets depending on the physics objectives: a liquid hydrogen target for studying baryon resonance spectroscopy and a solid carbon target for searching for in-medium meson mass modification. The liquid hydrogen target container consists of a cylindrical cell made of thin polyimide films in order to minimize material effects. It was filled with liquefied hydrogen, supplied from a closed-cycle refrigerator connected to hydrogen gas tanks. The effective thickness of the liquid hydrogen was $54$~$\mathrm{mm}$ with a density of $0.0708$~$\textrm{g/cm}^{3}$. The solid carbon target was also used to investigate meson production in nuclei. This target had a thickness of $20$~$\mathrm{mm}$ and a density of $1.73$~$\textrm{g/cm}^{3}$. The target holder was constructed by a $10$~$\mathrm{mm}$-thick polystyrene support inside a cylinder of $0.125$~$\mathrm{mm}$-thick Kapton foil. Physics data collection in the Phase-I experiment was carried out in 2014--2016 by alternately changing the above two targets.

In the region between the target and the BGOegg calorimeter, a detector system of plastic scintillators, called ``Inner Plastic Scintillator'' (IPS), was placed to identify charged particles incident on the calorimeter. It consists of 30 thin scintillator bars arranged cylindrically with the individual readout by multi-pixel photon counters (MPPCs). The IPS enabled the separation of charged tracks from neutral photon hits in the calorimeter, thereby being effective to reduce backgrounds in neutral-meson analyses. The information of $dE/dx$ at IPS was also useful to perform the particle identification of proton or pion, making different bands in the correlation with the kinetic energy measured at the BGOegg calorimeter.(See Fig.~\ref{pid2na}, for instance)

Charged-particle tracking at forward angles was achieved with a large drift chamber (DC) located $1.6$~$\mathrm{m}$ downstream of the target. The chamber was composed of six hexagonal wire planes with three different stereo angles, enabling three-dimensional reconstruction of straight tracks under no magnetic field. Each plane contains 80 sense wires with an $8$~$\mathrm{mm}$ pitch, resulting in a position resolution of about $300$~$\mu \textrm{m}$ per plane. The chamber covered polar angles below $21^{\circ}$, complementing the BGOegg calorimeter acceptance and providing the direction of a forward-going proton in the analyses described later.

At extremely forward polar angles less than $6.8^{\circ}$, an array of resistive plate chambers (RPCs) was set up for the time-of-flight measurement \cite{JINST.9.C10008, JINST.11.C11037}. Each RPC module was $1.0$~$\mathrm{m}$ long and $0.25$~$\mathrm{m}$ wide, containing 8 readout strips connected with amplifiers at both ends.
Here, the start timing of the time-of-flight measurement was obtained based on the accelerator RF signal, whose timing jitter was about $4$~$\mathrm{ps}$ at the LEPS2 experimental building.
The RPC system has achieved a timing resolution of $60$--$90$~$\mathrm{ps}$, which is the intrinsic resolution of the RPC modules themselves.
This resolution corresponds to a momentum resolution of around $1$\% for $2$~$\mathrm{GeV/c}$ protons. The capability of this precise timing determination was essential for separating protons from lighter charged particles and for adding extra information to the reaction kinematics even in a limited kinematical region.

In the world-wide situation of hadron photoproduction experiments, the LEPS2/BGOegg setup demonstrates distinctive and complementary features among them. Facilities such as MAMI (Mainz) \cite{PhysRevC.97.055212}, ELSA (Bonn) \cite{Elsner2009epja}, and CLAS at JLab \cite{SOBER2000263, MECKING2003513} have pioneered the hadron photoproduction physics using tagged bremsstrahlung photon beams. MAMI focuses on high-precision measurements at energies below $1.6$~$\mathrm{GeV}$ using calorimeter arrays such as Crystal Ball \cite{PhysRevD.25.2259} and TAPS \cite{GABLER1994168}. ELSA extends the accessible energy range to about $3$~$\mathrm{GeV}$ and places strong emphasis on polarization observables with both linearly and circularly polarized beams. In ELSA, experiments with calorimeters such as Crystal Barrel \cite{NIMA.321.69, EPJA.33.147} and BGO-OD \cite{Alef2020} have been mainly conducted. In JLab/CLAS, a tagged photon beam was available up to $6$~$\mathrm{GeV}$, and a large-acceptance tracking detector optimized for multi-particle final states was employed as a powerful tool for high-energy hadron spectroscopy. Recently, the energy of the electron accelerator in JLab was upgraded to $12$~$\mathrm{GeV}$, and GlueX \cite{ADHIKARI2021164807} started data collection and analyses by combining a linearly polarized photon beam around $9$~$\mathrm{GeV}$ with a general-purpose spectrometer inside a solenoid magnet. In contrast, LEPS2/BGOegg uses a tagged photon beam of $1.3$--$2.4$~$\mathrm{GeV}$, which is produced by laser Compton scattering. This experiment has an advantage in the high linear polarization of the photon beam thanks to the beam production method, as discussed in Sec.~\ref{sec:first}. Such a feature makes LEPS2/BGOegg unique in the high precision measurement of polarization observables related to the linear polarization in the energy region above $1.8$~$\mathrm{GeV}$, where the corresponding data are not available from bremsstrahlung beam experiments. Furthermore, LEPS2/BGOegg has another advantage in using a large acceptance calorimeter with world-leading resolutions in $\gamma \gamma$ invariant mass, as mentioned in Table~\ref{calcomp}.
Thus, LEPS2/BGOegg excels in the high precision analyses on mass line shapes, exotic hadrons, two-particle correlations, and so on in photoproduction of the mesons decaying into multiple photons. The measurement of differential cross sections can also be done precisely in a wide angular region, and particularly the most backward angles provide the highest precision data to date.

\subsubsection{Phase-II experimental setup}\label{subsubsec:second-seconde-second}

\begin{figure}[t]
 \centering
 \includegraphics[width=130truemm]{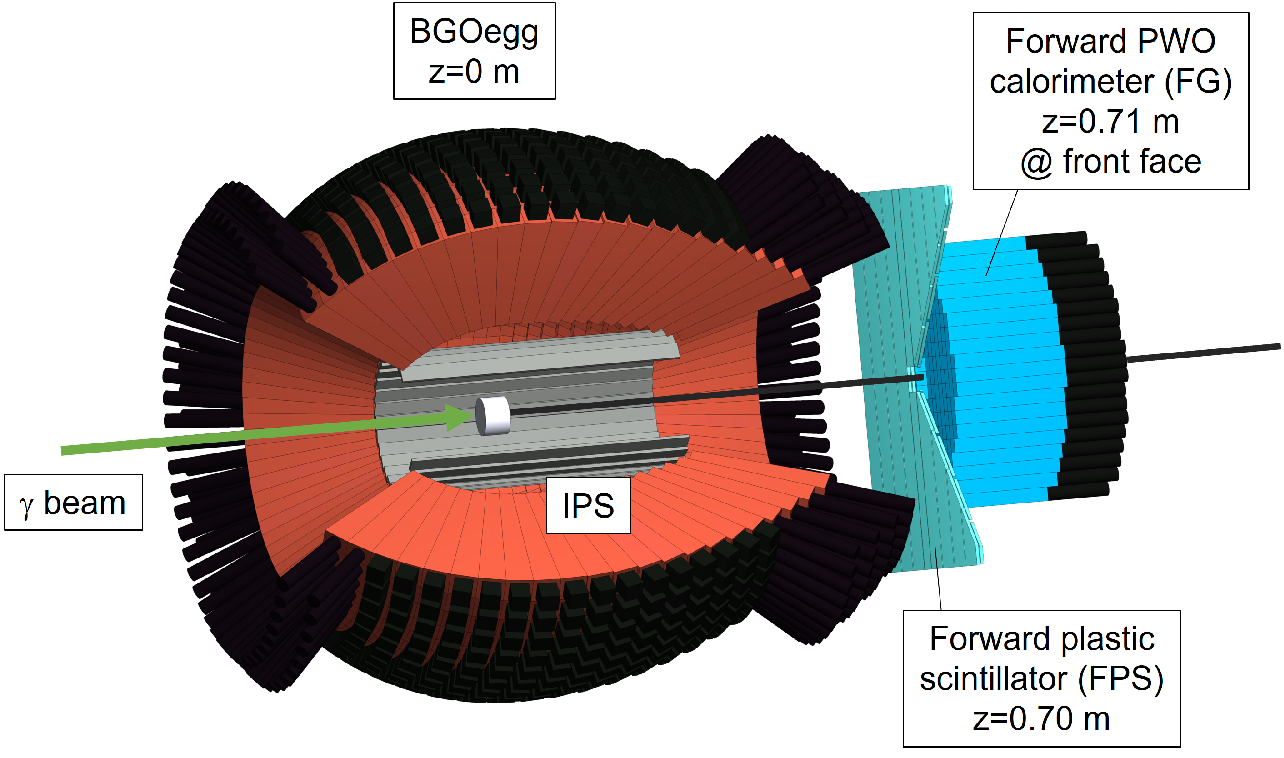}
 \caption{The setup of the BGOegg Phase-II experiment with the forward PWO calorimeter (FG) covering the polar angles less than $16^\circ$.}
 \label{ph2setup}
\end{figure}

As being discussed in Sec.~\ref{subsec:third-first}, the forward charged-particle detectors (DC and RPC) were replaced with an additional electromagnetic calorimeter system in the BGOegg Phase-II experiment by taking into account the Phase-I results. At first, the forward PWO calorimeter, used in past experiments at the LEPS beamline \cite{MURAMATSU2014184} and called ``Forward Gamma'' (FG), was placed downstream of the BGOegg calorimeter, covering the polar angle range of $3^\circ$--$16^\circ$, as shown in Fig.~\ref{ph2setup}. The FG consists of $252$ pure {\rm PbWO$_4$} (PWO) crystals, each of which has a rectangular-shape with a $22$~$\mathrm{mm}$-square cross section and a length of $180$~$\mathrm{mm}$, corresponding to $20$ radiation lengths. These crystals were assembled side by side to form a nearly circular area with a diameter of $418$~$\mathrm{mm}$. In the center part of FG, a pipe with an inner diameter of $54$~$\mathrm{mm}$ was placed for a photon beam path. The light output of each crystal was read out by a 3/4 inch photomultiplier tube with a Cockcroft-Walton type voltage supplier. The energy resolution of FG was measured using a positron beam at the test beamline of RARiS, Tohoku University, resulting in $2.7$\% for $1$~$\mathrm{GeV}$. In front of the FG, a wall of plastic scintillators, called ``Forward Plastic Scintillator'' (FPS), was installed to determine whether a calorimeter hit was charged or neutral. The FPS has a size of $600 \times 600$~$\mathrm{mm^2}$ with a square hole of $40 \times 40$~$\mathrm{mm^2}$ in the center, and consists of two layers, each of which is made by $5$~$\mathrm{mm}$-thick scintillators aligned in the $x$ or $y$ direction. Scintillation light was read out from both ends by MPPCs. A charged particle hit position is identified by the combination of hit channels in the two layers. The time resolution of an FPS hit has been estimated to be $0.3$~$\mathrm{ns}$ based on the beam test.

In the Phase-II experiment, the photon beam intensity was increased by introducing pulsed lasers as described in Sec.~\ref{subsec:second-first}. In addition, the nuclear target material was changed from carbon to copper for further studies on the in-medium meson mass modification. (See Sec.~\ref{subsec:third-first}.) Because the concrete shield of the beam dump in the LEPS2 experimental building was made thicker before starting the Phase-II experiment, the radiation safety regulation was relaxed to allow the target thickness of $0.5$ radiation lengths, which became five times more than that of the Phase-I nuclear target. Physics data collection of the Phase-II experiment has started in 2024 using a $7.5$~$\mathrm{mm}$-thick copper target. In the near future, the use of a niobium target, which has a larger mass number, is under consideration as well as the data collection with a liquid hydrogen target. In parallel, the size of FG is being expanded by adding more PWO crystals to fully cover the forward acceptance hole of the BGOegg calorimeter, corresponding to the polar angles less than $24^\circ$.

\subsection{LEPS2 Solenoid spectrometer}\label{subsec:second-third}
\subsubsection{Detector configuration}
The LEPS2 Solenoid spectrometer is installed downstream of the BGOegg calorimeter at the LEPS2 experimental building at SPring-8. It is designed for multi-purpose detection, capable of handling mixed charged and neutral final states \cite{LEPS2TDR}. A review of the recent status of the experiment can be found in \cite{Niiyama:2024wwg, Ryu:2020pjk}. 

\begin{figure}[htbp]
  \centering 
  \includegraphics[width=0.8\textwidth]{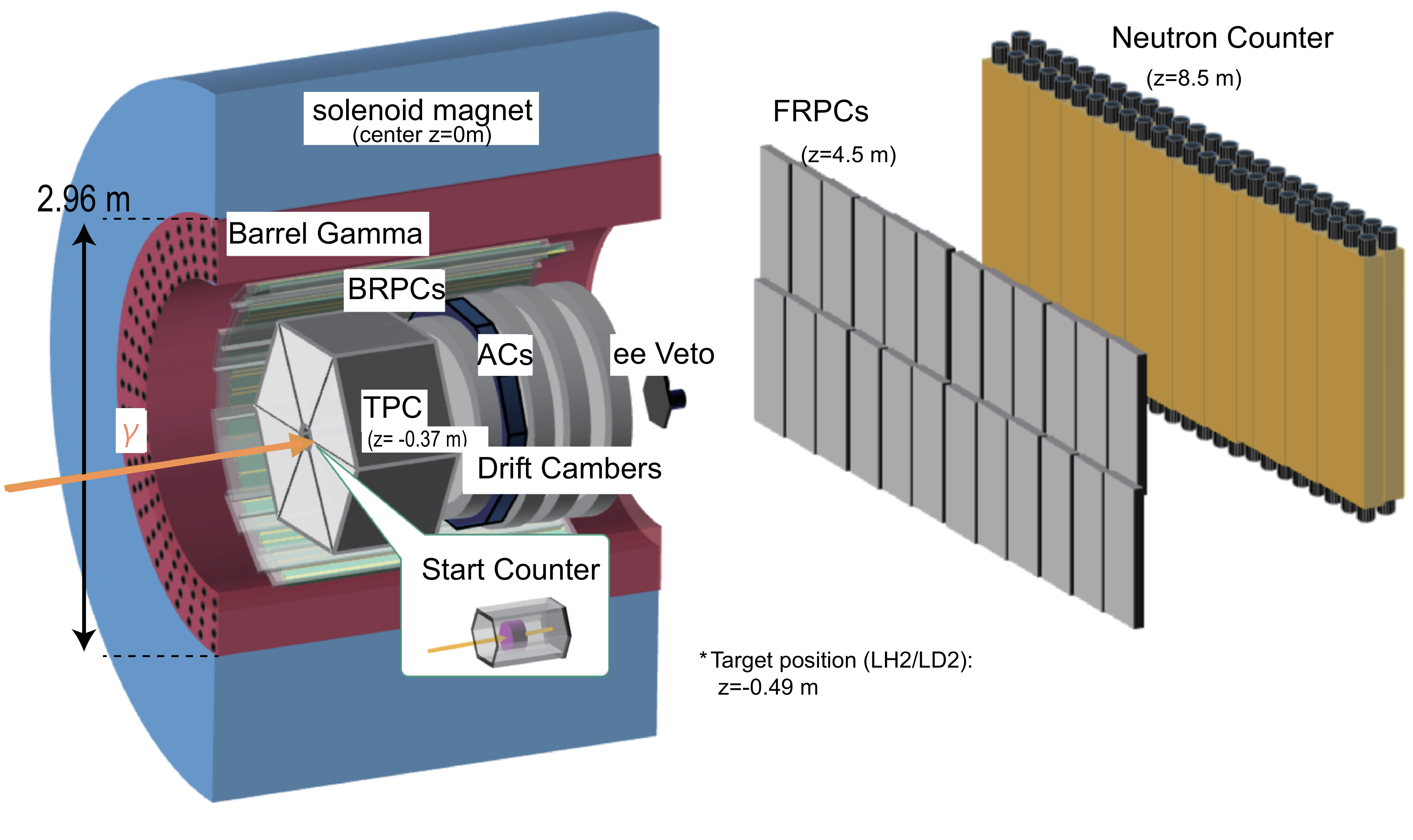}
  \caption{Schematic view of the LEPS2 Solenoid spectrometer \cite{LEPS2TDR}. The photon beam enters from the left side of the figure. The blue cylindrical structure represents the solenoid magnet, whose inner space has a diameter of 2.96~m and a length of 2.22~m.}
  \label{fig_solenoid_view} 
\end{figure}

As shown in Fig.~\ref{fig_solenoid_view}, the detectors are installed inside a solenoidal magnet with a bore diameter of 2.96~m and an inner-volume length of 2.22~m \cite{Atiya:1992vh}, providing a central magnetic field of 0.9~T. The detector system covers polar angles up to $120^\circ$ for charged particles and $40^\circ$--$110^\circ$ for photons. A cryogenic target system is inserted into the inner bore of the Time Projection Chamber (TPC). The target cell, made of Kapton film with a length of 15~cm, is filled with liquid hydrogen ($\mathrm{LH}_2$) or liquid deuterium ($\mathrm{LD}_2$) and cooled by a Gifford--McMahon refrigerator.

To efficiently collect hadronic events, the trigger uses a start counter (SC) near the target together with veto counters: an $e^+ e^-$ (ee-veto) counter and an upstream-veto (UPV) counter to suppress backgrounds from $e^+ e^-$ pairs. The SC provides the reference timing for time-of-flight (TOF) measurements. Segmented into a forward part (FSC) and a side part (SSC), it consists of 4~mm thick plastic scintillator (ELJEN EJ-212) read out by Multi-Pixel Photon Counters (MPPCs) through UV-filter light guides. This design operates well in the 0.9~T field and achieves a timing resolution better than 150~ps, which is required to determine a TOF start timing by identifying the beam bunch based on the SPring-8 RF structure with 2~ns intervals. Detector signals are acquired and processed using a Network-Distributed Data-Acquisition System \cite{Ryu:2025cjn}.
The event rate depends on the trigger and DAQ bandwidth. The total hadronic cross section for $\gamma + d$ collisions is estimated to be about 0.26--0.3~mb. For a 1~Mcps photon beam on a 15~cm $\mathrm{LD}_2$ target, the hadron production rate is about several hundred~cps. With the SC and veto counters, the typical DAQ trigger rate becomes about 100~cps.

Charged-particle tracking is performed in two angular regions to determine momentum from track curvature in the magnetic field. Side tracks ($40^\circ$--$120^\circ$) are measured by the TPC, which has a hexagonal shape with a drift length of 710~mm and an active radius of 655~mm. It is filled by a P10 gas, and the ionization signals from a charged track are read out by 10{,}830 pads located at the upstream end. To reduce multiple scattering for forward-going tracks, the downstream end-plate of the TPC is made with thin membranes. Forward tracks are measured by four planar drift chambers (DC0 to DC3 from upstream) using an Ar-based gas mixture. Each chamber features six wire planes with three stereo angles ($0^\circ$, $+60^\circ$, and $-60^\circ$) for full 3D track reconstruction, achieving a typical position resolution of about 150~$\mu$m. This combined tracking system ensures that the momentum resolution from hit-position measurements is close to the multiple-scattering limit, optimizing performance in the momentum region around 1~GeV/$c$ (e.g., for kaons) over the full angular acceptance \cite{LEPS2TDR}.

Several detectors are used for particle identification (PID). Resistive Plate Chambers (RPCs) provide precise TOF velocity measurements: forward RPCs (FRPCs) are placed about 4~m downstream from the target, and barrel RPCs (BRPCs) are at a radius of about 0.9~m. Both reach a timing resolution of about 100~ps. Aerogel Cherenkov counters (ACs) with a refractive index $n=1.03$ are installed between DC0 and DC1, utilizing fine-mesh Photomultiplier Tubes (PMTs) that operate in magnetic fields to separate forward kaons and pions. Neutral particles are detected mainly by a Barrel Gamma detector (B$\gamma$), a sampling calorimeter consisting of lead (1~mm) and scintillator (5~mm) layers. It is placed in the space outside a radius of 1~m to detect photons from decays like $\pi^0 \to \gamma \gamma$. Finally, a Neutron Counter (NC) located about 8~m downstream detects neutrons through recoil protons using two layers of 152~mm thick plastic scintillator bars.

\subsubsection{Detector performance}

Commissioning runs using nuclear targets began in FY 2019. Subsequently, physics runs using liquid hydrogen ($\mathrm{LH}_2$) and liquid deuterium ($\mathrm{LD}_2$) targets were initiated in FY 2021. Since then, data collection has continued annually up to FY 2026, with a few months dedicated to it each year.
Currently, detector calibration and data analysis for the respective run periods are underway, and new physics results are beginning to emerge. The overall detector performance is evaluated through established physics channels.

The TPC is used to reconstruct tracks in the side region for particles originating from the target. By measuring the track curvature in the magnetic field, the transverse momentum of a detected charged particle is obtained. Then, the particle momentum is reconstructed by correcting it with the measured polar angle. The typical momentum resolution is better than 5\% in the side region \cite{LEPS2TDR}.

\begin{figure}[htbp]
  \centering
 \includegraphics[width=1.0\linewidth]{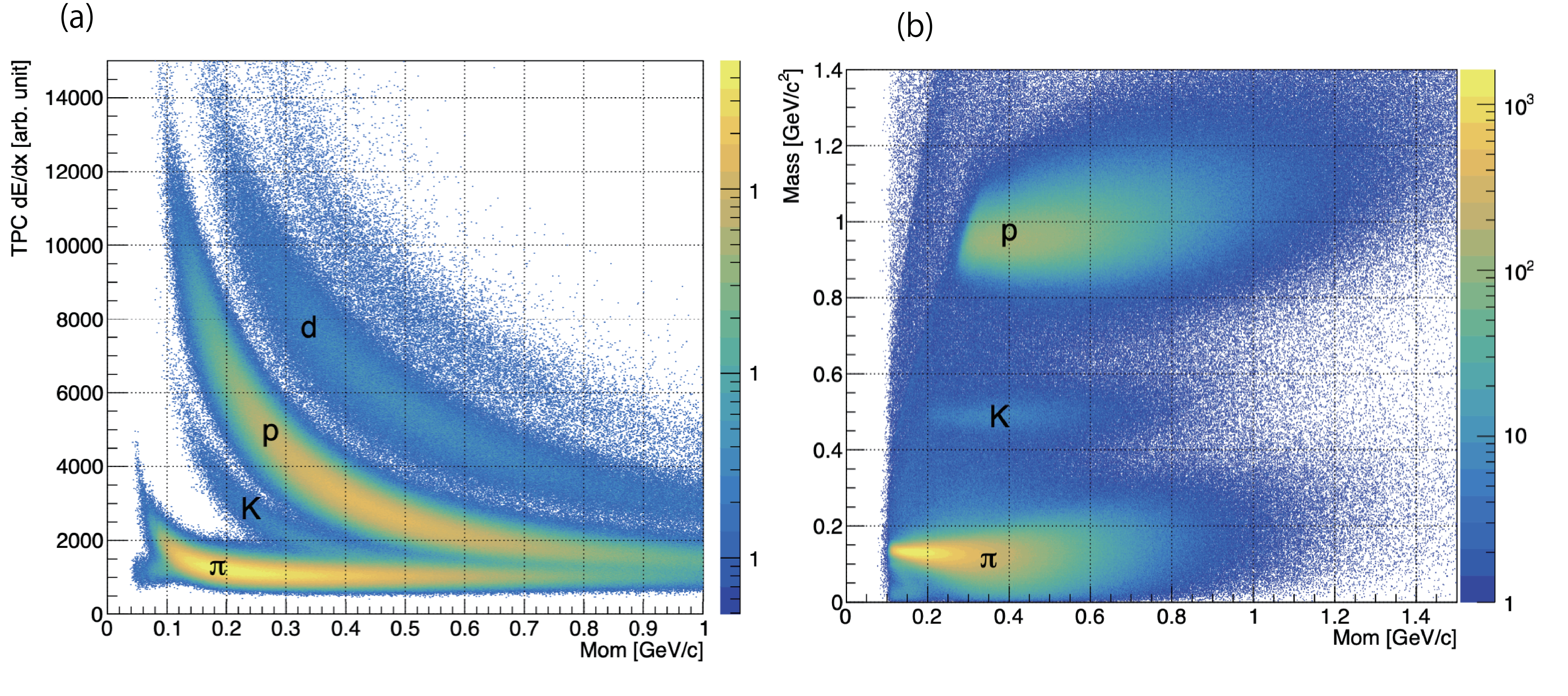}
  \caption{(a) Particle identification (PID) plot showing the correlation between momentum and energy loss ($dE/dx$) measured by the TPC. (b) PID plot showing the correlation between momentum and the calculated mass derived from the time-of-flight information measured by the BRPC.
  These plots are produced using data from a deuterium target within a photon beam energy range of 1.3--2.4 GeV and an effective angular coverage of $40^\circ$--$120^\circ$ \cite{LEPS2TDR}.}
  \label{fig_pid}
\end{figure}

The pad pulse height of the TPC, which is proportional to the energy deposit, is used to estimate $dE/dx$. To extract the valid track signals, a truncated-mean method is used: hits with the largest charges corresponding to a Landau tail are removed, and the average of the remaining charges is divided by the track path length on the pads. This method yields clear bands corresponding to pions, protons, and deuterons in momentum vs.\ $dE/dx$ plots, providing strong $\pi/p$ separation in the momentum region around 1~GeV/$c$ (Fig.~\ref{fig_pid}(a)).

Furthermore, by combining the TPC information with the time-of-flight measurements from the BRPCs, the particle identification becomes even more distinct. The particle velocity ($\beta$) is calculated using the time of flight obtained from the BRPCs and the track length derived from the TPC tracking. By combining this $\beta$ with the momentum measured by the TPC, the mass of the particles can be reconstructed. Utilizing the information of BRPCs allows kaons ($K$) to be successfully separated from the pion band in the momentum above 0.3~GeV/$c$ (Fig.~\ref{fig_pid}(b)).

\newpage
\section{Overview of physics programs}\label{sec:third}

A wide range of hadron physics programs have been promoted in the LEPS2 facility. This section overviews those physics programs with obtained results, while focusing on individual topics. Sections~\ref{subsec:third-first}, \ref{subsec:third-second}, and \ref{subsec:third-third} describe physics subjects, outcomes, and prospects in the BGOegg experiment. The on-going subjects carried out by the Solenoid experiment are described in Sec.~\ref{subsec:third-fourth}.

\subsection{Systematic studies for the in-medium $\eta^\prime$ mass modification in BGOegg experiment}\label{subsec:third-first}

The origin of hadron masses has been a fundamental research subject for a long time. For instance, the sum of bare quark masses, originating from the Higgs mechanism, can explain only about $1$\% of the nucleon mass \cite{pdg}. In the current discussions related to the QCD phase diagram \cite{RPP.74.014001}, the change of vacuum properties associated with the spontaneous breaking of chiral symmetry \cite{nambu1, nambu2} is widely considered to be a source of the dynamical hadron mass generation. This concept is an analogy of the superconductivity theory, where the appearance of Cooper pairs working as bosons lowers the filled energy level from the Fermi level in a metal and consequently generates the energy gap. In the case of hadrons, the vacuum of the hadron phase is possibly filled by quark condensates $\langle q \bar{q} \rangle$ and correspondingly produces non-zero mass levels. The evidence for this mass generation mechanism may be obtained experimentally by approaching the chiral phase transition boundary in the QCD phase diagram and observing the signal of the partial restoration of the chiral symmetry breaking. Thus, many experiments search for the in-medium mass modification by producing mesons inside nuclei, whose density of $\sim 2 \times 10^{14}$~$\mathrm{g/cm^3}$ is comparable to the early universe \cite{RMP.82.2949, IJMPE.19.147, PPNP.97.199}.

The BGOegg experiment has focused on photoproduction of the $\eta^\prime$ meson inside a nucleus. The $\eta^\prime$ meson is mainly composed of the singlet pseudoscalar meson $\eta_0$ and has an extraordinarily larger mass compared with the octet pseudoscalar mesons that correspond to Nambu-Goldstone bosons \cite{NuclPhysA.516.429}. Thus, the spontaneous breaking of chiral symmetry may play a key role in generating the $\eta^\prime$ mass. In addition, it has been suggested that its large mass is influenced by the $U_A$(1) quantum anomaly \cite{witten}. Here, the relation between the spontaneous chiral symmetry breaking and the $U_A$(1) quantum anomaly is still an interesting open question. A recent theoretical study has pointed out that a strong contribution of the $U_A$(1) anomaly can be a source of the spontaneous breaking of chiral symmetry \cite{kono}. Under such a situation, the study of the in-medium $\eta^\prime$ mass in the BGOegg experiment is expected to be a potential clue for solving their relation because the quantitative information of $\eta^\prime$ mass modification may provide constraints to the effective theories taking them into account.

\subsubsection{Theoretical and experimental situations of in-medium $\eta^\prime$ mass studies}\label{subsubsection:third-first-first}
Currently, three effective theories are known to describe the in-medium $\eta^\prime$ mass reduction with quantitative predictions, whereas its width modification is considered to be moderate based on the CBELSA/TAPS measurement ($\Gamma \approx 20$~$\mathrm{MeV}$) \cite{etapwid}. First, the Nambu-Jona-Lasinio (NJL) model, describing the spontaneous breaking of chiral symmetry, has been adopted in a calculation to construct an effective Lagrangian by further incorporating the Kobayashi-Maskawa-'t Hooft (KMT) term, which represents the $U_A$(1) quantum anomaly \cite{nagahiro}. This Lagrangian was expressed on the quark-field basis, and was used in the calculation of a quark-antiquark scattering amplitude to obtain the $\eta^\prime$ meson mass at finite densities. As a result, the mass reduction of about $-150$~$\mathrm{MeV}$ was predicted at the normal nuclear density. On the other hand, another calculation was done based on the linear sigma model to construct an effective Lagrangian by similarly taking into account both of the spontaneous chiral symmetry breaking and the $U_A$(1) quantum anomaly, but by emphasizing the hadron degree of freedom instead of describing an interaction model at the quark level \cite{sakai}. In this case, the medium effect on the $\eta^\prime$ meson mass was calculated with the nucleon one-loop approximation, and a smaller mass reduction of $-80$~$\mathrm{MeV}$ was predicted at the normal nuclear density. Finally, the other prediction has been available based on the Quark Meson Coupling (QMC) model, where the MIT bag model with a mean field of the $\sigma$ meson was considered \cite{bass}, suggesting the mass reduction of about $-40$~$\mathrm{MeV}$. As seen above, the three predictions contradict each other, so that the experimental measurement of the in-medium $\eta^\prime$ mass is important to progress the understanding of dynamical mass generation with effective models. In addition, the quantitative constraints to the effective theories have influence on various QCD calculations using those models.

The experimental efforts to measure in-medium $\eta^\prime$ mass have been made in several ways. In the COSY-11 experiment, the $\eta^\prime$-proton scattering length was extracted from the total cross section of the $p p \to p p \eta^\prime$ reaction near the threshold \cite{cosy11}. The real part of the scattering length was measured to be $0.00 \pm 0.43$~$\mathrm{fm}$, not supporting a large reduction of in-medium $\eta^\prime$ mass like the NJL and linear sigma model calculations. In the CBELSA/TAPS experiment, the information on the in-medium potential depth for $\eta^\prime$ was deduced from the excitation function and momentum distribution of that meson photoproduced with $^{12}$C and $^{93}$Nb nucleus targets \cite{elsa2013, elsa2016, elsa2018}. All of their measurements have indicated the potential depth of around $-40$~$\mathrm{MeV}$ at the normal nuclear density by comparing the data with collision model calculations for different potential parameters. Instead of these indirect estimations of in-medium $\eta^\prime$ mass, the search for $\eta^\prime$-meson bound nuclei was attempted by the $\eta$-PRiME/Super-FRS collaboration \cite{gsietap}. If a large reduction of the $\eta^\prime$ mass occurs inside a nucleus, it causes the in-medium potential of the meson to be bound. They measured the excitation spectrum of $^{11}$C in the $^{12}$C($p$,$d$) reaction near the $\eta^\prime$ production threshold, corresponding to the measurement of the sum of in-medium $\eta^\prime$ and $^{11}$C nucleus masses. No signal was observed below the threshold due to the dominance of backgrounds. A strongly attractive potential predicted by the NJL model calculation was disfavored by comparing the experimental excitation spectrum with its theoretical prediction, where the signal strength was determined based on the cross section of the elementary process $\gamma p \to \eta^\prime p$ and a nuclear response function.

\subsubsection{Search for $\eta^\prime$ bound nuclei}

\begin{figure}[t]
 \centering
 \includegraphics[width=14cm]{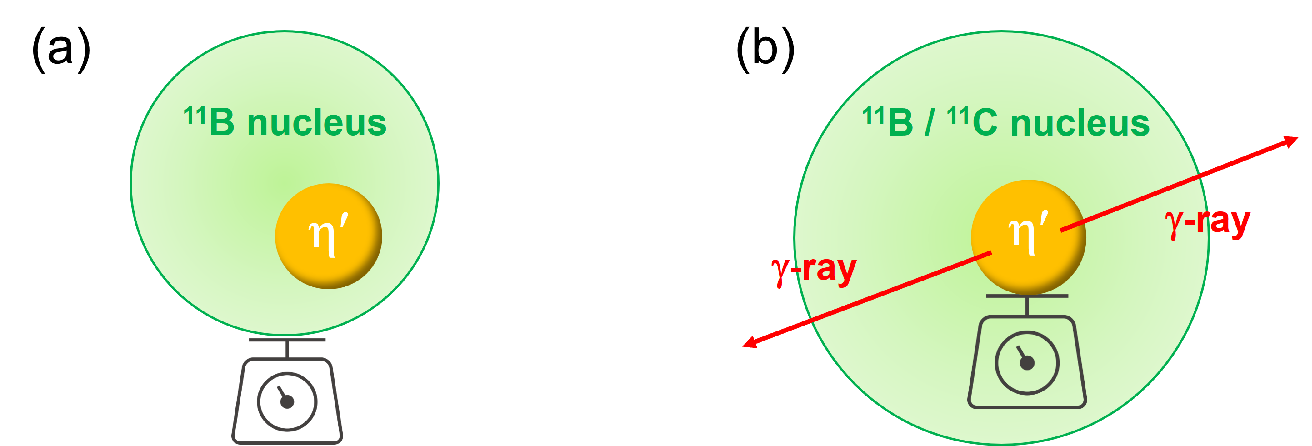}
 \caption{Conceptual descriptions of the in-medium $\eta^\prime$ mass measurements in the BGOegg Phase-I experiment: (a) the search for $\eta^\prime$ bound nuclei and (b) the direct measurement of in-medium $\eta^\prime$ mass through the $\gamma \gamma$ decay channel.}
 \label{egg2meas}
\end{figure}

In the situation where the understanding of in-medium $\eta^\prime$ mass is inconclusive both theoretically and experimentally, the BGOegg Phase-I experiment has systematically and complementarily carried out two ways of its measurement, as conceptually shown in Fig.~\ref{egg2meas}, via photoproduction reactions using a $20$~$\mathrm{mm}$-thick carbon target. First, the data collected in 2015 were analyzed to search for $\eta^\prime$ bound nuclei in the reaction $\gamma + ^{12}C \to (\eta^\prime \; ^{11}B)_{B.S.} + p_f$ with the detection of an extremely forward-going proton ($p_f$) \cite{etapnuc}. The momentum of the forward proton was measured based on the time-of-flight (TOF) at the wall of resistive plate chambers (RPC), covering the polar angles less than $6.8^\circ$. In this momentum calculation, the proton mass was assumed after removing the contamination from $e^+ e^-$ conversion by requiring $\beta < 0.98$. Then, an excitation energy distribution was obtained by subtracting the $^{11}$B and in-vacuum $\eta^\prime$ masses from the missing mass in the $^{12}$C($\gamma$,$p$) reaction. The excitation energy resolution of around $20$~$\mathrm{MeV}$ by the TOF measurement was unfortunately insufficient to observe the peaks corresponding to different bound orbits \cite{nagahiro}, so that the signal enhancement due to $\eta^\prime$ bound nuclei was searched for over backgrounds in the predetermined energy region around the threshold of quasi-free $\eta^\prime$ production. Here, the excitation energy distribution simply obtained from the missing mass of a forward proton is expected to be dominated by background processes because of the inclusive measurement. Thus, the final state of $\eta^\prime$ bound nuclei was tagged to improve the signal-to-noise ratio, for the first time. In the analysis of Ref.~\cite{etapnuc}, the simultaneous reaction $\eta^\prime p \to \eta p \to \gamma \gamma p$ was searched for using the BGOegg calorimeter as a one-nucleon absorption signal from the $\eta^\prime$ bound nucleus. This reaction channel was chosen by following a theoretical prediction that suggested the dominance of $\eta p$ conversion among one-nucleon absorption processes \cite{oset}.

\begin{figure}[t]
 \centering
 \includegraphics[width=18cm]{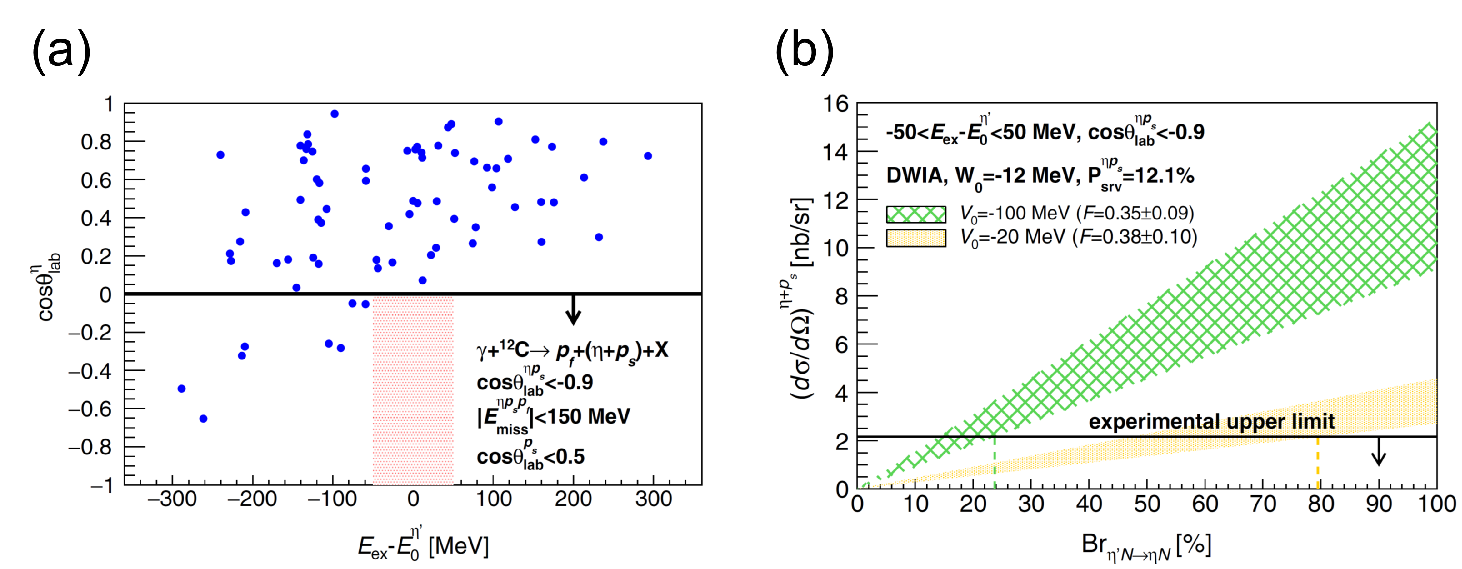}
 \caption{Results of the search for $\eta^\prime$ bound nuclei with the simultaneous detection of the one-nucleon absorption signal of $\eta^\prime p \to \eta p \to \gamma \gamma p$. (a) The scatter plot of $\eta$ polar angles vs.~excitation energies for the finally selected sample. (b) The experimental upper limit and theoretical predictions for the differential cross section of bound nuclei production with the absorption signal. The theoretical predictions were calculated for the in-medium $\eta^\prime$ potentials of $-100$ and $-20$~$\mathrm{MeV}$ depending on the branching fraction of the $\eta^\prime N \to \eta N$ reaction. {\it Source}: Both figures taken from Ref.~\cite{etapnuc}.}
 \label{res1na}
\end{figure}

The photon beam of around $2$~$\mathrm{GeV}$ can kinematically produce $\eta^\prime$ mesons with low recoil momenta (a few hundreds $\mathrm{MeV/c}$) if the mass reduction more than several tens $\mathrm{MeV}$ occurs, and has the potential to generate $\eta^\prime$ bound nuclei. For such signal events, the final-state particles of a one-nucleon absorption, a slow side-way proton ($p_s$) and an $\eta$ meson reconstructed from the $\gamma \gamma$ invariant mass, should be detected with a back-to-back relation at the BGOegg calorimeter. Therefore, the opening angle of them in the laboratory frame, $\theta^{\eta p_s}_{lab}$, was required to be $\cos \theta^{\eta p_s}_{lab} < -0.9$. In addition, the side-way proton was identified using the correlation between the kinetic energy measured by the BGOegg calorimeter and the $dE/dx$ obtained from the inner plastic scintillators. The selected event sample was still dominated by the backgrounds due to quasi-free $\eta$ and $\eta \pi$ photoproduction processes with the secondary interaction of a final-state particle. To completely exclude them, further kinematical conditions were applied so that the laboratory polar angles of the $\eta$ meson and the side-way proton should not be too forward. In the determination of the above event selection, the background level was controlled to be sufficiently smaller than one event by a blind analysis, masking the predetermined signal region. As a result of applying all the selection conditions, no events were observed in the signal region defined by the excitation energies from $-50$ to $50$~$\mathrm{MeV}$, as shown with the red hatched area of Fig.~\ref{res1na}(a). Note that usual quasi-free $\eta^\prime$ photoproduction events without a binding process are also reduced enough at the excitation energies around the $\eta^\prime$ production threshold by the one-nucleon absorption tag. Based on the Poisson distribution for null observation, the upper limit for the differential cross section of $\eta^\prime$ bound nuclei photoproduction associated with the one-nucleon absorption signal going to the $\eta p$ final state was evaluated to be $2.2$~$\mathrm{nb/sr}$ at the $90$\% confidence level.

\begin{figure}[t]
 \centering
 \includegraphics[width=12cm]{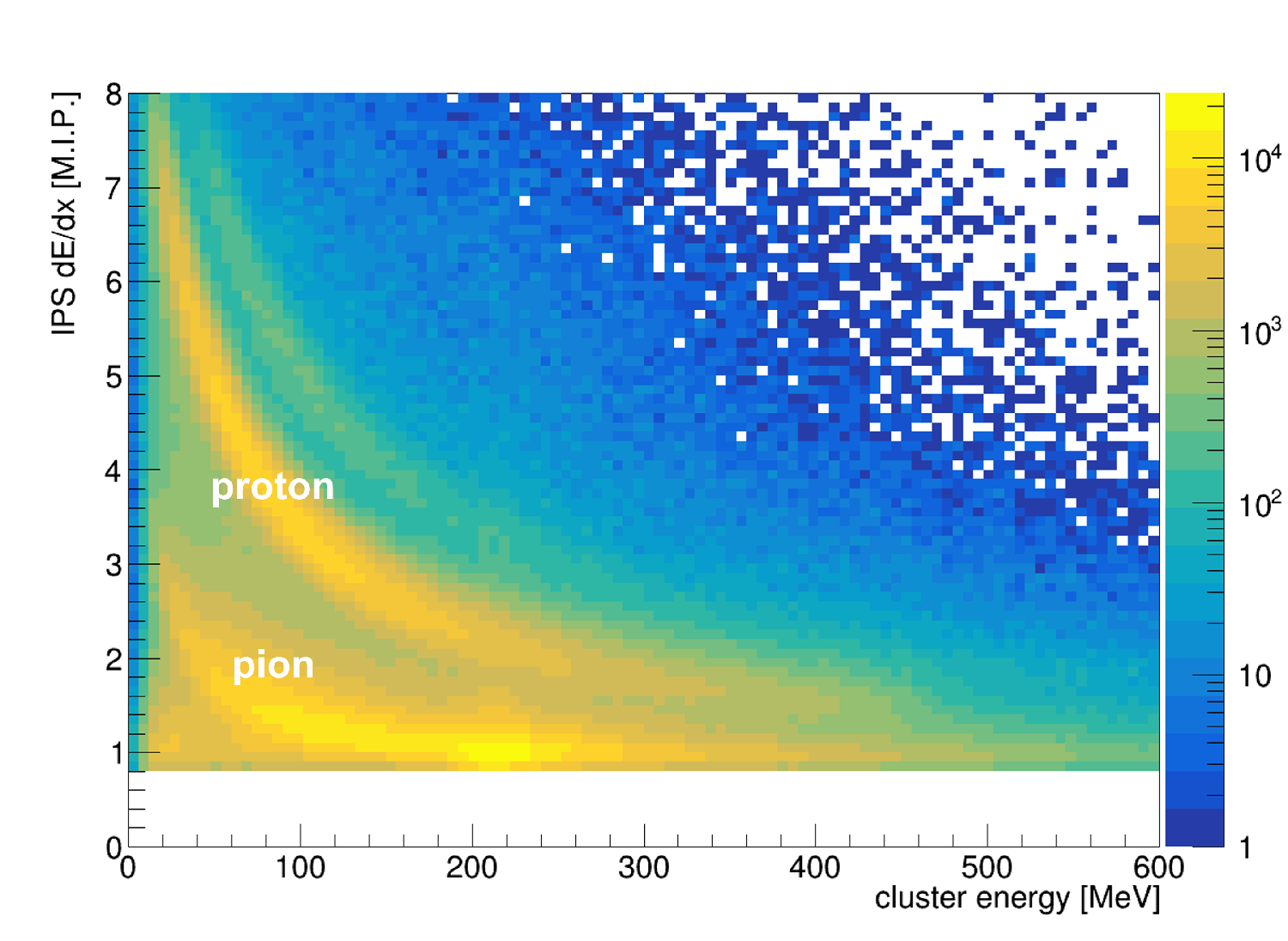}
 \caption{A scatter plot used for the identification of two charged particles detected at the BGOegg calorimeter. The vertical and horizontal axes show the $dE/dx$ at the inner plastic scintillators (IPS) and the charged cluster energy (kinetic energy) at the BGOegg calorimeter, respectively. Proton and pion bands are clearly seen. In this sample, an extremely forward proton was also detected at RPC.}
 \label{pid2na}
\end{figure}

This result was compared with the corresponding values estimated from the theoretical calculations of excitation energy spectra at the in-medium $\eta^\prime$ potential ($V_0$) of $-20$ and $-100$~$\mathrm{MeV}$. Here, the comparison was made depending on the branching fraction of $\eta^\prime N \to \eta N$, as shown in Fig.~\ref{res1na}(b). Note that the theoretical calculations of excitation energy spectra have been done by taking into account the realistic kinematics of $\eta^\prime$ photoproduction in a carbon target within the framework of distorted wave impulse approximation (DWIA) \cite{dwia}. In addition, those theoretical calculation results have been normalized with scale factors of about one third because the BGOegg experiment can simultaneously measure the cross section of $\gamma + ^{12}C \to \eta^\prime + p_f + X$ and compare it with the original theoretical estimation obtained in the same calculation process as for the bound signal. In the case of $V_0 = -100$~$\mathrm{MeV}$, the measured upper limit of $2.2$~$\mathrm{nb/sr}$ suggests the branching fraction of $\eta^\prime N \to \eta N$ to be $24$\% or less at the 90\% confidence level. Such a low branching fraction disfavors the in-medium $\eta^\prime$ mass reduction of $-100$~$\mathrm{MeV}$ or more if the one-nucleon absorption is considered to be dominant compared with the many-body nucleon absorption. This result does not support the current NJL model prediction as with the $\eta$-PRiME/Super-FRS collaboration's claim, but represents advances in the significant reduction of backgrounds with the one-nucleon absorption tag and the proper normalization of theoretical excitation energy spectra to be compared with measured data.

Recently, the analysis procedure to search for $\eta^\prime$ bound nuclei has been extended by tagging the two-nucleon absorption process $\eta^\prime p p \to p p$ as another possible final state of the bound signals \cite{okabe}. In this case, a back-to-back pair of two side-way protons with the kinetic energies around $480$~$\mathrm{MeV}$ is searched for at the BGOegg calorimeter. The side-way protons are being distinguished from charged pions by using the $dE/dx$ information at the inner plastic scintillators. (See Fig.~\ref{pid2na}.) The momentum of an extremely forward proton is simultaneously measured at RPC to obtain the excitation energy distribution, as done in the analysis with the tag of a one-nucleon absorption signal. The sample after selecting a forward proton at RPC and a pair of charged particles at the BGOegg calorimeter is dominated by $\pi^+ \pi^-$ photoproduction events. Those events are being eliminated so as to not remain in the masked signal region of excitation energies near the $\eta^\prime$ threshold by optimizing the above-mentioned conditions. After fixing the event selection that provides a reasonable signal-to-noise ratio, the masked region will be opened to examine the existence of signals. A large part of the branching fractions for bound-$\eta^\prime$ final states are expected to be covered by combining this result with the one-nucleon absorption analysis. A tighter constraint for the in-medium $\eta^\prime$ mass reduction must be available in the combined analysis increasing the sample statistics.

\subsubsection{Direct measurement of the in-medium $\eta^\prime$ mass}
In the BGOegg Phase-I experiment, direct measurement of the in-medium $\eta^\prime$ mass through the $\eta^\prime \to \gamma \gamma$ decay channel was also adopted as the other analysis method complementary to the bound nuclei search \cite{prlmetap}. In this method, the energies and directions of two final-state $\gamma$'s were measured at the BGOegg calorimeter to reconstruct the $\gamma \gamma$ invariant mass for the inspection of spectral change. The detected $\gamma$'s have not interacted with materials or lost their energies before reaching the calorimeter, so that this direct measurement represents the true nature of the $\eta^\prime$ meson in a nucleus. Particularly, the BGOegg calorimeter has an excellent energy resolution and granularity, achieving the world's highest mass resolutions for the mesons decaying into $\gamma$'s among the experiments handling such $\gamma$-rays at the energies around $1$~$\mathrm{GeV}$ or less. Thus, the BGOegg experiment uniquely has high sensitivity for detecting the in-medium $\eta^\prime$ mass reduction if its absolute amount is greater than $20$~$\mathrm{MeV}$, which corresponds to the $\gamma \gamma$ invariant mass resolution of $\eta^\prime$. In the BGOegg experiment, the direct measurement can cover a smaller mass reduction range compared with the search for $\eta^\prime$ bound nuclei.

The event selection required that only two neutral clusters, reconstructed for the electromagnetic showers due to $\gamma$ hits, were detected in addition to one or no charged particle corresponding to the emission of a nucleon ($p$ or $n$). A $\gamma \gamma$ invariant mass spectrum was examined for the sample where the momentum of a $\gamma \gamma$ system ($P_{\gamma \gamma}$) was lower than $1$~$\mathrm{GeV/c}$ to enhance in-medium $\eta^\prime$ decays before going out of a nucleus, while high-momentum $\gamma \gamma$ pairs ($1 < P_{\gamma \gamma} < 1.5$~$\mathrm{GeV/c}$) were gathered to use them as a reference sample. The $P_{\gamma \gamma}$ cut at $1$~$\mathrm{GeV/c}$ was determined before the real data analysis so as to maximize the significance of a possible mass reduction signal using Monte Carlo (MC) simulations for in-medium $\eta^\prime$ decay signal and other background processes. The examined distribution of $\gamma \gamma$ invariant mass mainly consists of (i) a quasi-free $\eta^\prime$ peak arising from the decays outside a nucleus, (ii) $\pi^0 \pi^0$ or $\pi^0 \eta$ photoproduction events going to the $4 \gamma$ final state but with the escape of two $\gamma$'s from the detector acceptance, and (iii) $\omega$ photoproduction events decaying into $\pi^0 \gamma \to 3 \gamma$ with a missing $\gamma$. In both low- and high-momentum samples, the mass spectra of the processes (i) and (iii) were well reproduced by the MC simulation where the detector geometries, resolutions, and responses were realistically implemented. The spectrum for the process (ii) was expressed by a free-parameter function $\exp(p_0 + p_1 x + p_2 x^2 + p_3 x^3)$ after the confirmation of its validity using the MC simulations for $\pi^0 \pi^0$ and $\pi^0 \eta$ photoproduction. The template mass spectra of (i) and (iii) prepared by the simulations plus the free-parameter function for (ii) were simultaneously fitted to the real-data sample in the low-momentum region for the purpose to find the in-medium modification of the $\eta^\prime$ mass, which could not be explained by the above-listed processes (i)--(iii).

\begin{figure}[t]
 \centering
 \includegraphics[width=18cm]{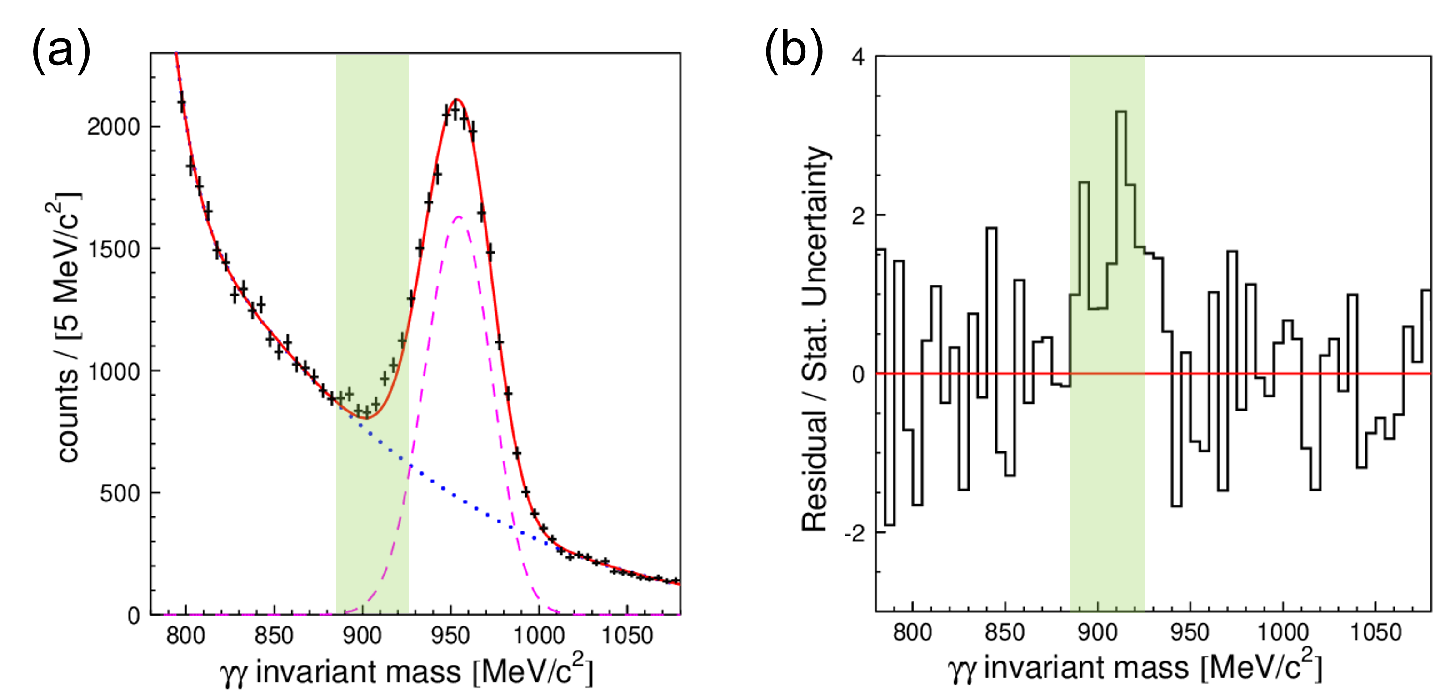}
 \caption{Results of the spectral fitting only with the processes (i)--(iii). The green hatched area indicates the region excluded from the fit. (a) The $\gamma \gamma$ invariant mass spectrum of the low-momentum sample in the carbon target data. The red-solid line shows the fitting result, while the magenta-dashed and blue-dotted lines represent the quasi-free $\eta^\prime$ component decaying in vacuum (the process (i)) and the sum of multi-meson production (the process (ii)) and $\omega$ photoproduction (the process (iii)) components, respectively. (b) The distribution of residuals over statistical uncertainties as a result of the fit in (a). {\it Source}: Figures taken from Ref.~\cite{prlmetap}.}
 \label{etapmass1}
\end{figure}

The analysis for the direct measurement was first done with the carbon target data collected in 2015 \cite{phdmatsu}, which was also used in the search for $\eta^\prime$ bound nuclei. In this initial analysis, the signal of in-medium $\eta^\prime$ mass reduction was indicated with a statistical significance more than $3 \sigma$. Therefore, a detailed confirmation analysis was further conducted by combining the carbon target data collected in 2015 and 2016 to increase the sample statistics nearly twice \cite{prlmetap}. The new analysis performed three complementary spectral fittings. First, the spectral functions of the processes (i)--(iii) were fitted to the $\gamma \gamma$ invariant mass distributions of the selected samples. In this inspection, only the low-momentum sample showed the significant deviation of a reduced $\chi^2$ from $1$, indicating an enhancement over the fit result in the lower tail region of the $\eta^\prime$ mass peak by the process (i). Thus, the same functions were fitted again to the $\gamma \gamma$ invariant mass distributions after excluding the range where the enhancement was observed. Figure~\ref{etapmass1}(a) and (b) show the result of this spectral fit for the low-momentum sample and the distribution of residuals divided by the statistical uncertainties of the individual bins, respectively. A clear enhancement is seen in the lower tail region with the statistical significance of $4.7 \sigma$. The obtained significance was unchanged even by varying the boundaries of the excluded region.

\begin{figure}[t]
 \centering
 \includegraphics[width=10cm]{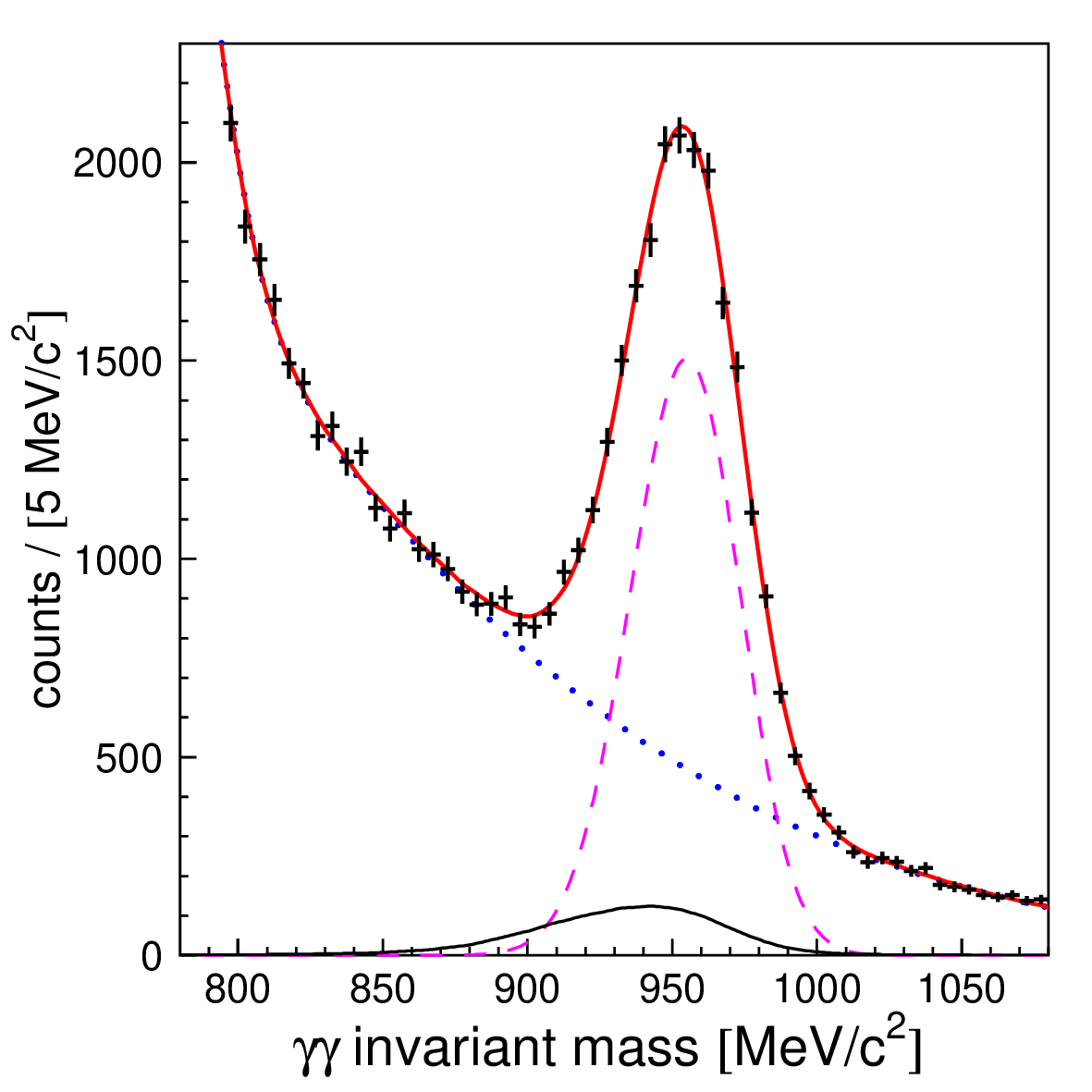}
 \caption{The same spectrum as Fig.~\ref{etapmass1} but with the fit using the functions that represent the simulated signal component and the processes (i)--(iii). The notation of lines is the same as Fig.~\ref{etapmass1} except for the black line, which indicates the signal function in the case of the best $\chi^2$ fit. {\it Source}: Figure taken from Ref.~\cite{prlmetap}.}
 \label{etapmass2}
\end{figure}

Secondly, spectral fits without the excluded region were performed using a signal function representing in-medium $\eta^\prime$ mass modification in addition to the functions representing the processes (i)--(iii) adopted in the first method. Here, the signal function was prepared as a template shape estimated by a MC simulation, where the production, propagation, and decay of an $\eta^\prime$ meson inside a nucleus occurred stochastically and the mass reduction at the decay point was assumed to be linearly dependent on the nuclear density following the Woods-Saxon distribution. Many template shapes were made by varying the mass reduction amount at the normal nuclear density as an input parameter of each simulation, whereas the in-medium width broadening was fixed at $20$~$\mathrm{MeV}$ based on the measurement by the CBELSA/TAPS experiment \cite{etapwid}. In the second method, $\chi^2$ difference tests comparing the fits with and without the signal function were done to evaluate the signal significances depending on the mass reduction amount at the normal nuclear density. As a result, the maximum significance of $4.1 \sigma$ was uniquely obtained when the in-medium $\eta^\prime$ mass reduction was set to $-(57.5^{+5.7}_{-27.8})$~$\mathrm{MeV}$. (See Fig.~\ref{etapmass2} for the best fit. Significance variation depending on the assumed mass reduction amount is shown in Fig.~3 of Ref.~\cite{prlmetap}.) Even if the mass reduction that was input for each signal simulation was conservatively treated as a free parameter of the spectral fit, the maximum statistical significance of the $\chi^2$ difference tests with one additional degree of freedom was still estimated at $3.7 \sigma$. Finally, the third fitting method was conducted in the manner similar to the second method, but the simulated signal function was replaced with a Gaussian shape to inspect a possible bias due to the signal shape assumption. This test supported the statistical significances obtained in the second method.

All the three methods of spectral fitting for the direct measurement show the statistical significances around $4 \sigma$ or more for the in-medium $\eta^\prime$ mass reduction signal in the low-momentum sample of the carbon target data. It has also been confirmed that the observed enhancement behaves like the true mass reduction signal by showing larger fractions of excess when the upper bound of $P_{\gamma \gamma}$ selection is lowered step-by-step from $1$~$\mathrm{GeV/c}$. In contrast, the $\gamma \gamma$ invariant mass distribution of the high-momentum sample was well reproduced only by the processes (i)--(iii). (See Fig.~1 of Ref.~\cite{prlmetap}.) In addition, other reference analyses were carried out using the low-momentum $\gamma \gamma$ samples of quasi-free $\eta$ photoproduction events in the carbon target data ($P_{\gamma \gamma} < 0.6$~$\mathrm{GeV/c}$, corresponding to the same $\beta \gamma$ range as the $\eta^\prime$ case) and $\eta^\prime$ photoproduction events in the liquid hydrogen target data ($P_{\gamma \gamma} < 1$~$\mathrm{GeV/c}$). (See Figs.~7 and 8 of Ref.~\cite{prlmetap}.) No enhancement was observed in the lower tail region of the quasi-free meson production peak for these reference samples, ensuring the validity of the adopted fitting methods without causing any significant systematic uncertainty in the interpretation of measured mass spectra. The obtained amount of in-medium $\eta^\prime$ mass reduction does not contradict the result from the bound nuclei search, suggesting a smaller reduction range of theoretical predictions. Such quantitative information is useful to develop effective QCD Lagrangians, as discussed in Sec.~\ref{subsubsection:third-first-first}. Details of the new results for the direct measurement are discussed in Ref.~\cite{prlmetap}.

Because the BGOegg Phase-I experiment has suggested evidence of in-medium $\eta^\prime$ mass reduction, the next experimental programs should focus on the direct measurement through the $\gamma \gamma$ decay channel. In the BGOegg Phase-II experiment, the charged particle detectors (RPC and DC) covering the forward acceptance hole of the BGOegg calorimeter have been replaced with an additional electromagnetic calorimeter (FG) made of 252 PWO crystals. The FG covers the polar angles less than $16^\circ$, and is being extended to erase the uncovered region up to $24^\circ$. Such detector upgrade is effective to reduce the $\pi^0 \pi^0$ or $\pi^0 \eta$ photoproduction backgrounds, namely the process (ii), down to $1/8$ ($1/40$) by setting the FG coverage to $\theta < 16^\circ$ ($\theta < 24^\circ$). The improvement of the background level should result in the increase of the signal significance and the decrease of systematic uncertainties. Then, the nuclear target was changed from carbon to copper to increase the nuclear radius $1.8$ times and raise the probability of in-medium decays. Simultaneously, the radiation shield of the LEPS2 beamline has been strengthened so as to allow $5$ times thicker target material, corresponding to $0.5$ radiation lengths ($X_0$). The $0.5 X_0$ copper target provides $1.8$ times more nucleons compared with the $0.1 X_0$ carbon target, contributing to the increase of data statistics. The thickness of a $0.5 X_0$ copper block is $7.5$~$\mathrm{mm}$, so that the $\gamma \gamma$ invariant mass resolution is also expected to be improved compared with a $0.1 X_0$ carbon block ($20$~$\mathrm{mm}$) thanks to the better vertex resolution for reconstructing a $\gamma$ direction. As mentioned in Sec.~\ref{subsec:second-first}, the $\gamma$-ray beam intensity of the LEPS2 beamline has been increased by introducing new pulsed lasers, also contributing to increase the data statistics. The data collection of the BGOegg Phase-II experiment has started from 2024 with the FG coverage of $\theta < 16^\circ$. More precise analyses are expected in the near future to explore and establish the in-medium $\eta^\prime$ natures.

\subsection{Baryon resonance spectroscopy in BGOegg experiment}\label{subsec:third-second}
Understanding the excited mass spectra of baryons remains a central objective in strong-interaction physics.
In the naive constituent–quark model,
baryon excitations are often described in terms of single-particle orbital motions of three valence quarks,
leading to a rather regular pattern of level spacings and spin multiplets.
While this picture reproduces the masses and decay properties of several low-energy states,
it becomes increasingly unreliable in the high-energy regions, especially for $W \geq 1.8$ GeV.
In this energy region, discrepancies between the observed and predicted spectra become apparent.
Some of the levels expected in the three-quark basis have not yet been associated with convincing experimental evidence (so-called "missing resonances").
Whether these discrepancies point to a limitation of the constituent-quark picture, or reflect the fact that states weakly coupled to the $\pi N$ channel have escaped detection in the reactions analyzed so far, is not yet settled.
In addition, the discrepancy between theoretical and experimental mass spectra includes the level reorderings such as the Roper resonance and the appearance of possible exotic configurations.
They indicate that the baryon excitation spectrum is formed by nontrivial QCD dynamics beyond a simple valence-quark potential.
To investigate this problem, it is necessary to experimentally clarify the properties of higher resonance states, including their masses, spins, and parities.
However, the identification of individual resonances in this energy region cannot be achieved by any single experiment.
The role of the measurements presented below is to supply cross sections and polarization observables in kinematic regions where no data exist, which enter global partial-wave analyses together with the data accumulated by various experiments so far.

\begin{figure}[tbp!]
 \centering
\includegraphics[width=0.75\textwidth]{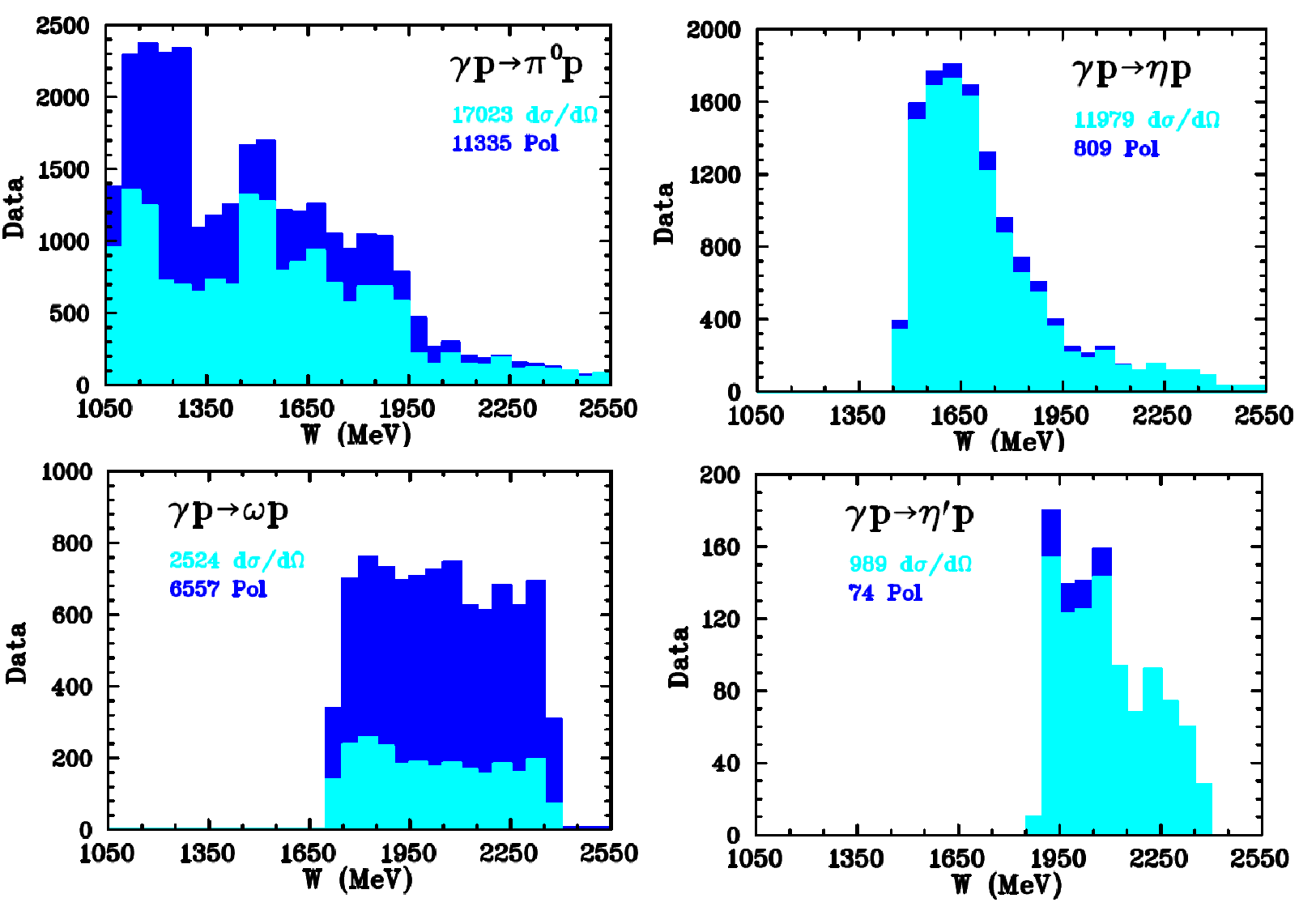}
 \caption{Number of experimental data in the SAID database \cite{SAID.web} for $\gamma p \rightarrow \pi^{0}p$(top-left), $\gamma p \rightarrow \eta p$(top-right), $\gamma p \rightarrow \omega p$(bottom-left), and $\gamma p \rightarrow \eta^{\prime} p$(bottom-right).
 Light-shaded histograms show the data of cross sections, and  dark-shaded histograms show the data of polarization observables.
 \textit{Source}: Figure taken from Ref.~\cite{IRELAND2020103752}.}
 \label{photopro_data}
\end{figure}

Experimentally, the most efficient reaction to probe these resonances is photoproduction.
When a real photon interacts with a proton, the electromagnetic interaction imposes well–defined initial quantum numbers, so that the reaction amplitude can be organized in a multipole basis in which each multipole corresponds to a definite spin-parity $J^{P}$.
This decomposition does not by itself isolate individual resonances.
Non-resonant contributions, in particular the $t$- and $u$-channel exchanges, become increasingly important in the high-energy region.
In addition, the resonances in the high-energy region have large widths and overlap with each other,
so that these resonances interfere complexly.
The decomposition of resonance contributions therefore requires energy-dependent partial wave analyses constrained by as many independent observables as possible.
The observables measured from various angular distributions—particularly those involving the polarization of a beam, a target, or a recoil baryon—directly encode bilinear interference terms among the multipoles.
Through such interference, small resonance components that are invisible in an unpolarized cross section measurement may become detectable in a partial wave analysis,
making photoproduction an especially sensitive filter in baryon spectroscopy.
Figure \ref{photopro_data} shows the number of available data for each neutral meson photoproduction reaction on the proton \cite{SAID.web, IRELAND2020103752}; $\gamma p \to \pi^{0} p$, $\gamma p \to \eta p$, $\gamma p \to \omega p$, and $\gamma p \to \eta^{\prime} p$.
In energy regions above $W \sim 2$ GeV, experimental data for polarization observables are scarce, particularly for $\eta$ and $\eta^{\prime}$.
The LEPS2/BGOegg experiment covers the energy range $1.8 < W < 2.3$ GeV and can provide experimental data in this region.

\subsubsection{Polarization observables}
\label{polarization_observables}
Pseudoscalar meson photoproduction is described by four complex amplitudes, and 16 observables can in principle be measured: the unpolarized cross section, the three single-polarization observables ($\Sigma$, $T$, $P$)
, and 12 double-polarization observables ($E$, $G$, $F$, $H$, $C_{x}$, $C_{z}$, $O_{x}$, $O_{z}$, $L_{x}$, $L_{z}$, $T_{x}$, $T_{z}$) \cite{PhysRevC.46.2430, Sandorfi_2011}.
In this section, we discuss the relations between the four complex amplitudes, such as the Chew-Goldberger-Low-Nambu(CGLN) amplitudes, and the multipole amplitudes that describe baryon resonances, and further the relations of these to the unpolarized differential cross section and the photon beam asymmetry, which are the observables mainly measured in the BGOegg experiment.

The differential cross section for pseudoscalar meson photoproduction in the center-of-mass system can be written as
\begin{eqnarray}
  d\sigma/d\Omega &=& \frac{1}{4}\sum_{m_{s_{i}}=\pm 1/2}\sum_{m_{s_{f}}=\pm1/2}\sum_{\gamma-\mathrm{spins}}\left(\frac{1}{4\pi}\right)^{2}\frac{q}{k}\frac{m_{p}^{2}}{W^{2}}|\bm{J}|^{2}, \\
  \bm{J} &=& \frac{4\pi W}{m_{p}}\langle m_{s_{f}} | F_{\mathrm{CGLN}} | m_{s_{i}} \rangle,
  \label{eq:dcs_CGLN}
\end{eqnarray}
 where $| m_{s} \rangle$ is the eigenstate of the Pauli operator $\sigma_{z}$ and $F_{\mathrm{CGLN}}$ is the CGLN amplitude \cite{PhysRev.106.1345}:
 \begin{equation}
  F_{\mathrm{CGLN}} = \sum_{i = 1}^{4}O_{i}F_{i}(\theta, E)
  \label{eq:CGLN1}
 \end{equation}
 with
 \begin{eqnarray}
  O_{1} &=& i\bm{\sigma} \cdot \bm{\varepsilon}, \\
  O_{2} &=& [\bm{\sigma} \cdot \hat{q}][\bm{\sigma} \cdot (\hat{k} \times \bm{\varepsilon})], \\
  O_{3} &=& i[\bm{\sigma} \cdot \hat{k}][\hat{q} \cdot \bm{\varepsilon}], \\
  O_{4} &=& i[\bm{\sigma} \cdot \hat{q}][\hat{q} \cdot \bm{\varepsilon}],
  \label{eq:Pauli}
 \end{eqnarray}
 where $\bm{\sigma}$ are the Pauli matrices, and $\hat{k}$ and $\hat{q}$ are the unit vectors of the photon and meson momenta, respectively.
Although the CGLN amplitudes are closely related to physical observables, they do not explicitly reflect the quantum numbers of the intermediate resonance.
Expressing the reaction in terms of electromagnetic multipoles makes the underlying partial-wave structure explicit and allows a direct identification of resonances with definite spin-parity $J^{P}$.
In this representation, the multipole amplitudes are written as $E_{l\pm}$ and $M_{l\pm}$, where $l$ denotes the orbital angular momentum and the $\pm$ corresponds to total spin $J = l \pm 1/2$ of the intermediate state.
Electric and magnetic multipoles produce states of opposite parity, establishing a one-to-one correspondence between each multipole and a resonance with definite $J^{P}$.
 The CGLN amplitude expanded by electromagnetic multipole amplitudes can be expressed as follows:
\small
 \begin{eqnarray}
  F_{1} &=& {\displaystyle \sum_{l=0}}[(E_{l+} + lM_{l+})P^{\prime}_{l+1}(\cos{\theta}) + \{E_{l-} + (l+1)M_{l-}\}P^{\prime}_{l-1}(\cos{\theta})], \\
  F_{2} &=& {\displaystyle \sum_{l=1}}[(l+1)M_{l+}P^{\prime}_{l}(\cos{\theta}) + lM_{l-}P^{\prime}_{l}(\cos{\theta})], \\
  F_{3} &=& {\displaystyle \sum_{l=1}}[(E_{l+} - M_{l+})P^{\prime\prime}_{l+1}(\cos{\theta}) + (E_{l-} + M_{l-})P^{\prime\prime}_{l-1}(\cos{\theta})], \\
  F_{4} &=& {\displaystyle \sum_{l=2}}(-E_{l+} + M_{l+} - E_{l-} - M_{l-})P^{\prime\prime}_{l}(\cos{\theta}),
  \label{eq:CGLN2}
 \end{eqnarray}
\normalsize
 where $P_{l}(\cos{\theta})$ are Legendre polynomials.
 Since electromagnetic multipole amplitudes are associated with the different spin-parities of resonance states,
 we can derive the contribution of each resonance state by solving the CGLN amplitudes.
The differential cross section $d\sigma/d\Omega$ and the photon beam asymmetry $\Sigma$ are represented as follows:
\small
\begin{eqnarray}
 \label{eq:dcs_cgln}
 d\sigma/d\Omega &=& \frac{q}{k}\mathrm{Re}\{\ff{1}{1} + \ff{2}{2} + \frac{1}{2}(\ff{3}{3} + \ff{4}{4})\sin^{2}{\theta} \notag \\
 &+& (\ff{2}{3} + \ff{1}{4} + \ff{3}{4}\cos{\theta})\sin^{2}{\theta} - 2\ff{1}{2}\cos{\theta}\}, \\
 \label{eq:ba_cgln}
 \Sigma &=& -\frac{q}{k}\mathrm{Re}\{\frac{1}{2}(\ff{3}{3} + \ff{4}{4}) + \ff{2}{3} + \ff{1}{4} + \ff{3}{4}\cos{\theta}\}\sin^{2}{\theta} / \left(d\sigma/d\Omega\right).
\end{eqnarray}
\normalsize
We can obtain the information of multipole amplitudes from angular-dependent differential cross sections and photon beam asymmetries.

The power to decompose amplitudes is amplified by the availability of a linearly polarized photon beam.
The photon beam asymmetry $\Sigma$, for instance, measures the azimuthal modulation of the cross section relative to the polarization plane.
It is an important observable because proposed complete experiments typically include its measurement.
In the regions around $W \sim 2$ GeV, there are contributions from both $s$-channel baryon excitations and $t$-channel meson exchange. Therefore, deviations of $\Sigma$ from Reggeized behavior suggest the persistence of $s$–channel strength.
In photoproduction of the vector meson whose spin is 1, spin–density matrix elements provide even more detailed information about the intermediate state,
encoding the correlations between the photon polarization and the vector–meson decay angular distribution.
Such quantities offer a stringent test of $s$–channel helicity conservation and natural/unnatural–parity exchange,
thereby discriminating reaction mechanisms that would remain indistinguishable in an unpolarized measurement.

\subsubsection{\texorpdfstring{Experimental results for $\gamma p \to \pi^{0} p$, $\eta p$, and $\omega p$ reactions}{}}
\label{light_meson_photoproduction}
The LEPS2/BGOegg experiment at SPring-8 has reported measurements of  
$\gamma p \to \pi^{0} p$ \cite{PhysRevC.100.055202}, $\gamma p \to \omega p$ \cite{PhysRevC.102.025201}, and $\gamma p \to \eta p$ \cite{PhysRevC.106.035201}, obtained under nearly identical event selection conditions.
All three experiments utilize Compton-backscattered photons from the 8 GeV storage ring,
reaching a maximum beam energy of $E_{\gamma}\sim 2.4$ GeV with a linear polarization approaching 90\% near the Compton edge,
and they employ the large-acceptance BGOegg calorimeter together with forward tracking detectors.
Data collection has been done using a $54$ mm-thick liquid hydrogen target at the center of BGOegg calorimeter.
As a result, these measurements cover a remarkably wide angular range, including extremely backward meson angles where previous data have been sparse or inconsistent.
The $\pi^{0}$, $\eta$, and $\omega$ mesons have several features that make them useful probes for baryon resonances.
All three are neutral mesons with multi-photon decay modes, allowing clean measurements.
Their photoproduction reactions are therefore well suited to studying intermediate baryon excitations.
Moreover, the three mesons provide complementary filters on the baryon spectrum.
The $\pi^{0}$ couples to both $N^{*}$ and $\Delta^{*}$ ($I = 1/2$ and $3/2$),
while the $\eta$ is an isoscalar and restricts the intermediate states to $I = 1/2$.
The $\omega$ is also isoscalar, but as a vector meson it provides access to additional polarization
observables such as spin-density matrix elements (SDMEs).
These channels offer complementary information on baryon excitations.

In the BGOegg experiment, neutral mesons ($\pi^{0}, \eta, \omega$) are identified through the detection of decay photons by using the BGOegg electromagnetic calorimeter.
The mesons are reconstructed from the invariant mass distributions of the detected photons.
A recoil proton is measured either by the BGOegg or the DC, which in total cover most of the solid angle.
In principle, only the angle of the recoil proton can be measured.
When it reaches the RPC, its momentum can be obtained from the time-of-flight and used to check the validity of the event reconstruction.
Since all the decay photons from a meson and the recoil proton are measured, the reaction can be identified with high confidence.
To suppress background contributions, we apply a kinematic fit.
The input kinematic variables are the four-momenta of $\gamma$’s, the polar and azimuthal angles of the recoil proton, the photon beam energy, and a vertex z-position.
The recoil proton momentum was treated as an unmeasured variable.
The kinematic fit imposes four-momentum conservation and constrains the mass of mesons reconstructed via two-photon decays ($\pi^{0}$ and $\eta$) to their nominal values, while no mass constraint is imposed on the $\omega$ meson (only the intermediate $\pi^{0}$ is constrained).

\begin{figure}[tbp!]
 \centering
\includegraphics[width=0.64\textwidth]{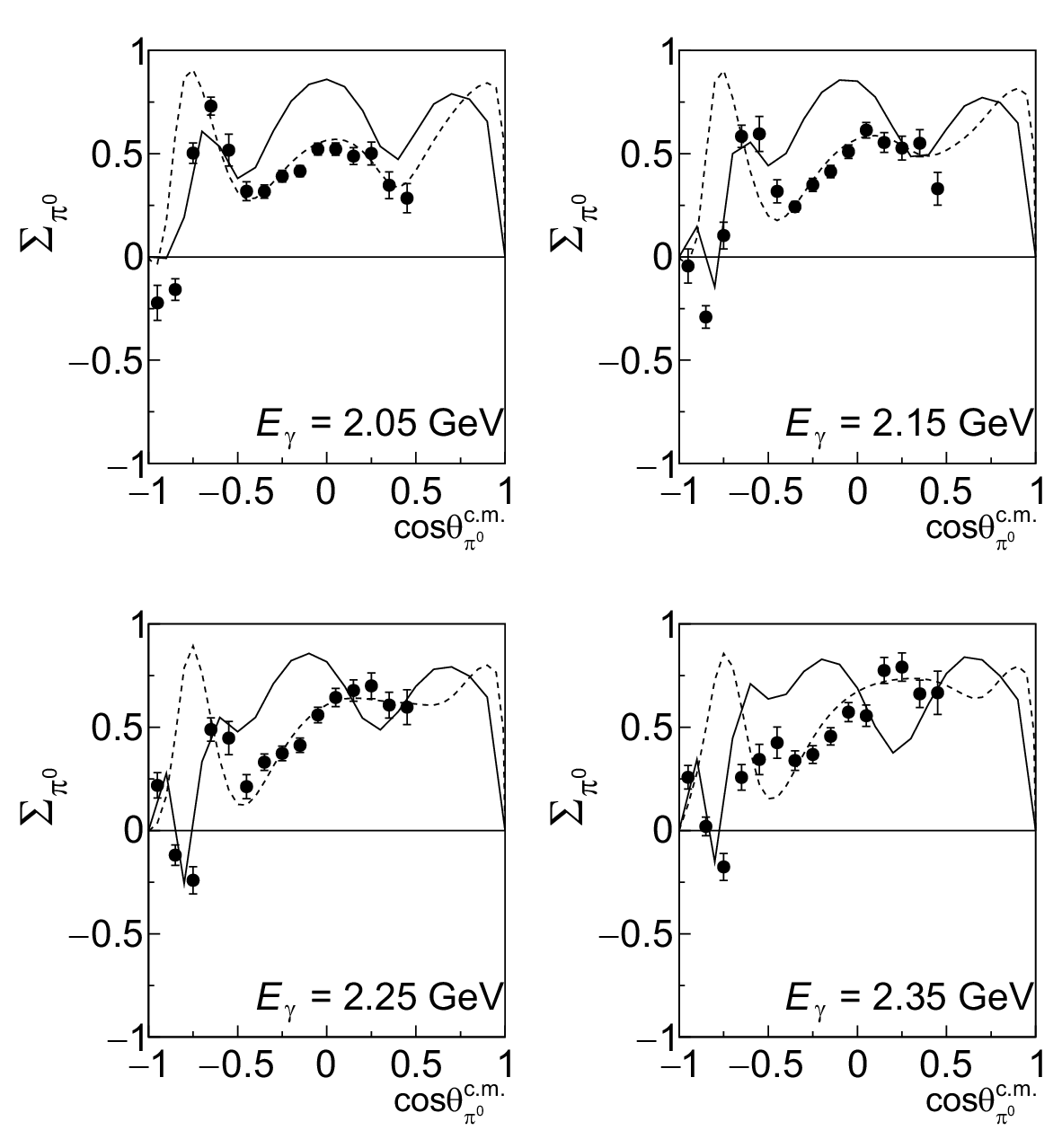}
 \caption{Photon beam asymmetries for the $\pi^{0}$ photoproduction reaction in the region of $E_{\gamma} > 2.0$ GeV. The solid and dashed curves show the model calculations by the Bonn-Gatchina \cite{BnGa.web} and the SAID \cite{SAID.web} groups, respectively.
          \textit{Source}: Figure taken from Ref.~\cite{PhysRevC.100.055202}.}
 \label{pi0_bma}
\end{figure}

Neutral-pion photoproduction has historically served as the baseline channel for baryon spectroscopy because the $\pi^{0}$,
as an isotriplet, couples not only to $N^{*}$ but also to $\Delta^{*}$ resonances,
and because its relatively low production threshold has enabled extensive measurements at many experimental facilities.
The $\pi^{0}$ is identified through the $2\gamma$ decay channel: $\pi^{0} \to \gamma\gamma$.
The branching ratio of this decay is $98.82 \pm 0.03$\%.
The BGOegg measurement provides differential cross sections and photon beam asymmetries for $-1 < \cos{\theta^{\mathrm{c.m}}_{\pi^{0}}} < 0.7$ and $E_{\gamma} = 1.3 - 2.4$ GeV,
including the first wide-angle asymmetry data above $W \sim 1.9$ GeV \cite{PhysRevC.100.055202}.
The differential cross section for the $\pi^{0}$ photoproduction reaction has been measured in many experiments \cite{PhysRevLett.94.012003, PhysRevC.84.055203, PhysRevC.76.025211, PhysRevC.80.052201}, and the BGOegg results are consistent with the other data.
A central finding from BGOegg is that none of the existing partial-wave analyses \cite{SAID.web, BnGa.web} (including the fits tuned to earlier CLAS \cite{PhysRevC.88.065203} and CBELSA/TAPS \cite{PhysRevC.81.065210} data) successfully reproduce the measured $\Sigma$ in the high-energy and the wide-angle region.
The angular dependence of the $\Sigma$ showed a significantly different behavior: the $\Sigma$ data display the oscillatory patterns characteristic of interference between higher-$L$ multipoles.
Figure \ref{pi0_bma} shows the measured photon beam asymmetries in the higher energy region of $E_{\gamma} > 2.0$ GeV.
The BGOegg data suggested that orbital angular-momentum components as high as $L$ = 5, i.e., amplitudes in the G- and H-wave sectors, may be required to explain the observed asymmetries  \cite{PhysRevC.100.055202}.
Although the $N(1680)$ and $\Delta(1700)$ remain essential in the conventional third-resonance region, the rise of $L = 5$ strength may imply that additional high-spin states ($J \geq 7/2$) contribute in the $1.9 < W < 2.1$ GeV region.

The $\eta$ meson has an isoscalar pseudoscalar nature and therefore couples only to $N^{*}$ ($I$ = 1/2), while also being sensitive to possible $s\bar{s}$ admixtures.
Previous measurements by CLAS \cite{PhysRevC.80.045213, CLAS:2017rxe}, CBELSA/TAPS \cite{PhysRevC.80.055202, PhysRevLett.125.152002}, and LEPS \cite{PhysRevC.80.052201} had reported a backward-angle enhancement above $W \sim 2$ GeV, but there were significant inconsistencies among their differential cross sections.
In the BGOegg experiment, the $\eta$ was identified through the $2\gamma$ decay channel: $\eta \to \gamma\gamma$.
The branching ratio of this decay is $39.41 \pm 0.20$\%.

\begin{figure}[tbp!]
 \centering
 \includegraphics[height=1.00\textwidth,angle=-90]{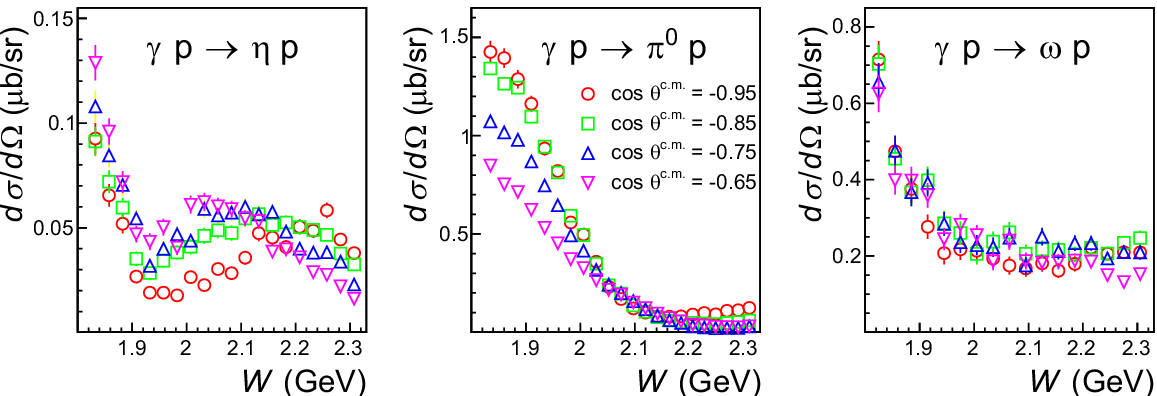}
 \caption{The differential cross sections of the $\eta$ (left), $\pi^{0}$ (middle), and $\omega$ (right) photoproduction processes as a function of the center-of-mass energy \textit{W} in the backward meson angle regions \cite{PhysRevC.100.055202, PhysRevC.102.025201, PhysRevC.106.035201}.
 Only the differential cross sections of $\eta$ photoproduction show a clear bump structure.
 \textit{Source}: Figure taken from Ref.~\cite{PhysRevC.106.035201}}
 \label{dcs_comp}
\end{figure}

The resulting differential cross sections in the range $-1 < \cos{\theta^{\mathrm{c.m}}_{\eta}} < 0.6$ and $W = 1.82 - 2.32$ GeV confirm a pronounced backward-angle bump at $W = 2.0 - 2.3$ GeV, as shown in the left panel of Fig.~\ref{dcs_comp} \cite{PhysRevC.106.035201}.
The BGOegg result was consistent with the differential cross sections measured by CLAS in the overlapping angular region $\cos{\theta^{\mathrm{c.m}}_{\eta}} > -0.8$,
and provided new precise values in the extremely backward angles.
At the backward $\eta$ angles, the enhancement of differential cross sections can originate
either from $u$-channel exchanges or from the decay of high-spin $s$-channel resonances.
This point was discussed in Ref.~\cite{PhysRevC.106.035201}.
Regge phenomenology suggests that the $u$-channel cross section has a smooth energy dependence of the form $s^{2\alpha(u)-2}$, with $2\alpha(u)-2$ expected to be negative at small $|u|$; such a term cannot generate an enhancement restricted to the narrow range $2.0 < W < 2.3$ GeV.
Furthermore, the EtaMAID2018 analysis \cite{Tiator2018}, which parameterizes the non-resonant background by $s$- and $u$-channel Born terms together with $t$-channel vector-meson exchanges, yields only a small $u$-channel amplitude.
The $t$-channel meson exchange, in turn, is strongly suppressed at backward meson angles.
Therefore, the backward enhancement is likely attributed to the $s$-channel resonances.
Since the helicity of the initial $\gamma p$ state is limited to $|h|\leq3/2$, a resonance with $J\geq5/2$ preferentially emits the $\eta$ meson toward extremely backward or forward angles in a two-body decay, following the corresponding Wigner $d$ functions;
the backward region is thus selectively sensitive to high-spin states.

In Fig.~\ref{dcs_comp}, the differential cross section of backward $\eta$ photoproduction is compared with the results of other reaction channels.
The $\pi^{0}$ and $\omega$ cross sections, obtained from the same data set, show no comparable structure, as shown in the middle and right panels.
In the flavor SU(3) classification, the $\eta$ has an $s\bar{s}$ component while the $\pi^{0}$ and $\omega$ are built from $u\bar{u}$ and $d\bar{d}$ alone.
This may suggest that the high-spin resonances responsible for the observed enhancement have stronger coupling with $\eta N$ channel, possibly possessing an $s\bar{s}$ component.
More importantly, the $\eta$ beam-asymmetry data above $E_{\gamma} \sim 2.1$ GeV, obtained for the first time with a highly polarized photon beam, provide new constraints on the relative contributions of such high-spin resonances.
The backward enhancement is interpreted as evidence for high-spin states that appear in the $s$-channel, possibly including poorly known states in the $2.1 - 2.3$ GeV region(e.g.,  $N(2120) \frac{3}{2}^{-}$, $N(2190) \frac{7}{2}^{-}$,
$N(1895) \frac{9}{2}^{+}$, $N(2250) \frac{9}{2}^{-}$).

The $\omega$ is a vector meson ($J=1$) with a high mass, and therefore, it may be sensitive to resonance states different from those coupling more with the pseudoscalar mesons $\pi^{0}$ and $\eta$. 
Photoproduction of a vector meson is described by 12 complex amplitudes because of its three helicity states.
Thus, more observables compared with pseudoscalar meson photoproduction must be measured by various experiments complementarily.
In the BGOegg experiment, the $\omega$ meson is reconstructed via the radiative decay channel $\omega \to \pi^{0}\gamma \to 3\gamma$.
The branching ratio of the $\omega \to \pi^0 \gamma$ decay is $8.33 \pm 0.25$\%.
The BGOegg measurement reports differential cross sections, photon beam asymmetries, and SDMEs for $-1 < \cos{\theta^{\mathrm{c.m}}_{\omega}} < 0.8$ and $W = 1.8 - 2.3$ GeV \cite{PhysRevC.102.025201}.
In particular, the photon beam asymmetries and the polarized SDME were measured for the first time in the energy region above $W=2$ GeV.

\begin{figure}[tbp!]
 \centering
 \includegraphics[height=0.8\textwidth,angle=-90]{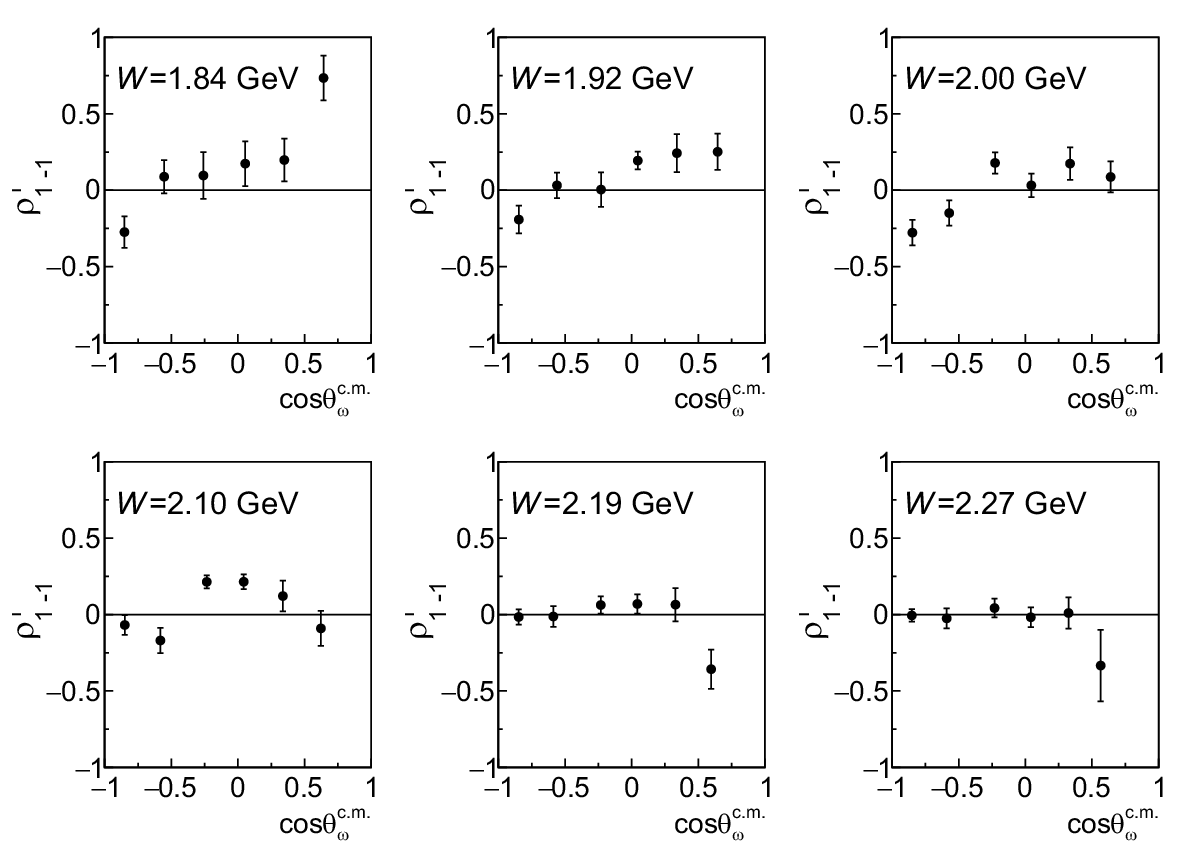}
 \caption{Spin density matrix elements $\rho^{1}_{1 -1}$ of the reaction $\gamma p \to \omega p \to \pi^{0}\gamma p$.
 \textit{Source}: Figure taken from Ref.~\cite{PhysRevC.102.025201}}
 \label{omg_polda}
\end{figure}

The BGOegg result obtained for $\omega$ photoproduction show the following features.
\begin{enumerate}
  \item The measured differential cross sections are consistent with other experimental results except for the CBELSA/TAPS measurement \cite{Dietz2015, Wilson2015407},
which shows higher values at backward $\omega$ angles.
The BGOegg result solves a problem of the long-standing discrepancy.
  \item The measured photon beam asymmetries are consistent with other experimental results in the overlapping energy region.
        In the newly obtained high-energy region, the photon beam asymmetry at backward $\omega$ angles shows discrepancies from existing partial wave analyses \cite{PhysRevC.102.025201}.
  \item The earlier $\omega$ photoproduction measurements with unpolarized beams (SAPHIR \cite{Barth2003}, CLAS \cite{PhysRevC.80.065208}, CBELSA/TAPS \cite{Wilson2015407}) determined the unpolarized SDMEs $\rho^{0}_{0 0}$, $\rho^{0}_{1 -1}$, and Re$\rho^{0}_{1 0}$.
        The BGOegg experiment obtained $\rho^{0}_{0 0}$ and $\rho^{0}_{1 -1}$, which are sensitive to double and single helicity-flip amplitudes in $\omega$ photoproduction, respectively.
        They were consistent with the previous results.
  \item The polarized SDME $\rho^{1}_{1 -1}$ was measured by the BGOegg experiment in the total energy range $1.81 < W < 2.32$ GeV, as shown in Fig.~\ref{omg_polda}.
        It corresponds to the pion asymmetry $\Sigma_{\pi}$ obtained by the CBELSA/TAPS collaboration \cite{PhysRevD.78.117101},
        but their measurement is limited to the lower energy region.
        The $\rho^{1}_{1 -1}$ takes the values $+0.5$ and $-0.5$ for pure pomeron and pure pion exchange in the $t$-channel, respectively, whereas $s$- and $u$-channel processes lead to $\rho^{1}_{1 -1} = 0$.
        The measured values are close to zero over the entire angular range, although a slight angular dependence is observed.
        For $W \geq 2.1$ GeV, they remain close to zero except at forward angles, where they approach $-0.5$.
        The angular dependence of the measured $\rho_{1 -1}^{1}$ indicates that while $s$-channel processes dominate in the backward and middle angular range, some contributions from $t$-channel mechanisms appear at forward angles.
\end{enumerate}
As a possible physical interpretation, the $\omega$ channel may lie in a transitional regime in which non-Reggeized dynamics still play an important role.
The combination of isospin filtering, photon beam asymmetries, and SDMEs provides new inputs to partial wave analyses in the higher energy region, where the contributing $N^{*}$ states are not well established. 

In summary, the three channels probe the 2 GeV region of baryon resonances in complementary ways.
In $\pi^{0}$ photoproduction, the angular dependence of $\Sigma$ likely requires higher multipoles, indicating that high-spin strength.
In $\omega$ photoproduction, the polarized SDME $\rho_{1 -1}^{1}$ stays close to zero over the backward and middle angular range, suggesting the importance of $s$-channel processes even at $W \approx 2.3$ GeV.
In $\eta$ photoproduction, the backward enhancement, absent in the other two channels measured with the same data set, indicates the possibility of high-spin resonances, which may be more than one according to the angular dependence of enhancement.
These observed features may suggest the existence of resonances that have not been fully taken into account in the existing PWAs.
Combined with the accumulated data from other experiments,
the BGOegg results provide complementary input for the partial wave analysis.

\subsubsection{\texorpdfstring{Experimental results for $\gamma p \to \eta^{\prime}p$ and $\pi^{0} \eta p$ reactions}{}}
\label{preliminary_photoproduction}
The issues addressed in Sec.~\ref{light_meson_photoproduction} are also closely related to the photoproduction of heavier mesons and multi-meson final states.
In particular, the $\gamma p \to \eta^{\prime} p$ and $\gamma p \to \pi^{0} \eta p$ reactions provide access to complementary information on baryon excitations in the higher total energy regions.
Results from these channels are described elsewhere \cite{s7m3-8xjq, Hashimoto:20258M}.

The $\eta^{\prime}$ meson is a pseudoscalar particle with the same quantum numbers as the $\eta$ meson.
However, its mass of about $958 \mathrm{MeV}/c^{2}$ is significantly larger than that of the $\eta$ meson.
This unusually large mass is associated with the axial ${U_{A}(1)}$ anomaly and the gluonic degrees of freedom.
In photoproduction, the $\eta^{\prime}$ can couple with the nucleon resonances contributing to the $\eta$ case, but its higher production threshold makes it particularly sensitive to high-mass states.
The GRAAL collaboration reported the first $\Sigma$ data in $\eta^{\prime}$ photoproduction near threshold and observed a characteristic angular dependence that had not been reported by earlier partial-wave solutions and was discussed in association with narrow resonances \cite{LeviSandri2015}.
This is a good example in which a highly polarized beam is essential since the effect appears in a polarization observable rather than in the cross section.

\begin{figure}[tbp]
 \centering
 \includegraphics[width=1.0\textwidth]{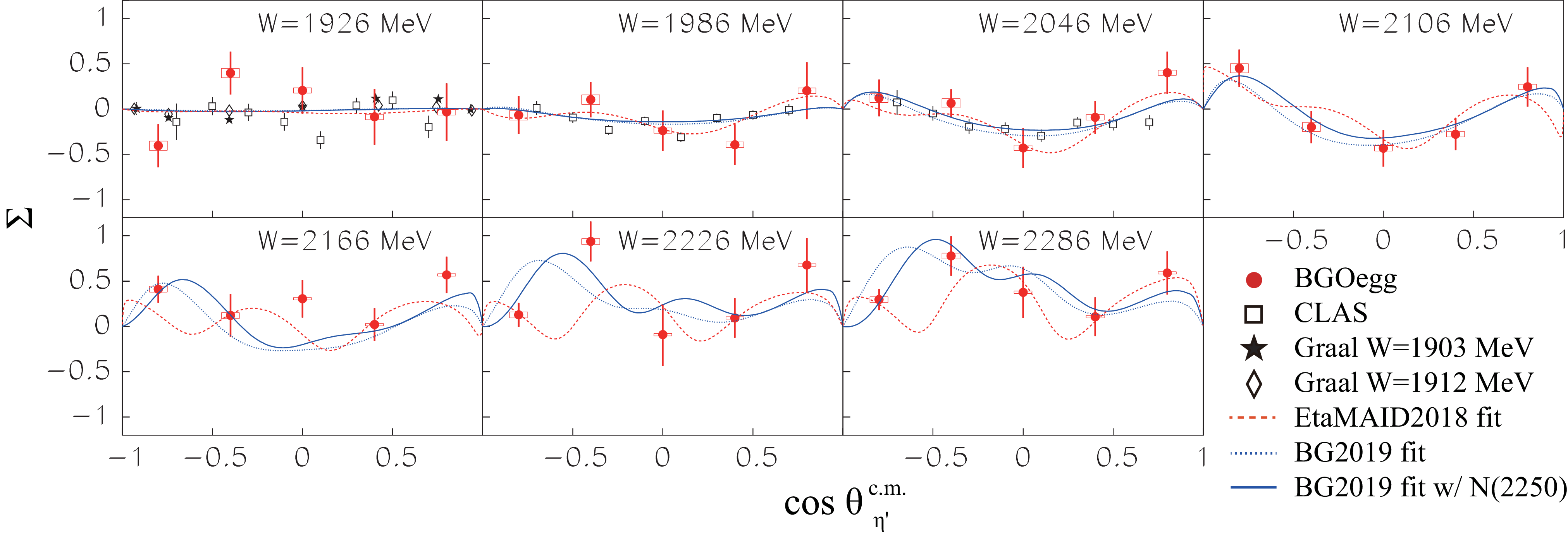}
 \caption{Photon beam asymmetries $\Sigma$ of the  $\gamma p \to \eta^{\prime} p$. Data points with statistical uncertainties for the BGOegg analysis and the previous results (GRAAL and CLAS) are shown simultaneously. Two partial wave analysis fits using the BGOegg result are also overlaid.
 \textit{Source}: Figure taken from Ref.~\cite{s7m3-8xjq}.}
 \label{pba_etap}
\end{figure}

In the BGOegg experiment, two decay channels were adopted in the data analysis.
The decay $\eta^{\prime} \to \gamma \gamma$ with a branching ratio of $2.31 \pm 0.03$\% enables a straightforward two-photon identification analogous to that of $\pi^{0}$ and $\eta$.
However, the small branching ratio of this decay limits the achievable statistics.
To increase the statistical sensitivity, the hadronic decay chain $\eta^{\prime} \to \pi^{0} \pi^{0} \eta$, which has a much larger branching ratio of $22.4 \pm 0.5$\%, was therefore included in the analysis.
This decay channel leads to a six-photon final state through subsequent decays $\pi^{0} \to \gamma \gamma$ and $\eta \to \gamma \gamma$.
The acceptance for detecting all the six photons is small, so the final statistics are comparable to those for the two photon decay mode.
The signal statistics was doubled by combining them after a consistency check between the two channels having different backgrounds and systematics.
The $\eta^{\prime}$ mass resolutions were $10.5$ and $6.3$ $\mathrm{MeV}/c^{2}$, for two- and six-photon channels, respectively.
We measured differential cross sections and photon beam asymmetries covering $1.9 < W < 2.3$ GeV.
In the differential cross section measurement, the most precise data to date were obtained at extremely backward $\eta^\prime$ angles.
Total cross sections were also derived from the differential cross sections whereas the only previous data was available from CBELSA/TAPS \cite{PhysRevC.80.055202}.
As shown in Fig.~\ref{pba_etap}, the measured photon beam asymmetries were consistent with the preceding results from GRAAL \cite{LeviSandri2015} and CLAS \cite{CLAS:2017rxe} at lower energies.
It was impossible to examine the precise angular-dependent behavior of $\Sigma$ with fine energy binning, reported by the GRAAL collaboration, because the statistics were not enough.
On the other hand, the $\Sigma$'s for $W > 2.1$ GeV were measured for the first time by BGOegg,
providing new information for partial wave analyses.
The EtaMAID2018 \cite{Tiator2018} and BG2019 \cite{2020135323} models were fitted to the BGOegg data in partial wave analyses with increased weights.
The fitting results for differential cross sections were essentially unchanged by including the BGOegg data,
while $\Sigma$ curves substantially moved by the new fits, particularly at backward angles and high energies.
The new constraint thus acts on the polarization observable rather than on the cross section.
In the EtaMAID2018 fit, where nine $N^{*}$ states were taken into account, a significant change was observed in the $\eta^{\prime} N$ coupling of $N(2250) (J^{P}=9/2^{+})$.
The BG2019 model had not contained this state originally, but adding a $J^{P}=9/2^{+}$ resonance to it improved the reduced $\chi^{2}$ for the measured $\Sigma$ from $1.8$ to $0.9$ (See also Fig.~\ref{pba_etap}).
Therefore, both fit results possibly imply the importance of $N(2250)$ in $\eta^{\prime}$ photoproduction.

Multi-meson photoproduction channels becomes important above $W = 2$ GeV, where their cross sections increase.
The reaction with the largest cross section is charged double-pion photoproduction, $\gamma p \to \pi^{+} \pi^{-} p$, but charged final states are dominated by large non-resonant contributions from $t$-channel $\rho$ production, from meson exchange and Born terms feeding $\Delta \pi$, and from the Kroll–Ruderman contact term \cite{TEJEDOR1994667, GOMEZTEJEDOR1996413}.
These mechanisms require a charged pion at the photon vertex and
are therefore strongly suppressed in neutral final states such as $\pi^{0} \pi^{0} p$ and $\pi^{0} \eta p$, where in addition $\rho \to \pi^{0} \pi^{0}$ is forbidden.
Thus, neutral final states are expected to be more sensitive to $s$-channel resonance excitations \cite{Zehr2012}.
In addition, a single polarization observable with the linear polarization of a photon beam can be extended to quasi two-body and genuine three-body asymmetries in multi-meson photoproduction.
For a two-body final state, a linearly polarized beam defines a single reaction plane, and the azimuthal modulation of the yield is described by one observable $\Sigma$.
In the case of a three-body final state, the asymmetry definition is richer; the final-state momenta define two independent planes (i.e., reaction and decay planes). The azimuthal dependence of a cross section is described by two asymmetries, $I^{\mathrm{C}}$ and $I^{\mathrm{S}}$ \cite{PhysRevC.71.055201},
 \begin{equation}
  \sigma = \sigma_{0}[1 - P_{\gamma}(I^{\mathrm{C}}\cos{2\phi} + I^{\mathrm{S}}\sin{2\phi})],
 \end{equation}
where $\sigma_{0}$ is the \textit{unpolarized} differential cross section, $P_{\gamma}$ is the degree of linear polarization of a photon beam, $\phi$ is the angle between the polarization direction and the reaction plane.
Precisely saying, $I^{\mathrm{C}}$ and $I^{\mathrm{S}}$ depend on the angle between the reaction and decay planes.
$I^{\mathrm{C}}$ is the natural extension of $\Sigma$, and can be evaluated for the quasi two-body configurations of $\gamma p \to (\pi^0 \eta) p$, $(\pi^0 p) \eta$, and $(\eta p) \pi^0$.
$I^{\mathrm{S}}$, in contrast, is odd under the reflection of a final state through the reaction plane and vanishes for two-body kinematics.
It is non-zero only through the interference of amplitudes carrying different phases,
and therefore constrains the relative phases of the contributing partial waves — information that is averaged out in two-body measurements \cite{Sokhoyan201515095x, GUTZ2014140741}.
As a result, the measurement of the three-body asymmetries provides additional constraints for the partial-wave analysis. 

In the BGOegg experiment, the $\gamma p \to \pi^{0} \eta p$ reaction was measured at $1.8 < W < 2.3$ GeV.
Both $\pi^{0}$ and $\eta$ were identified through their $2\gamma$ decay channels.
To identify the $\gamma p \to \pi^{0} \eta p \to \gamma\gamma\gamma\gamma p$ reaction,
events with four neutral clusters and one charged particle were selected.
To reduce background events, a kinematic fit was performed under the constraints of four-momentum conservation and nominal $\pi^{0}$ and $\eta$ masses.
After applying a 99\% confidence level cut in the kinematic fit,
the typical background contribution from the $\gamma p \to 3\pi^{0} p$ reaction with two undetected photons is suppressed to the 10\% level.
We measured the quasi-two-body and three-body photon beam asymmetries of the $\gamma p \to \pi^{0} \eta p$ reaction in $1.3 < E_{\gamma} < 2.4$ GeV.
In the lower energy region $E_{\gamma} < 1.65$ GeV,
the BGOegg results are consistent with the previous results measured by CBELSA/TAPS collaboration \cite{GUTZ2014140741}.
In the higher energy regions, the first-time data for this reaction were obtained as new inputs to the partial wave analysis.

In the reactions whose final states contain many photons, such as the $\eta^{\prime}$ and $\pi^{0}\eta$ channels,
the detector acceptance for photons becomes particularly important.
In the BGOegg Phase-II experiment, the introduction of the forward PWO calorimeter (FG) will increase the acceptance for these reactions by several times, enabling to provide data with much better statistical precision in the near future.

\subsection{Other physics analyses in BGOegg experiment}\label{subsec:third-third}

\subsubsection{$f_0$(980) photoproduction}
The data collected in the BGOegg experiment are usable for a wide range of physics analyses. One of such analyses was done for photoproduction of the scalar meson $f_0$(980) off the proton target. The structure of the $f_0$(980) meson has been controversial for a long time, being discussed as a possible candidate of the tetraquark state or the $K\bar{K}$ molecule \cite{JPG.28.R249, PhysRep.389.61, PhysRep.454.1}. Recently, Ref.~\cite{f0theory} has suggested that the differential cross section $d\sigma / dt$ of $f_0$(980) photoproduction at a small momentum transfer $-t$ may depend on the structure of this scalar meson. In this article, a Regge model calculation with the $t$-channel exchange of $\rho$ and $\omega$ mesons shows the larger $d\sigma / dt$ for the diagram implying the stronger contribution of a non-exotic $q\bar{q}$ component in the $f_0$(980) meson. On the other hand, the diagram containing an intermediate kaon loop, which more likely couples with the exotic structures, must return much lower differential cross sections. Thus, the measurement of $d\sigma / dt$ was carried out in the BGOegg experiment for the comparison with the theoretical predictions using the different models sensitive to the $f_0$(980) structure.

\begin{figure}[t!]
      \centering
      \includegraphics[width=10cm,height=9cm]{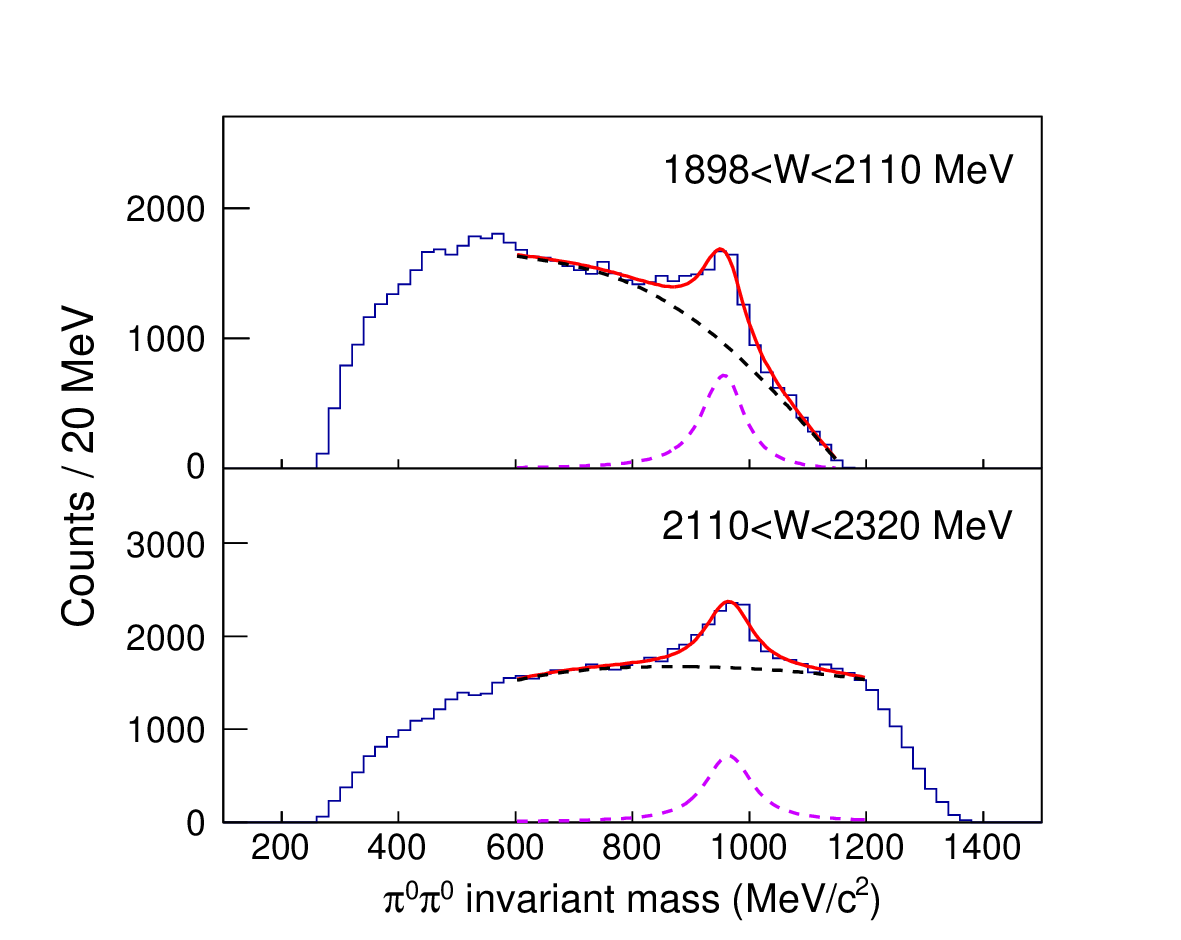}
      \caption{Invariant mass distributions of $\pi^0 \pi^0$ in two total energy regions. Voigt functions are fitted with polynomial background functions to extract $f_0$(980) signals. {\it Source}: Figure taken from Fig.~1 of Ref.~\cite{rf2023}.}
      \label{f0mass}
\end{figure}
\begin{figure}[t!]
  \centering
  \begin{tabular}{cc}
    \begin{minipage}{0.45\textwidth}
      \centering
      \includegraphics[width=7cm,height=8cm]{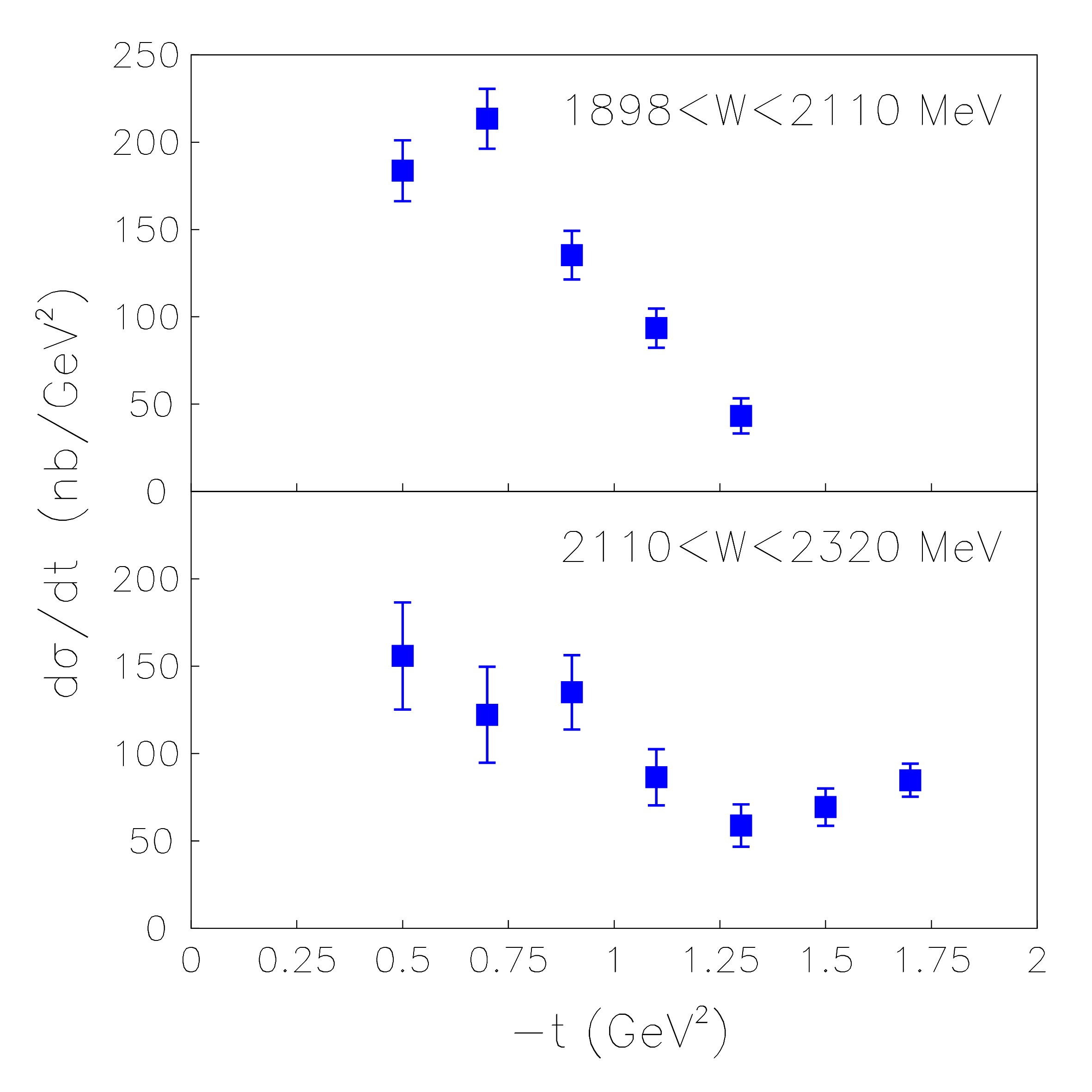}
      \caption{Differential cross sections $d\sigma / dt$ of the reaction $\gamma p \to f_0$(980)$p \to \pi^0 \pi^0 p$. {\it Source}: Figure taken from Fig.~3 of Ref.~\cite{rf2023}.}
      \label{f0dcs}
    \end{minipage}
    \hspace{5.0mm}
    \begin{minipage}{0.45\textwidth}
      \centering
      \includegraphics[width=7cm,height=8cm]{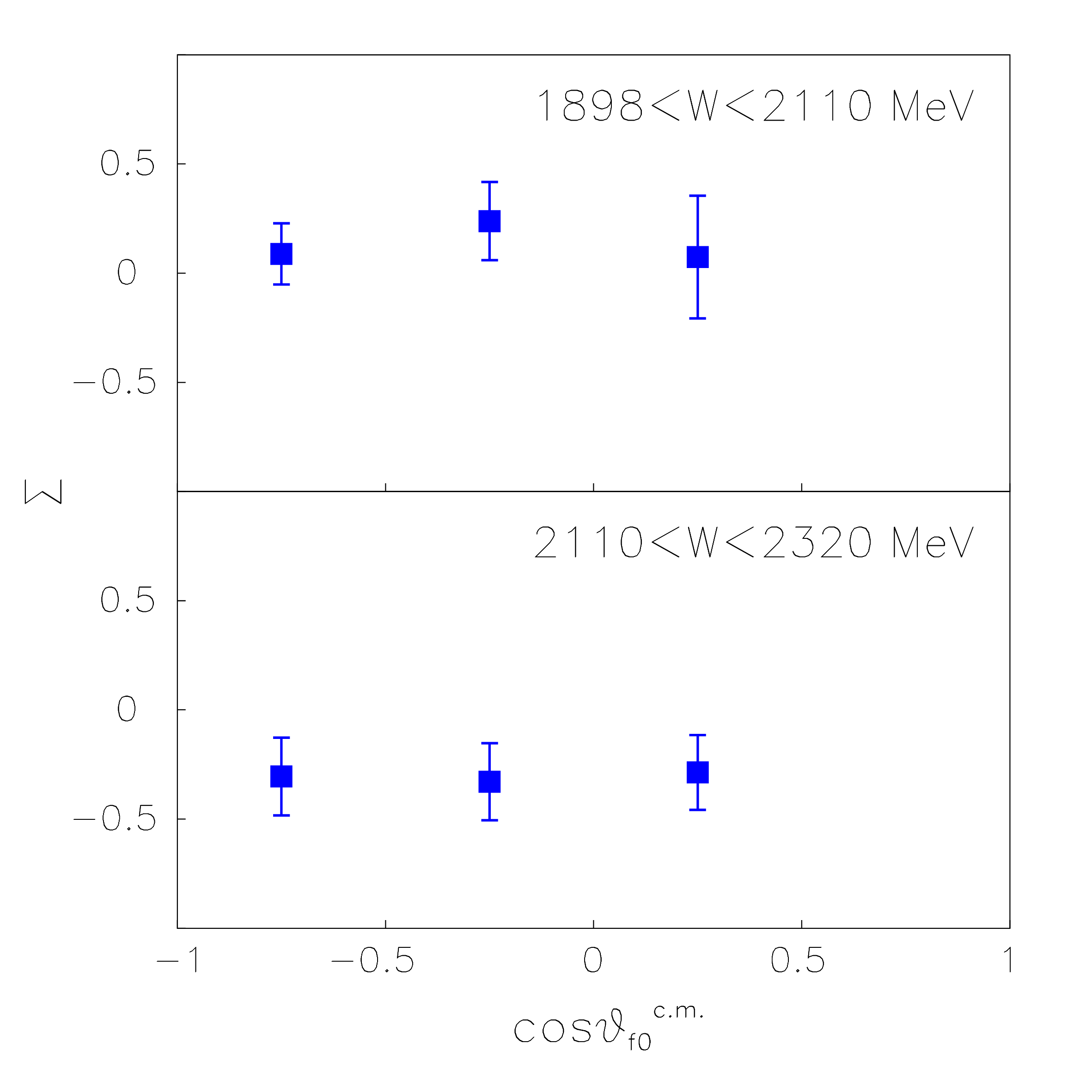}
      \caption{Photon beam asymmetries $\Sigma$ of the reaction $\gamma p \to f_0$(980)$p$ as a function of the $f_0$(980) polar angle. {\it Source}: Figure taken from Fig.~4 of Ref.~\cite{rf2023}.}
      \label{f0pba}
    \end{minipage}
  \end{tabular}
\end{figure}

In the BGOegg experiment, the $f_0$(980) meson was identified by detecting its decay into $\pi^0 \pi^0 \to \gamma \gamma \gamma \gamma$. Although the differential cross sections of $f_0$(980) photoproduction have been measured using the $f_0$(980) decays into $\pi^+ \pi^-$ by the CLAS collaboration \cite{f0clas}, this analysis has suffered from the large contribution of $\rho$ photoproduction and the influence of so-called $S$-$P$ interference. In contrast, the $\pi^0 \pi^0$ decay mode has no influence of them because the $\rho \to \pi^0 \pi^0$ decay is prohibited. In the BGOegg experiment, the $f_0$(980) peak in the $\pi^0 \pi^0$ invariant mass distribution was clearly observed with simple event selection \cite{f0egg}, for the first time in photoproduction, as shown in Fig.~\ref{f0mass}. Here, a kinematic fit was applied to ensure the reaction $\gamma p \to \pi^0 \pi^0 p \to \gamma \gamma \gamma \gamma p$ and improve the resolutions of $\gamma$ four-momenta. Signal amounts were extracted by fitting the sum of Voigt (signal) and fourth-order polynomial (background) functions to the $\pi^0 \pi^0$ invariant mass distributions.

Differential cross sections $d\sigma / dt$ and photon beam asymmetries $\Sigma$ were measured in the two total energy ranges, $1898 < W < 2110$ and $2110 < W < 2320$~$\mathrm{MeV}$, as shown in Figs~\ref{f0dcs} and \ref{f0pba}. The $\Sigma$'s of $f_0$(980) photoproduction were, for the first time, measured to investigate its reaction mechanism in the available range of $f_0$(980) polar angles, $\cos \theta^{c.m.}_{f_0}$. They are close to zero or positive in the lower energy region, whereas the higher energy region shows negative $\Sigma$ values around $-0.3$, indicating the increase of $t$-channel vector meson (natural parity) exchange in the scalar meson photoproduction \cite{sigmasign}. Note that Ref.~\cite{f0theory} has proceeded with the calculation using the $t$-channel exchange of $\rho$ and $\omega$ mesons. The $d\sigma / dt$'s of the reaction $\gamma p \to f_0$(980)$p \to \pi^0 \pi^0 p$ were obtained as a function of $-t$ in the same energy ranges. The measurement in the higher energy region, where a large contribution from the $t$-channel vector meson exchange was confirmed, resulted in the values around $0.15$~$\mathrm{\mu b / GeV^2}$ at $-t<1$~$\mathrm{GeV^2}$. These values are comparable to the prediction in Fig.~4 of Ref.~\cite{f0theory}, which is calculated for the photon beam energy of $3.5$~$\mathrm{GeV}$. The experimental result indicates that the measured differential cross sections relatively favor the model implying a non-exotic $q\bar{q}$ structure for $f_0$(980).

\subsubsection{Bose-Einstein correlation of $\pi^0 \pi^0$}
Another example of unique analyses using the BGOegg data is the study of Bose-Einstein correlation (BEC) for $\pi^0 \pi^0$ pairs, contained in the final state of the $\gamma p  \to \pi^0 \pi^0 p$ reaction at $E_\gamma = 1.3$--$2.4$~$\mathrm{GeV}$. This study was carried out using the proton target data to investigate the space-time properties of a reaction volume, taking into account $\pi^0 \pi^0$ strong final-state interaction (FSI) through $f_0$(500) and $f_0$(980) \cite{bec1}. The BEC effect was measured based on the correlation function (CF), which was dependent on the Lorentz-invariant relative momentum $Q$ ($Q^2=-(p_1-p_2)^2$ for two identical particles possessing the four-momenta $p_1$ and $p_2$) and was expressed by the ratio of the count of $\pi^0 \pi^0$ pairs in the experimental data to that of $\pi^0 \pi^0$ combinations from different events in the event mixing method \cite{bec2, bec3, bec4, bec5}. In the BGOegg data, enhancement due to the quantum statistics arising from BEC was observed at a small $Q$ region, and the Lednick\'{y} parameterization \cite{bec6, bec7} was considered for correlation functions to incorporate the BEC effect with strong FSI:
\begin{equation} \label{Lednicky}
 C_{Lednicky}(k^*) = 1 + \lambda_2 e^{-4 {k^*}^2 {r_0}^2} + \lambda_2 \alpha \left[ \left| \frac{f(k^*)}{r_0} \right|^2 + \frac{4 \Re f(k^*)}{\sqrt{\pi} r_0} F_1(2 k^* r_0) - \frac{2 \Im f(k^*)}{r_0} F_2(2 k^* r_0) \right] ,
\end{equation}
where the quantity $k^* = Q/2$ is the momentum of one particle in the $\pi^0 \pi^0$ rest frame, $r_0$ the Gaussian radius of the pion emitting source, and $\lambda_2$ the correlation strength varying from $0$ to $1$, corresponding to totally coherent or chaotic emission. The functions contained in Eq.~\ref{Lednicky} are defined by $F_1(z) = \int_0^z \frac{e^{x^2 - z^2}}{z} dx$ and $F_2(z) = \frac{1 - e^{-z^2}}{z}$, while $f(k^*) = \frac{f_0(k^*) + f_1(k^*)}{2}$ represents the scattering amplitude, where $f_0(k^*)$ and $f_1(k^*)$ originate from $f_0$(500) and $f_0$(980), respectively. The parameter $\alpha$ is set to $0.5$ by assuming the symmetry in $\pi^0 \pi^0$ production.

\begin{figure}[t!]
  \centering
  \includegraphics[width=16cm]{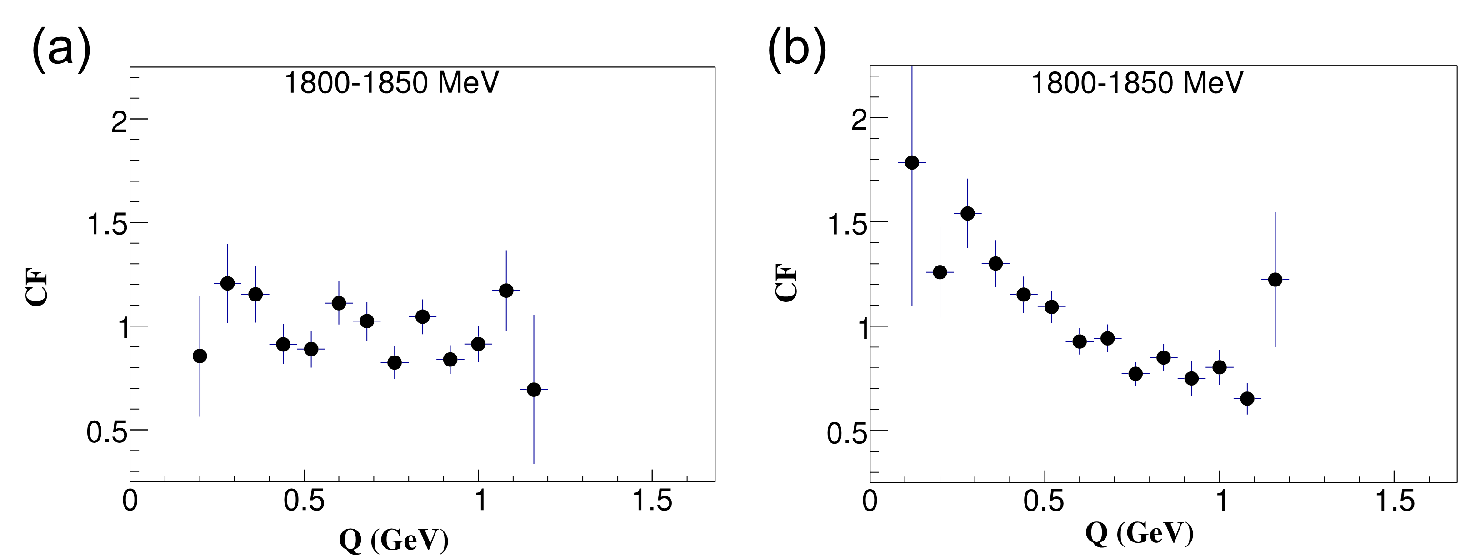}
  \caption{Correlation functions of $\pi^0 \pi^0$ in the photoprodution $\gamma p \to \pi^0 \pi^0 p$ at $E_\gamma = 1800$--$1850$~$\mathrm{MeV}$ \cite{bec1}. Panels (a) and (b) show the first scheme enhancing the $\pi^0 \Delta$ sequential decay contribution and the second scheme suppressing the $\pi^0 \Delta$ contribution, respectively.}
  \label{becfig}
\end{figure}

The $\pi^0 \pi^0$ correlation was measured in two different event selection schemes to compare the correlation strengths in the different levels of $\pi^0 \Delta$ sequential decay contribution. In Fig.~\ref{becfig}, the first scheme shown in panel (a) focused on the $\pi^0 \Delta$ sequential decay by requiring $\left| m(p, \pi^0_{low}) - 1232 \right| < 50$~$\mathrm{MeV}$, where $\pi^0_{low}$ represents the pion with a relatively lower momentum among $\pi^0 \pi^0$ pairs. In contrast, the second scheme shown in panel (b) tried to suppress the $\pi^0 \Delta$ contribution with the condition $\left| m(p, \pi^0_{low}) - 1232 \right| > 50$~$\mathrm{MeV}$. A preliminary result indicated that the strength of the $\pi^0 \pi^0$ correlation in the second scheme (enhancement at small $Q$ values) is greater than that in the first scheme, indicating the BEC effect of $\pi^0 \pi^0$ arising from $\pi^0 \Delta$ sequential decays is weak \cite{bec1}. In the second scheme, the Lednick\'{y} parameterization fit provides the values of the Gaussian radius $r_0$ and the correlation strength $\lambda_2$ to be $r_0 = 0.33298 \pm 0.00004$ and $\lambda_2 = 0.06 \pm 0.01$ \cite{bec1}.

\subsection{Exotic hadron studies in Solenoid experiment}\label{subsec:third-fourth}
The LEPS2 solenoid experiment can detect both neutral and charged particles over a large solid angle, which opens new opportunities for hadron physics with photoproduction reactions. In particular, because charged tracks can be measured with high acceptance, the solenoid experiment plays a complementary role to the BGOegg experiment, which is optimized for the precise measurement of neutral final states. Searches for exotic hadrons and hadronic states are among the most important physics goals of the solenoid experiment. 

\subsubsection{$\bar{K}NN$ bound state}
The $\bar{K}N$ interaction is known to be strongly attractive, especially in the $I=0$ channel, suggesting the existence of meson-baryon bound states or systems where a meson is bound to a nucleus. As a first step toward exploring such phenomena, the three-body $\bar{K}NN$ system has been a primary target of investigation. Whether it forms a bound state has been a long-standing central question in few-body physics within low-energy QCD. 
If such a state is confirmed, it would represent a new type of nucleus in which a meson is an essential constituent. If the antikaon is deeply bound and pulls nucleons into a compact configuration, it could also imply unusually dense baryonic matter. Such systems may become a unique laboratory to investigate the behavior of hadrons in dense matter.

Among $\bar{K}NN$ candidates, the simplest is the ``$K^-pp$'' system, often discussed with quantum numbers $I=1/2$ and $J^P=0^-$. Early theoretical work predicted binding energies ($B_K$) of a few tens of MeV and decay widths ($\Gamma$) of a similar scale---for example, $B_K \simeq 50$~MeV and mesonic decay widths of $\mathcal{O}(10$--$100)$~MeV (see, e.g., \cite{Akaishi:2002bg}). More recent calculations based on chiral effective approaches often predict shallower binding and/or broader widths. The predicted binding energy and width are still not settled (see reviews such as \cite{Hyodo:2022xhp}). Experiments are eagerly awaited to establish the existence or non-existence of $K^-pp$ and, if it exists, to determine its binding energy and width, thereby testing the validity of the theoretical calculations.

Experimental searches for $K^-pp$ grew rapidly in the early 2000s. The system is expected to contain a sizable $\bar{K}N$ component with $I=0$, and several decay modes such as $\Lambda + p$ (and possibly $\Sigma^0 + p$) are experimentally accessible. If multi-nucleon absorption does not completely dominate, a peak-like structure may appear in invariant-mass spectra of the decay particles such as hyperons and nucleons, which has encouraged many searches using different probes.
One of the earliest notable reports came from FINUDA at DA$\Phi$NE \cite{FINUDA:2005lqd}. They used low-energy $K^-$ from $\phi$ decays on nuclear targets ($^{6}$Li, $^{7}$Li, $^{12}$C, $^{27}$Al, and $^{51}$V) and studied back-to-back $\Lambda + p$ pairs. They reported a peak-like structure in the $\Lambda + p$ invariant-mass distribution corresponding to a binding energy of $B_K = 115 ^{+9}_{-5} \text{ (stat.)} ^{+3}_{-4} \text{ (sys.)}$~MeV and a decay width of $\Gamma = 67 ^{+14}_{-11} \text{ (stat.)} ^{+2}_{-3} \text{ (sys.)}$~MeV. This triggered many follow-up searches.

Later, high-intensity hadron beams at J-PARC led to further progress in the search for $K^-pp$. The E27 experiment studied $\pi^+ + d \to K^+ + X$ at $p_{\pi^+} \simeq 1.69$~GeV/$c$ \cite{Ichikawa:2014ydh}, and the E15 experiment studied $K^- + {}^{3}\mathrm{He} \to \Lambda + p + n$ at $p_{K^-} \sim 1$~GeV/$c$ \cite{J-PARCE15:2018zys}. Both reported structures below the $K^-pp$ threshold. The extracted binding energies and widths differ between experiments: J-PARC E15 reported $B_K = 47 \pm 3 \text{ (stat.)} ^{+3}_{-6} \text{ (sys.)}$~MeV with a decay width of $\Gamma = 115 \pm 7 \text{ (stat.)} ^{+10}_{-20} \text{ (sys.)}$~MeV, whereas E27 reported $B_K = 95 \pm 18 \text{ (stat.)}$~MeV with $\Gamma = 162 \pm 87 \text{ (stat.)}$~MeV. It remains an open question whether these signals represent a single reaction-dependent state or multiple distinct states. This motivates further systematic studies.

Photoproduction provides an important complementary approach because the production mechanisms and background sources differ from those in hadron-induced reactions. The first photoproduction search was performed by LEPS \cite{LEPS:2013dqu}. They studied $\gamma + d \to K^+ + \pi^- + X$ and searched for a peak structure in the inclusive spectrum of $X$. Since photoproduction cross sections are generally smaller than in hadron-induced reactions, LEPS combined data from multiple running periods (2002, 2003, 2006, and 2007) to increase statistics. They did not observe a clear peak and set an upper limit at the level of a few percent relative to reference channels such as $\gamma + d \to K^+ + \pi^- + Y^\ast$ \cite{LEPS:2013dqu}. A limitation of this method is the requirement of an associated pion, which can increase backgrounds from excited-hyperon decays and other sources.

To overcome these limitations, the LEPS2 Solenoid experiment extends this study by aiming at a more direct, exclusive, and kinematically clean search for the $K^-pp$. The large acceptance of the LEPS2 solenoid spectrometer is highly advantageous for detecting multi-particle final states with high efficiency.
The main reaction is
\begin{equation}
  \gamma + d \to K^0 + (K^-pp), \qquad K^-pp \to \Lambda + p,
\end{equation}
where the $\Lambda + p$ final state is measured explicitly to enhance the signal and reduce backgrounds. 

\begin{figure}[h]
  \centering
  \includegraphics[width=15cm]{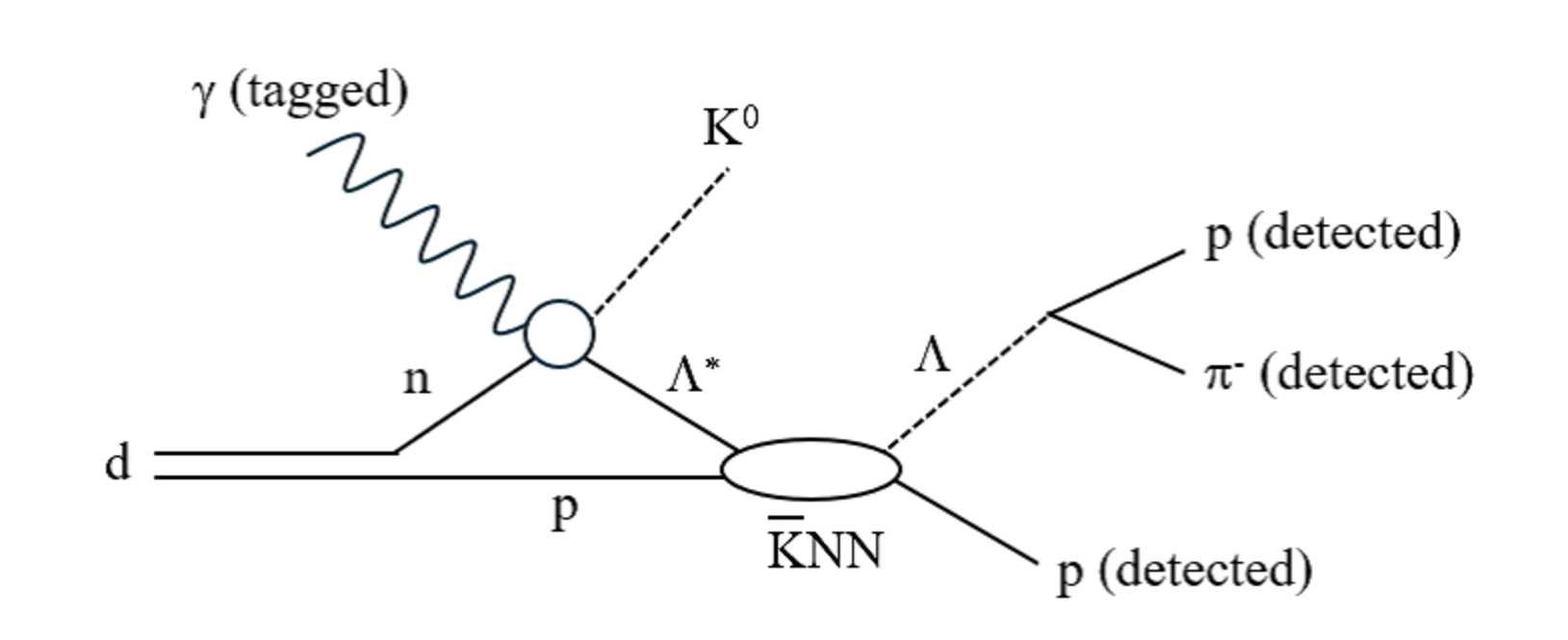}
  \caption{A schematic diagram of the $K^-pp$ state production to be investigated in the LEPS2 Solenoid experiment.}
  \label{fig:diagram_kpp}
\end{figure}

The production mechanism can be understood through the ``$\Lambda(1405)$ doorway'' picture, as illustrated in Fig.~\ref{fig:diagram_kpp}. In this process, the incident photon interacts with a neutron in the deuteron to produce a forward-going $K^0$ alongside an intermediate $\Lambda(1405)$ state. Because the $\Lambda(1405)$ acts as a spatially extended $\bar{K}N$ cluster, it effectively supplies a virtual antikaon. Driven by the strongly attractive $\bar{K}N$ interaction, this virtual antikaon rapidly pulls the nearby spectator proton into the cluster, facilitating the fusion into the deeply bound $K^-pp$ state. 

In the forward-angle kinematics of $K^0$ where the momentum transfer is small, the reaction is dominated by a $t$-channel process involving the exchange of a virtual meson between the incident photon and the target nucleon. 
A characteristic theoretical advantage of this $\gamma \to K^0$ photoproduction is that the electromagnetic $\gamma K^0 K^0$ coupling is forbidden by parity conservation. 
Consequently, the exchange of a spin-0 meson ($K$) is suppressed, meaning that a spin-1 meson exchange ($K^\ast$) dominate the reaction. 
Furthermore, if the $K^-pp$ state possesses the theoretically favored quantum numbers of $J^P = 0^-$, angular-momentum matching in this two-body scattering picture naturally disfavors spin-0 exchange, making the spin-1 $K^\ast$ exchange the most natural mechanism.

To experimentally verify this theoretical picture and lock down the spin-parity of the $K^-pp$ state, the use of a linearly polarized photon beam is crucial. The polarization observable, known as beam asymmetry, manifests as a difference in the $K^0$ production yield depending on whether the $K^0$ is emitted parallel or perpendicular to the photon's electric field vector. This asymmetry is highly sensitive to the spin of the exchanged virtual meson. By measuring the beam asymmetry, we can experimentally distinguish whether the $K$ or $K^\ast$ exchange dominates. Because the spin and parity of the exchanged meson tightly restrict the allowed quantum numbers of the final $K^-pp$ cluster, this polarization measurement serves as a definitive probe to indirectly establish the spin-parity assignment of the newly observed state.

A detailed analysis of the dataset and the search results are reported in Ref.~\cite{LEPS2Solenoid:2026obe}. In this analysis, the $\Lambda$ is reconstructed via its $\Lambda \to p + \pi^-$ decay using tracking detectors, combined with secondary-vertex reconstruction and an invariant-mass selection. The $K^0$ kinematics can be constrained via the missing mass. In the present approach, the simultaneous production of $\Lambda$ and $K^0$ is identified using a two-dimensional distribution of the $p+\pi^{-}$ invariant mass and the missing mass in the $\gamma + d \to \Lambda + p + X$ reaction. Thus, the event selection requires two protons and one $\pi^-$ detected in the Time Projection Chamber (TPC). The $\Lambda$ baryon is reconstructed by demanding that the decay vertex formed by a $p\pi^-$ pair is spatially detached from the beam axis.

\begin{figure}[h]
  \centering
  \includegraphics[width=15cm,angle=-90]{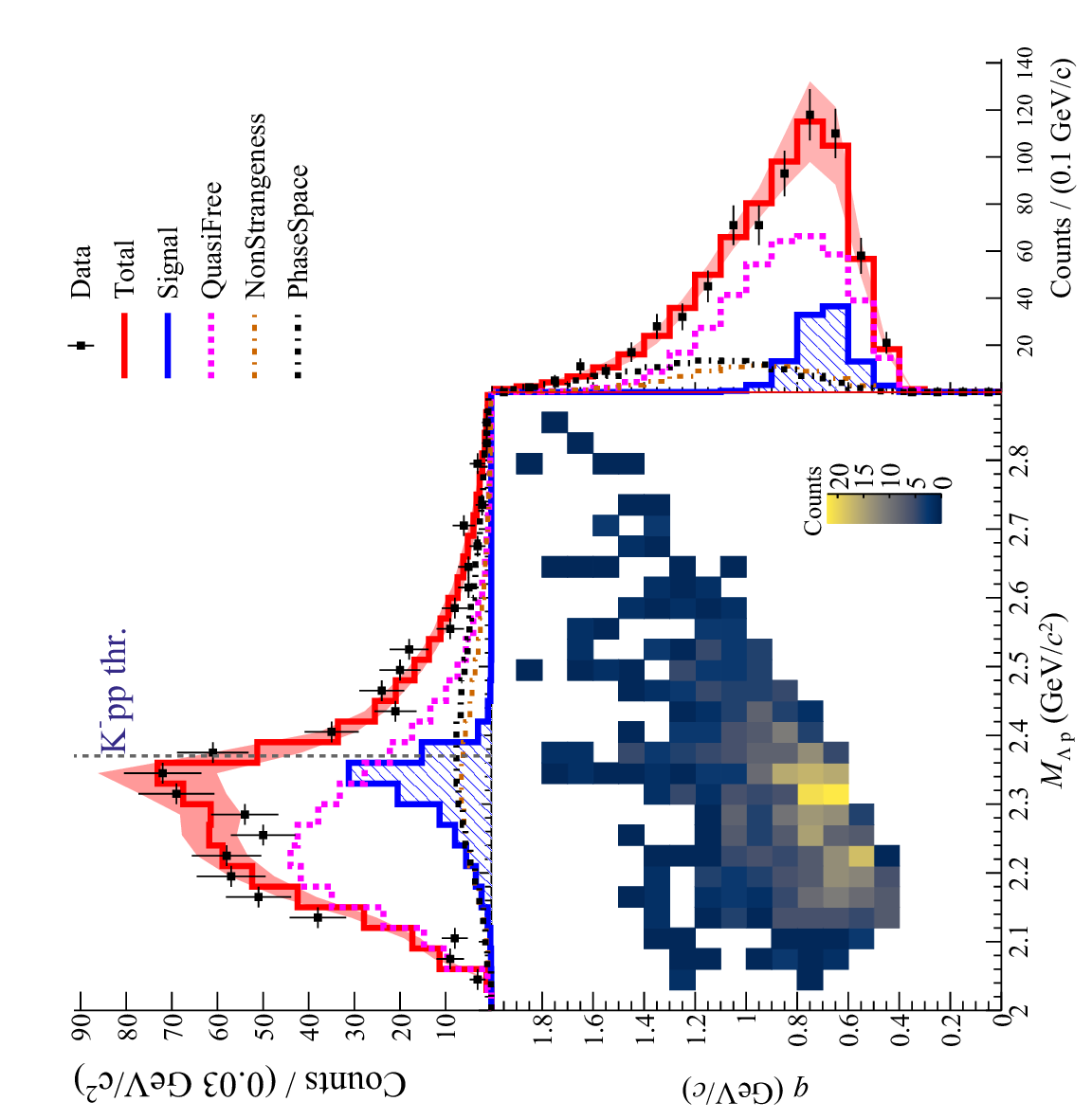}
  \caption{A two-dimensional distribution of the $\Lambda p$ invariant mass ($M_{\Lambda p}$) and the momentum transfer ($q$) to the $\Lambda p$ system and the fitted result from the LEPS2 Solenoid experiment. A localized enhancement corresponding to the $K^-pp$ signal is observed below the mass threshold. {\it Source}: Figure taken from Ref.~\cite{LEPS2Solenoid:2026obe}.}
  \label{fig:kpp_fit}
\end{figure} 

To refine the event selection and improve the momentum resolution, a kinematic fit is applied. The fit imposes two constraints: (i) four-momentum conservation for the $\gamma d \to \pi^- p p X$ reaction, with the missing mass fixed to the $K^0$ mass ($M_X = M_{K^0}$), and (ii) an invariant-mass constraint for the $\Lambda$ baryon ($M_{p\pi^-} = M_\Lambda$). According to Monte Carlo simulations, this kinematic fit significantly improves the $\Lambda p$ invariant mass resolution from 30~MeV/$c^2$ to 11~MeV/$c^2$, which is crucial for searching for the $K^-pp$ state.

Building upon this exclusive event selection, a $K^-pp$ signal is searched for by analyzing the two-dimensional distribution of the $\Lambda p$ invariant mass ($M_{\Lambda p}$) and the momentum transfer ($q$) to the $\Lambda p$ system, defined as $q \equiv |\vec{p}_\gamma - \vec{p}_{K^0}| = |\vec{p}_{\Lambda p}|$. Quasi-free background processes on the nucleons contained in the deuteron (such as $\gamma N \to K^0 Y$ accompanied by a spectator nucleon) exhibit a broader distribution. By utilizing these $M_{\Lambda p}$ and $q$ distributions, the background processes were carefully evaluated and disentangled. The two-dimensional distribution of $M_{\Lambda p}$ and $q$, obtained after the kinematic fit, is shown in Fig.~\ref{fig:kpp_fit}. 

In this two-dimensional plot, a distinct localized cluster of events appears below the $K^-pp$ mass threshold at a low momentum transfer region ($q \simeq 0.6$--$0.8$~GeV/$c$). This specific localized enhancement cannot be explained by standard background processes---such as quasi-free $K^0\Lambda$ photoproduction or purely phase-space distributions---which would spread over much broader regions in $M_{\Lambda p}$ and $q$.
To quantitatively evaluate this structure and disentangle the signal from backgrounds, an extended maximum-likelihood fit is applied to the two-dimensional histogram in the region of $2.0 < M_{\Lambda p} < 2.9$~GeV/$c^2$ and $0 < q < 2.0$~GeV/$c$. The fit model utilizes four Monte Carlo-generated templates: 
(i) quasi-free $\gamma n \to K^0 \Lambda$ production with a spectator proton, 
(ii) a purely three-body phase-space process for $\gamma d \to K^0 \Lambda p$, 
(iii) a non-strange background estimated from the $\Lambda$ sidebands, and 
(iv) a signal component representing the $K^-pp$ formation. 
The signal template is modeled as a Voigt profile in $M_{\Lambda p}$ generated by reweighting the template of the process (ii).

The fit result yields a $K^-pp$ signal with a mass of $M = 2.354 \pm 0.011 \text{ (stat.)} ^{+0.009}_{-0.005} \text{ (syst.)}$~GeV/$c^2$ (corresponding to a binding energy of $16$~MeV) and a width of $\Gamma = 0.055 \pm 0.023 \text{ (stat.)} ^{+0.035}_{-0.009} \text{ (syst.)}$~GeV/$c^2$, achieving a local statistical significance of $7.3 \sigma$. 
Compared to the previous measurement by the J-PARC E15 experiment, the obtained mass is closer to the $K^-pp$ threshold, while the width are consistent within errors. It remains an open question for future theoretical and experimental studies whether this mass difference originates from reaction-dependent dynamics, interference effects, or other underlying mechanisms.

In future, the measurement of photon beam asymmetry will be done, as its importance has been discussed earlier. As a next step, a systematic study of the $\bar{K}NN$ state is planned. Taking advantage of the barrel detector's high efficiency for gamma-ray detection, future analyses will include searches in the $\Sigma^0 + p$ decay channel (measured via the $\Lambda + \gamma + p$ final state) as well as mesonic decay modes such as $\Lambda + p + \pi^0$. Furthermore, the search for the isospin partner state, $K^-pn$, via the $\gamma + d \to K^+ + X$ reaction is also anticipated. These current findings for the $K^-pp$ state, together with multifaceted searches across different production and decay modes, are highly complementary to future planned experiments such as J-PARC E80~\cite{J-PARCE80:2026jjt}. Comprehensive investigations remain crucial for achieving a unified understanding of antikaon-nuclear interactions.

\subsubsection{Searches for other exotic hadrons}

In addition to the aforementioned topics, the LEPS2 is planning other physics programs to study exotic hadrons. For instance, studies on the $\Lambda(1405)$ structure and searches for the $\Theta^+$ pentaquark are considered as highly important topics. While these subjects have been previously investigated in the LEPS experiment~\cite{LEPS:2003wug,LEPS:2008ghm,Niiyama:2008rt}, clarifying their nature requires the data collection with higher statistics and a large acceptance covering a wide kinematic region. Therefore, they are planned as major subjects in LEPS2.

The $\Lambda(1405)$ is a representative exotic hadron that has been continuously studied for decades. Its exceptionally low mass cannot be easily explained by an ordinary three-quark configuration. 
Instead, it is widely described by a two-pole structure arising from the coupled-channel $\bar{K}N$ and $\pi\Sigma$ scattering dynamics, often interpreted as having a large hadronic molecule or pentaquark-like component. 
The current understanding of this state, from both theoretical and experimental viewpoints, is comprehensively reviewed in Ref.~\cite{Hyodo:2020czb}. 
A key feature of the $\Lambda(1405)$ is that its observed lineshape in the $\pi\Sigma$ invariant mass spectrum is strongly dependent on the reaction process. 
Specifically, the isospin interference term between the $I=0$ and $I=1$ amplitudes causes distinct differences in the lineshapes of the charged $\pi^+\Sigma^-$ and $\pi^-\Sigma^+$ final states, whereas the neutral $\pi^0\Sigma^0$ channel is completely free from this interference contribution.

Experimentally, differences in the lineshapes among the different final states were first observed by the LEPS collaboration~\cite{Niiyama:2008rt} and later confirmed with high statistics by the CLAS experiment~\cite{CLAS:2013rjt}. 
However, the experimental results from both collaborations contradict earlier theoretical predictions. 
Although it is possible that this discrepancy arises from the different kinematic regions covered by the experiments, the situation remains unsettled. 
Because different theoretical models predict varying shifts in these spectra, a systematic study to resolve these discrepancies is highly motivated.
The LEPS2 Solenoid experiment is uniquely equipped to address this issue. Its advanced combination of charged-particle tracking and wide-acceptance gamma-ray counters enables the explicit detection of the purely neutral decay channel, $\pi^0\Sigma^0$. Since the $I=1$ $\Sigma(1385)$ cannot decay into a $\pi^0\Sigma^0$ pair, this neutral spectrum acts as a pure $I=0$ amplitude, completely isolating the $\Lambda(1405)$ from the dominant $I=1$ background. By systematically measuring this clean, background-suppressed channel and combining it with polarization observables such as beam asymmetry, LEPS2 aims to rigorously constrain the underlying reaction dynamics and reveal the true nature of the $\Lambda(1405)$.

As another attractive exotic hadron, the $\Theta^+$ pentaquark is also a major target.
To settle the long-standing debate over the existence of the $\Theta^+$ pentaquark, the LEPS2 Solenoid setup provides a decisive capability for the examination. A pentaquark is an exotic hadron consisting of four quarks and one antiquark. 
In the case of $\Theta^+$, its quark composition is $uudd\bar{s}$.
Unlike recently discovered heavy-flavor pentaquarks at LHCb (e.g., $P_c$ states~\cite{LHCb:2015yax,LHCb:2019kea})---which are characterized by hidden charm and are often interpreted as hadronic molecules near thresholds---the $\Theta^+$ pentaquark poses a distinctly different challenge. 
While pentaquark candidates with hidden strangeness (containing an $s\bar{s}$ pair) are also discussed in the light-quark sector, the $\Theta^+$ is a ``genuine'' pentaquark because its minimal quark content ($uudd\bar{s}$ with strangeness $S=+1$) explicitly forbids a standard three-quark configuration via $q\bar{q}$ annihilation.
If such a narrow, genuine exotic state exists, it would strongly suggest short-range multiquark dynamics, such as diquark correlations, offering a crucial probe for understanding color confinement and non-perturbative QCD.

The existence of $\Theta^+$ has extensively been discussed so far.
The LEPS experiment first reported evidence for a narrow structure near $M \simeq 1540~\mathrm{MeV}/c^2$ in $\gamma n \to K^- \Theta^+$~\cite{LEPS:2003wug}, and later supported it with a deuterium measurement~\cite{LEPS:2008ghm}. However, the high-statistics CLAS experiment at JLab did not observe the signal~\cite{CLAS:2003yuj,CLAS:2006czw}, leaving the situation unsettled. This discrepancy has been widely debated in terms of the differences in angular acceptance (LEPS covers extreme forward angles, while CLAS covers wider angles) and the analysis methods for handling the spectator nucleon, such as the reliance on a Fermi-motion correction.

The LEPS2 Solenoid experiment is designed to overcome these historical limitations. It combines a large-acceptance detector covering the angular ranges of both LEPS and CLAS, providing high statistics and excellent mass resolution. 
This unique coverage enables us to systematically investigate the discrepancies between the previous LEPS and CLAS reports.
The main LEPS2 searches utilize reactions where all related final-state particles can be fully reconstructed as charged tracks. 
The primary targeted processes are $\gamma + n \to K^- + \Theta^+$ and $\gamma + p \to \bar{K}^{*0} + \Theta^+$. 
In both cases, the $\Theta^+$ is reconstructed via the $\Theta^+ \to K_S^0 + p \to \pi^+\pi^- p$ decay chain. 
Detecting a $K^-$ in the final state is particularly important because it directly tags the production of an $S=+1$ baryon at the event-selection level. 
Furthermore, many previous searches relied on detecting $K^+K^-$ pairs, which inherently suffer from massive backgrounds arising from $\gamma + p \to \phi + p$ ($\phi \to K^+K^-$). 
By explicitly using the $K_S^0 + p$ mode as the primary channel, LEPS2 can effectively avoid this $\phi$-meson background.

Crucially, this complete kinematic reconstruction of all final-state particles allows the analysis to bypass the ambiguities of Fermi-motion corrections, offering a much cleaner signal. 
Assuming a production cross section of $4~\mathrm{nb/sr}$, a 100-day run is expected to yield several hundred events. 
The anticipated $K_S^0 p$ invariant mass resolution is about $6~\mathrm{MeV}/c^2$, which is highly sufficient to efficiently search for such a narrow state~\cite{Nakano:2025ilt}. 
Data acquisition with a deuterium target, specifically optimized for the $\Theta^+$ search, has already commenced, and physics results are expected in the near future.

\newpage
\section{Development for future photoproduction experiments}\label{sec:fourth}
   As overviewed in the previous section, the LEPS2 project has carried out various physics programs for photoproduction reactions mainly in the center-of-mass energy range of $1.82$--$2.32$~$\mathrm{GeV}$. For example, in the case of two-body meson photoproduction $\gamma p \to X p$, a producible meson mass $M_X$ is limited up to $1.38$~$\mathrm{GeV/c^2}$. The introduction of a pulsed laser with the deep UV wavelength of $266$~$\mathrm{nm}$ has recently enabled to increase the center-of-mass energies of reactions up to $2.51$~$\mathrm{GeV}$ providing a moderately high intensity $\gamma$-ray beam, but further energy increase is not likely expected because the shorter wavelength laser with a high power (at least $1$~$\mathrm{W}$) and frequency (a few tens $\mathrm{MHz}$ to CW) output is not available. The plans to construct higher energy storage rings that are open to wide purposes by various users are also limited. On the other hand, it is desired to significantly extend the $\gamma$-ray beam energy range toward the higher side in the next-generation experiments aiming the studies of higher-mass particles like exotic hadrons. Particularly, the storage ring energy of SPring-8 is scheduled to be lowered to $6$~$\mathrm{GeV}$ from the fiscal year of 2028 (SPring-8-II \cite{spring8ii}), so an innovative idea to overcome the current restricted situation has been strongly needed.

\begin{figure}[h]
   \centering
   \includegraphics[width=10cm]{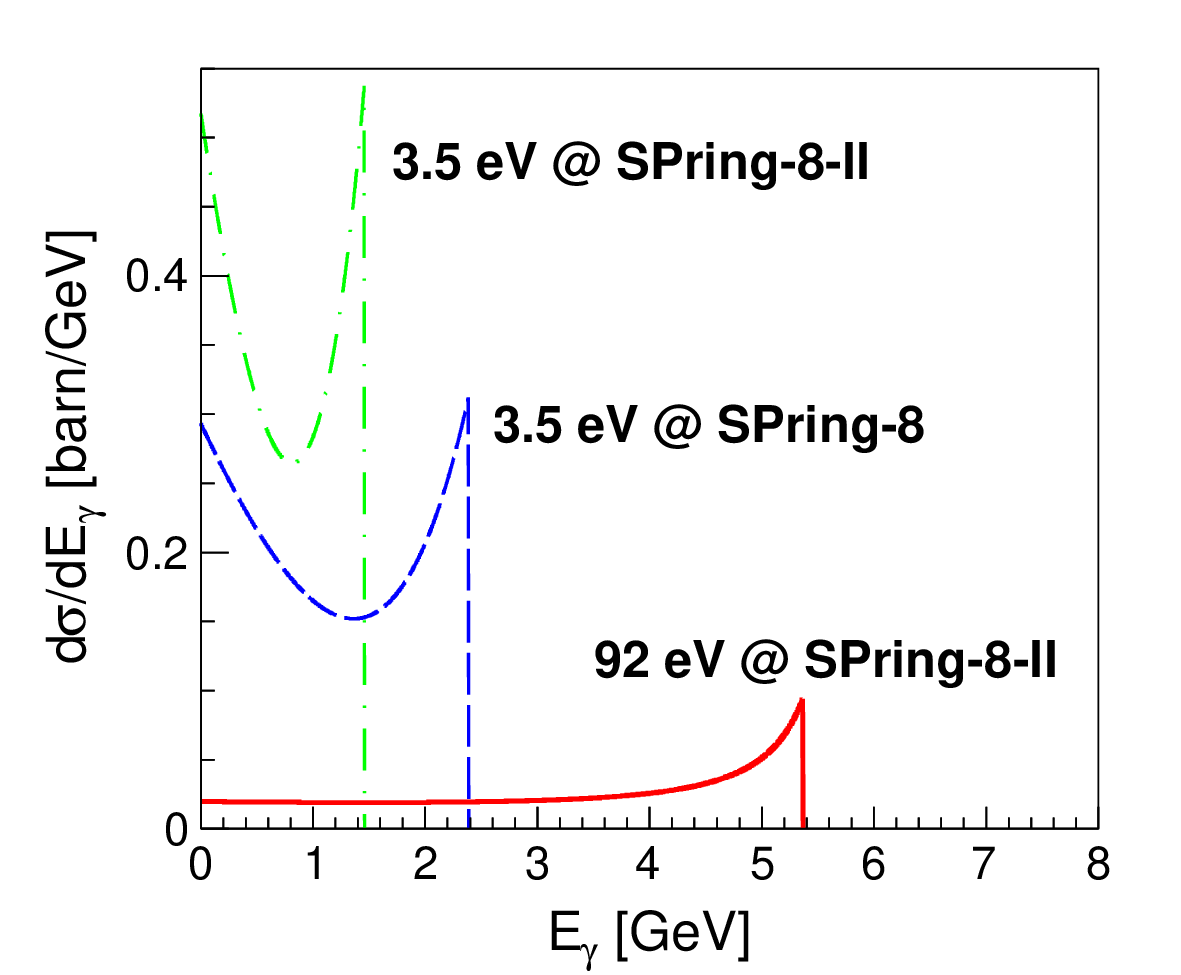}
   \caption{Energy spectra of Compton scattering $\gamma$-rays from leading-order QED calculations. Red solid, blue dashed, and green dash-dotted lines show the injection of $92$~$\mathrm{eV}$ extreme ultraviolet light into SPring-8-II, $3.5$~$\mathrm{eV}$ ultraviolet laser light into SPring-8, and $3.5$~$\mathrm{eV}$ ultraviolet laser light into SPring-8-II, respectively. The vertical axis indicates the scale of differential cross sections.}
   \label{comp_lcs_xcs}
\end{figure}

   To achieve the production of a significantly higher energy $\gamma$-ray beam, the idea of backward Compton scattering with the injection of extreme ultraviolet (EUV) light instead of laser light into an electron storage ring was pursued in the collaboration of experts in the fields of hadron physics via photoproduction experiments, accelerator research and development, and materials science using synchrotron radiation \cite{JPASJ.16.154}. This idea has focused on the recent availability of the Mo/Si multilayer mirror \cite{mosimirror}, which has a high reflectance for $92$~$\mathrm{eV}$ EUV light with the incident angle of $0^\circ$. The $92$~$\mathrm{eV}$ photons can be obtained from an undulator installed in a storage ring, so that it is possible to reflect the radiated photons backwardly to the original ring for the $\gamma$-ray beam production via Compton scattering. This method is unique and cost-effective by completing all the processes at one beamline of a storage ring that is commonly used for many research activities. If such a new $\gamma$-ray beam source is achieved, the maximum energy of $\gamma$-rays ($E_\gamma^{max}$) approaches the storage ring energy as shown in Fig.~\ref{comp_lcs_xcs}. In this figure, the current usual operation of $355$~$\mathrm{nm}$ ($3.5$~$\mathrm{eV}$) laser Compton scattering (LCS) at SPring-8 is indicated to provide $\gamma$-ray energies only up to $2.4$~$\mathrm{GeV}$, while the $E_\gamma^{max}$ for the Compton scattering of $92$~$\mathrm{eV}$ EUV light (EUVCS) reaches $5.4$~$\mathrm{GeV}$ even at SPring-8-II. In the case of $355$~$\mathrm{nm}$ LCS at SPring-8-II, $E_\gamma^{max}$ drops to $1.5$~$\mathrm{GeV}$.

\begin{figure}[h]
   \centering
   \includegraphics[width=16cm]{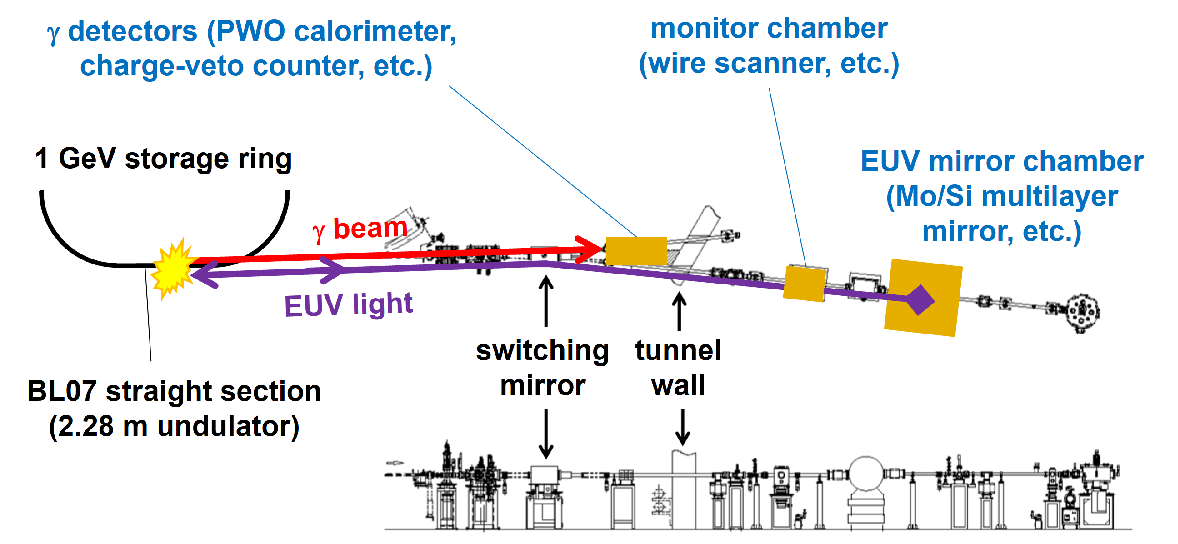}
   \caption{Experimental setup for the Compton scattering of EUV light at NewSUBARU BL07A.}
   \label{xcs_setup}
\end{figure}

   The technical development of the new $\gamma$-ray beam source was conducted in the $1$~$\mathrm{GeV}$ electron storage ring NewSUBARU \cite{newsubaru}, which is easy to use for tests and is located on the same campus as SPring-8. The electron beam current of NewSUBARU reaches $350$~$\mathrm{mA}$ at the beam energy of $0.949$~$\mathrm{GeV}$. Details of this development and experimental results are described elsewhere \cite{prlxcs}. In the beamline BL07, a $2.28$~$\mathrm{m}$-long undulator is equipped and provides the first harmonic radiation of $92$~$\mathrm{eV}$ photons with a flux of $4.11 \times 10^{14}$~$\mathrm{s^{-1} \cdot (0.1\%bandwidth)^{-1}}$. BL07 has two branches for the transportation of X-rays or EUV photons, and either branch is operated at a user experiment by changing the beam path with a switching mirror. The switching mirror is made of a spherical silicon plate with a platinum coating. Its reflectance at $92$~$\mathrm{eV}$ is $88$\% with the incident angle of $87^\circ$. The EUVCS development project has used the branch "A" (BL07A \cite{bl07a}) by constructing the experimental setup as shown in Fig.~\ref{xcs_setup}. All the handling of EUV light at BL07A is done in a ultra-high vacuum of $10^{-6}$~$\mathrm{Pa}$.

   Radiated photons travel BL07A until reaching a large chamber, where a Mo/Si multilayer mirror is placed. The Mo/Si mirror substrate was made of a silicon plate with an area of $50 \times 50$~$\mathrm{mm^2}$, and its reflection surface was polished with the surface roughness enough lower than the EUV wavelength by magnetorheological finishing. The surface shape was made vertically cylindrical with the curvature radius of $16.7$~$\mathrm{m}$, corresponding to the distance to the Compton scattering point. This shape was determined by a ray-tracing simulation so as to make a focus of returning light for the efficient collision with electrons. The reflection surface was coated with 40 periodic layers of molybdenum and silicon pairs. The reflectance of the produced Mo/Si multilayer mirror was measured to be nearly $70$\% for $92$~$\mathrm{eV}$ EUV light. The mirror substrate was attached to a water-cooled pure copper holder for the removal of heat load due to higher harmonic radiation. The holder was further mounted on two precision rotary stages to remotely adjust the vertical and horizontal angles of a reflected EUV photon beam. The optimization of the reflected beam direction can be done by monitoring the positions of both radiated and reflected photon beams with a wire scanner and overlapping the two beam positions precisely. The wire scanner measures the increase of micro-current due to the photoelectric effect by EUV photon hits using tungsten wires which moves in the vertical and horizontal directions.

   Reflected EUV photons are injected into the $6$~$\mathrm{m}$-long straight section of the storage ring where the undulator is set up. Because both sizes of electron and EUV photon beams do not strongly vary for this $6$~$\mathrm{m}$ distance, Compton scattering occurs over the whole range of the straight section. Backwardly scattered $\gamma$-rays in EUVCS are directed to the downstream of BL07, penetrating the switching mirror and the beamline structure to go out into the atmosphere. The $\gamma$-ray detectors are thus placed in an open space where two branch beamlines start to separate from each other. The energies of $\gamma$-rays were measured by an electromagnetic calorimeter made of $3 \times 3$ PWO crystals which have $22.5$ radiation lengths. The scintillation light outputs from the nine crystals were simultaneously read out by a two-inch photomultiplier tube (PMT) for a better energy resolution. In the upstream of the PWO calorimeter, a $3$~$\mathrm{mm}$-thick plastic scintillator was installed to veto charged particle contamination in the $\gamma$-ray beam. The data collection of the $\gamma$-ray detectors were done by the self-triggering signal of the PWO calorimeter.

\begin{figure}[h]
 \centering
 \includegraphics[width=12cm]{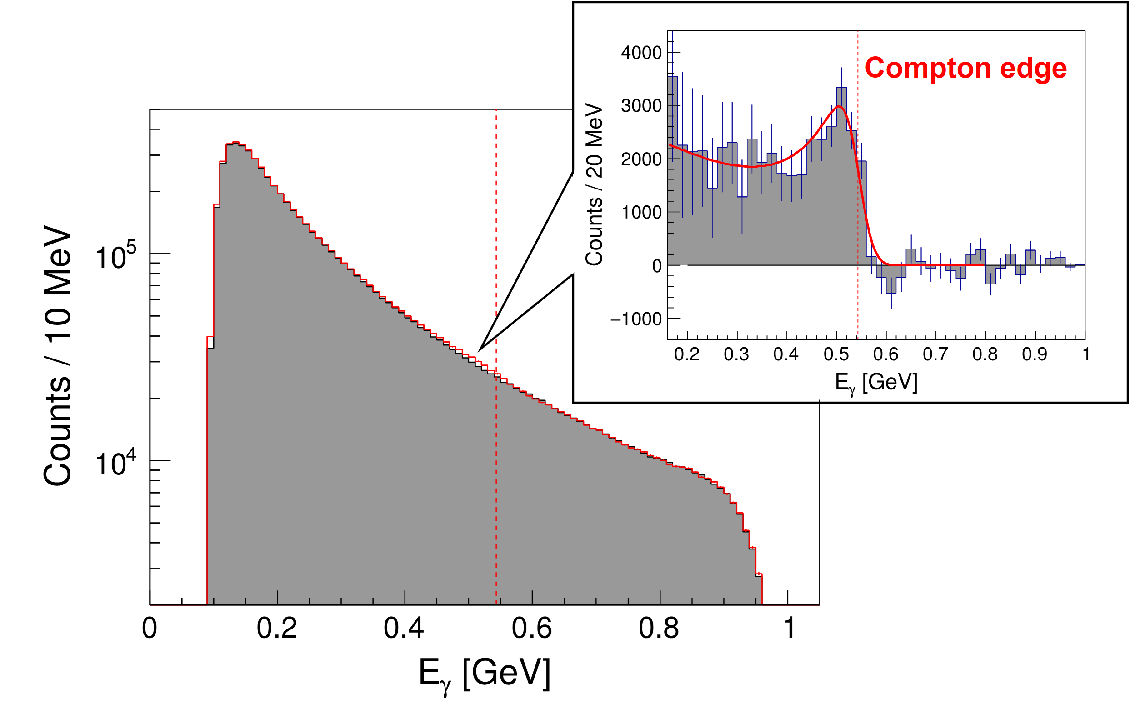}
 \caption{Main panel: Red open and gray filled histograms show the $\gamma$-ray energy spectra measured in the demonstration experiment with and without EUV light reflection, respectively. The gray-filled histogram is scaled to match the event entries in the range above the Compton edge with the corresponding number of the red open histogram. Upper-right panel: A gray-filled histogram shows the difference between the two spectra in the main panel. A red thick curve is a fit result of the theoretical spectrum smeared by the calorimeter energy resolution. {\it Source}: Figures taken from Ref.~\cite{prlxcs}.}
 \label{xcs_results}
\end{figure}

   After developing the above experimental setup with trial and error, demonstration experiments were carried out to prove the feasibility of $\gamma$-ray beam production. In the end of a series of the demonstration experiments (on 19 Dec.~2024), the $\gamma$-ray beam production via EUVCS was observed for the first time at a synchrotron radiation facility, as shown in Fig.~\ref{xcs_results}. The main panel of Figure~\ref{xcs_results} shows the $\gamma$-ray energy spectra with and without EUV light injection into the storage ring (red open and gray filled histograms, respectively). The latter case was examined to know background contribution by disturbing the EUV light reflection process with an optical shutter. As seen in the figure, a large background contribution from the bremsstrahlung of an electron beam through residual gas inside the ring unavoidably exists. The energy calibration of ADC data was done by using the highest edge of this bremsstrahlung, corresponding to the ring energy. A kinematical calculation for the head-on collision of $92$~$\mathrm{eV}$ photons and $0.949$~$\mathrm{GeV}$ electrons predicts the maximum $\gamma$-ray energy (Compton edge) of $0.543$~$\mathrm{GeV}$ at the scattered angle of $0^\circ$, so that the normalization of the two measured spectra was performed using the event entries in the energy region above this Compton edge. The upper-right panel of Figure~\ref{xcs_results} shows the difference between the two spectra in the main panel. A clear $\gamma$-ray energy spectrum corresponding to EUVCS was observed with the shape predicted by a leading-order QED calculation. If the theoretical spectrum with the convolution of the energy resolution of the PWO calorimeter was fitted, the experimental data were well reproduced as shown in the upper-right panel.

   The $\gamma$-ray production rate was measured to be $1.4$~$\mathrm{kcps}$ based on the number of observed excess events, the DAQ efficiency, the data collection time, and the material amount due to the beamline structure between the Compton scattering point and the PWO calorimeter. On the other hand, an ideal production rate can be estimated based on the performance of individual optical components. First, a flux of injected EUV light is evaluated by multiplying the undulator radiation flux by the reflectances of the switching and Mo/Si multilayer mirrors as well as the bandwidth of the Mo/Si mirror reflectance. The luminosity is calculated to be $1.45 \times 10^{32}$~$\mathrm{m^{-2}s^{-1}}$ by taking into account the EUV light flux, the electron beam current, the transverse size of EUV photon and electron beam bunches ($\sigma_x \approx \sigma_y \approx 0.6$~$\mathrm{mm}$), and their collision rate with $2$~$\mathrm{ns}$ intervals. Finally, this luminosity is multiplied by the total cross section of $92$~$\mathrm{eV}$ EUV light Compton scattering from $0.949$~$\mathrm{GeV}$ electrons, which is estimated as $330$~$\mathrm{mb}$. As a result, the ideal rate is evaluated to be $3.2$~$\mathrm{kcps}$, which is comparable to the measured rate. The $\gamma$-ray production rate in the demonstration experiment is quantitatively explainable, so the further development toward a higher intensity should be well controllable.

   This successful observation is significant as the first achievement to accelerate the realization of a $\gamma$-ray beam source by the Compton scattering of EUV light. The new $\gamma$-ray beam source applicable to any storage ring paved the way to approach high energies close to the ring energy. For future use in the next-generation hadron photoproduction experiments, the $\gamma$-ray beam intensity must be raised up 100 times or more. Newly constructed storage rings with low emittance will help increase the luminosity by realizing narrow beams. In the case of SPring-8-II, vertical and horizontal electron beam sizes are designed to be $4$ and $20$~$\mathrm{\mu m}$, which are 150 and 30 times smaller than $0.6$~$\mathrm{mm}$ in the demonstration experiment at NewSUBARU and can contribute to the increase of luminosity in inverse proportion \cite{prlxcs}. Currently, the developed method uses $92$~$\mathrm{eV}$ EUV light, but the photon energy can be appropriately lowered by changing the multilayer mirror coating from Mo/Si to SiC/Mg, for example. In the case of SPring-8-II with $31$~$\mathrm{eV}$ EUV light, the $E_\gamma^{max}$ still reaches $4.4$~$\mathrm{GeV}$ while the linear polarization increases up to $40$\%. This high linear polarization is beneficial because there are no facilities to provide such a polarized $\gamma$-ray beam at the energies above the LEPS / LEPS2 energies except for GlueX, which focuses on the energies around $9$~$\mathrm{GeV}$. A unique facility can be constructed in the future by using the developed method.

\newpage
\section{Summary}\label{sec:99th}

Understanding the non-perturbative nature of low-energy Quantum Chromodynamics (QCD) remains a central challenge in modern physics, particularly regarding hadron structure, the origin of hadron masses, and the existence of exotic multiquark states. The LEPS2 facility at SPring-8 was constructed to address these fundamental questions by utilizing a highly polarized GeV-scale photon beam produced via Laser Compton Scattering (LCS). By injecting ultraviolet laser light($\lambda = 355$~$\mathrm{nm}$) into the 8 GeV electron storage ring, LEPS2 generates a photon beam with energies ranging from 1.3 to 2.4 GeV. 
A distinctive feature of this facility is the high degree of linear polarization, which is crucial for disentangling complex reaction mechanisms and isolating small resonance contributions through interference effects. 

In addition, in LEPS2, by implementing a four-laser injection scheme, it becomes possible to operate with a high-intensity photon beam upto 5\,Mcps. 
Specifically, the collimated beam is transported far downstream (to approximately 130 m) so that a dedicated experimental hall can be constructed, enabling the installation of a large-acceptance detector system that covers a wide solid angle.
To maximize the physics output, the facility operates two complementary detector systems, BGOegg and Solenoid, which are run alternately to cover a wide range of photoproduction physics.

The BGOegg experiment utilizes a large-acceptance electromagnetic calorimeter composed of 1,320 BGO crystals, achieving a world-leading energy resolution with the highest granularity in the detected photon energy region around 1 GeV and below.
It focuses on neutral-meson photoproduction to investigate the in-medium hadron mass and to study baryon resonance spectroscopy.
 A primary objective has been the search for the in-medium mass modification of the $\eta'$ meson, which is linked to the $U_A(1)$ quantum anomaly and the spontaneous breaking of chiral symmetry.
 The Phase-I experiment pursued two complementary approaches using a carbon target: the search for $\eta'$-bound nuclei with the first-time tag of a nucleon absorption signal and the direct measurement of the in-medium $\eta'$ mass via its two-photon decay. The search for bound states yielded upper limits on the formation cross section, disfavoring the mass reduction larger than 100 MeV. On the other hand, the direct measurement in the $\gamma \gamma$ invariant mass spectrum of the low-momentum sample provided evidence of $\sim 4\sigma$ or more for a mass reduction of approximately 60 MeV at the normal nuclear density. Based on these Phase-I results, the Phase-II experiment has recently commenced with an upgraded setup, employing a copper target whose nuclear radius is larger and an additional forward calorimeter to reject the background coming from multi-pion production.
The increased effective target thickness, the enhanced photon-beam intensity by the introduction of a pulsed laser, and the deployment of new detector systems make it possible to investigate mass reduction with improved sensitivity. We are currently taking data and carrying out the analysis.

Parallel to the in-medium $\eta^\prime$ mass studies, the BGOegg experiment has advanced baryon resonance spectroscopy by providing precise data on differential cross sections and polarization observables. 
These results contain the first data on the polarization observables, particularly photon beam asymmetries, for the photoproduction of $\pi^0$, $\eta$, $\eta^\prime$, $\omega$, and multi-mesons in the high-energy region above 2 GeV. The measurements reveal their polar angle dependent behaviors not reproduced by existing partial-wave analyses.
Furthermore, a unique enhancement of differential cross sections was observed in $\eta$ and $\eta^\prime$ photoproduction at higher energies, possibly suggesting high-mass resonances with high spins.
These results improve the current understanding of baryon resonances based on partial wave analyses and emphasize the necessity of polarization data for their establishment.

Complementing the neutral-meson focus of the BGOegg experiment, the LEPS2 Solenoid experiment employs a large-acceptance magnetic spectrometer primarily to investigate the final states with multiple charged particles. The research program places a strong emphasis on the study of exotic hadrons and hadronic states, such as the $\bar{K}NN$ bound state, the $\Theta^+$ pentaquark, and the $\Lambda(1405)$ hyperon.
Using the datasets collected in 2022 and 2023, we have successfully observed the $K^-pp$ quasi-bound state for the first time in photoproduction. This was achieved through the analysis of the exclusive $\gamma d \to K^0 (K^-pp)$ reaction, followed by its decay into $\Lambda + p$. Detailed comparisons with results from other hadron-beam experiments and comprehensive theoretical interpretations of this new observation are anticipated to further elucidate its nature.
Another critical goal of the experiment is to settle the long-standing debate regarding the $\Theta^+$ pentaquark. The existence of the $\Theta^+$ has remained a highly controversial topic since the initial positive reports by the LEPS collaboration and the subsequent contradictory results from the CLAS experiment. The LEPS2 Solenoid setup aims to provide a definitive conclusion on this issue while covering wide polar angles, which contain the sensitive regions of both experiments. By focusing on exclusive reactions such as $\gamma n \to K^- \Theta^+ \to K^- K_S^0 p$ (measured via the $K^- \pi^+ \pi^- p$ final state), the complete final-state kinematics can be reconstructed without relying on Fermi-motion corrections for the target neutron. The data acquisition and physics analysis for the $\Theta^+$ search are currently underway, paving the way to finally resolve this decades-old puzzle.

Looking toward the future, a new $\gamma$-ray beam source to address the energy limitations inherent in the standard laser Compton scattering procedure is being created by developing a novel method based on the backward Compton scattering of extreme-ultraviolet (EUV) light at an undulator beamline of NewSUBARU. In a demonstration experiment, we confirmed $\gamma$-ray production for the first time at a synchrotron radiation facility, successfully validating the method. This breakthrough technology paves the way for extending the maximum photon-beam energy nearly up to the storage ring energy and remaining the LEPS2 facility at the frontier of hadron photoproduction research in the coming decades.

\newpage
\section*{Acknowledgements}
The BGOegg and Solenoid experiments were performed at the BL31LEP (LEPS2 beamline) of SPring-8 with the approval of the Japan Synchrotron Radiation Institute (JASRI) as the project in a contract beamline (Proposal Nos. BL31LEP/6101, 6102, and 6103). The authors thank the members of the LEPS2 collaboration for cooperating the project in the LEPS2 facility. In addition, the authors gratefully acknowledge to the supports of the staff at SPring-8 for providing excellent experimental conditions. The experiments for the XCS development were performed at the BL07A of NewSUBARU with the approval of the Laboratory of Advanced Science and Technology for Industry (LASTI), University of Hyogo. It is a great pleasure to thank the collaborators in this project and the technical staff at NewSUBARU for their cooperation and help. This research was supported in part by the Ministry of Education, Culture, Sports, Science and Technology of Japan. The authors have obtained financial supports from JSPS KAKENHI Grant Nos. 24244022 (NM), 16K13806 (NM), 17H02892 (NM), 18H05325/20K20344 (NM, AT), 20H01933 (NM), 21H04986 (NM, AT, TH), 22K18707 (NM), 22H01225/23K22496 (NM, AT), 23H01201/23K25897 (NM).

\bibliography{refs}


\end{document}